\documentclass[twocolumn]{aastex62}

\newcommand{\Msun}{${\rm M}_{\odot}$}

\newcommand{\lya}{Ly$\alpha$}
\newcommand{\lyman}{Lyman-$\alpha$}
\newcommand{\hi}{\ion{H}{1}}
\newcommand{\hii}{\ion{H}{2}}

\newcommand{\civ}{\ion{C}{4}}

\newcommand{\nv}{\ion{N}{5}}

\newcommand{\siii}{\ion{Si}{2}}
\newcommand{\siiii}{\ion{Si}{3}}
\newcommand{\siiv}{\ion{Si}{4}}

\newcommand{\mgii}{\ion{Mg}{2}}

\newcommand{\ovi}{\ion{O}{6}}

\newcommand{\mhalo}{M_{\rm halo}}

\newcommand{\kms}{km\,s$^{-1}$}
\newcommand{\km}{km\,s$^{-1}$}
\newcommand{\nref}{n_{\rm ref}}

\newcommand{\spectacle}{\texttt{spectacle}}
\newcommand{\nhi}{$N_{\rm H\,I}$}
\newcommand{\mcloud}{m_{\rm cloud}}
\newcommand{\mlnhi}{\log N_{\rm H\,I}}
\newcommand{\ckpch}{$h^{-1}$ckpc}

\newcommand{\nmin}{N_{\rm min}}

\newcommand{\Ncells}{N_{\rm cells}}
\newcommand{\sigth}{\sigma_{\rm th}}
\newcommand{\dvcloud}{\Delta v_{\rm cloud}}

\defcitealias{corlies18}{Paper~II}

\usepackage{verbatim}
\received{November 7, 2018}
\revised{January 23, 2019}
\accepted{January 26, 2019}
\submitjournal{ApJ}

\shorttitle{FOGGIE I: Resolving Circumgalactic Absorption at $z\geq 2$}
\shortauthors{Peeples et~al.}

\begin{document}

\title{Figuring Out Gas \& Galaxies in Enzo (FOGGIE). I.\\ Resolving Simulated Circumgalactic Absorption at $2\leq z \leq 2.5$}

\correspondingauthor{Molly S.\ Peeples}
\author[0000-0003-1455-8788]{Molly S.\ Peeples}
\affiliation{Space Telescope Science Institute, 3700 San Martin Drive, Baltimore, MD, 21218}
\affiliation{Department of Physics \& Astronomy, Johns Hopkins University, 3400 N.\ Charles Street, Baltimore, MD 21218}
\email{molly@stsci.edu}

\author[0000-0002-0646-1540]{Lauren Corlies}
\affiliation{Department of Physics \& Astronomy, Johns Hopkins University, 3400 N.\ Charles Street, Baltimore, MD 21218}
\affiliation{Large Synoptic Survey Telescope, Tucson, AZ 85719}

\author[0000-0002-7982-412X]{Jason Tumlinson}
\affiliation{Space Telescope Science Institute, 3700 San Martin Drive, Baltimore, MD, 21218}
\affiliation{Department of Physics \& Astronomy, Johns Hopkins University, 3400 N.\ Charles Street, Baltimore, MD 21218}

\author[0000-0002-2786-0348]{Brian W.\ O'Shea}
\affiliation{Department of Computational Mathematics, Science and Engineering, Department of Physics and Astronomy, National Superconducting Cyclotron Laboratory, Michigan State University, East Lansing, MI 48823}

\author[0000-0001-9158-0829]{Nicolas Lehner}
\affiliation{Department of Physics, University of Notre Dame, Notre Dame, IN 46556}

\author[0000-0002-7893-1054]{John M.\ O'Meara}
\affiliation{Department of Physics, St.\ Michael's College, Colchester, VT 05439}
\affiliation{W.\ M.\ Keck Observatory, Waimea, HI 96743}

\author[0000-0002-2591-3792]{J.\ Christopher Howk}
\affiliation{Department of Physics, University of Notre Dame, Notre Dame, IN 46556}

\author[0000-0003-1714-7415]{Nicholas Earl}
\affiliation{Space Telescope Science Institute, 3700 San Martin Drive, Baltimore, MD, 21218}

\author[0000-0002-6804-630X]{Britton D.\ Smith}
\affiliation{San Diego Supercomputer Center, University of California, San Diego, 10100 Hopkins Drive, La Jolla, CA 92093}

\author[0000-0003-1173-8847]{John H.\ Wise}
\affiliation{Center for Relativistic Astrophysics, School of Physics, Georgia 
Institute of Technology, Atlanta, GA 30332}

\author[0000-0002-3817-8133]{Cameron B.\ Hummels}
\affiliation{Department of Astronomy, California Institute of Technology, Pasadena, CA 91125}

\begin{abstract}
\noindent
We present simulations from the new ``Figuring Out Gas \& Galaxies in Enzo'' (FOGGIE) project. In contrast to most extant simulations of galaxy formation, which concentrate computational resources on galactic disks and spheroids with fluid and particle elements of fixed mass, the FOGGIE simulations focus on extreme spatial and mass resolution in the circumgalactic medium (CGM) surrounding galaxies. Using the Enzo code and a new refinement scheme, FOGGIE reaches spatial resolutions of $381$ comoving $h^{-1}$ pc and resolves extremely low masses ($\lesssim 1$--$100$\,\Msun) out to 100 comoving $h^{-1}$ kpc from the central halo. At these resolutions, cloud and filament-like structures giving rise to simulated absorption are smaller, and better resolved, than the same structures simulated with standard density-dependent refinement. Most of the simulated absorption arises in identifiable and well-resolved structures with masses $\lesssim 10^4$\,\Msun, well below the mass resolution of typical zoom simulations. However, integrated quantities such as mass surface density and ionic covering fractions change at only the $\lesssim 30$\% level as resolution is varied. This relatively small changes in projected quantities---even when the sizes and distribution of absorbing clouds change dramatically---indicate that commonly used observables provide only weak constraints on the physical structure of the underlying gas. Comparing the simulated absorption features to the KODIAQ (Keck Observatory Database of Ionized Absorption toward Quasars) survey of $z \sim 2$--3.5 Lyman limit systems, we show that high-resolution FOGGIE runs better resolve the internal kinematic structure of detected absorption, and better match the observed distribution of absorber properties. These results indicate that CGM resolution is key in properly testing simulations of galaxy evolution with circumgalactic observations. 
\end{abstract}
\nopagebreak

\section{Introduction} \label{sec:intro}
In the galactic ecosystem, the majority of baryons and heavy elements are in the diffuse gas outside of galaxies rather than in the relatively dense gas and stars comprising the disks and central spheroids \citep{pagel08,stocke13,werk14,peeples14,prochaska17}. Hydrodynamic simulations of galaxy evolution have met with mixed success in reproducing the observed column densities of low- or high-ionization circumgalactic gas, despite creating realistic-looking galaxies (\citealp{hummels13,ford16,oppenheimer16,suresh17}; see also \citealp*{tumlinson17}). The focus of these theoretical studies has usually been the improvement of stellar feedback models: how the energy and momentum from star formation and supermassive black holes are coupled with the interstellar medium (ISM) and propagated into the circumgalactic medium (CGM) and beyond (e.g., \citealp{oppenheimer08}; \citealp{trujillo15}; \citealp*{salem16}; \citealp{christensen16,christensen18}). While feedback is certainly important for understanding how galaxies and the CGM co-evolve, we posit here that numerical resolution in the CGM is a critically important factor in the comparison between CGM data and cosmological simulations that influences results for any subgrid physical models. We show here that improving the resolution of circumgalactic gas strongly affects the simulated CGM structures that give rise to detected absorption; in \citetalias{corlies18} of this series \citep{corlies18}, we show that a resolved circumgalactic medium is crucial for predicting and subsequently understanding the sources of circumgalactic emission.

Observational evidence has existed for decades that the CGM is structured on scales smaller the $\gtrsim 1-10$ kpc scales that are typically resolved in cosmological simulations. Absorbers seen along lines of sight towards multiply-lensed quasars have structure on the scale of $\lesssim 1$--10\,kpc, with lower ionization species (e.g., \mgii\ or \siii, thought to trace cooler, higher density gas) showing structure on smaller scales than more highly ionized species (e.g., \civ\ or \ovi, \citealp{rauch01a, rauch01b, ellison04,rubin18a}). Milky Way high velocity clouds (HVCs) with well-constrained distances show \hi\ masses of $\sim 10^5$--$5\times 10^6$\,\Msun\ \citep[e.g.,][]{wakker08,putman12} and sizes in the few to $\sim 15$\,kpc range. Given how relatively close to the disk most HVCs are \citep[e.g.,][]{lehner12,lehner11,wakker07,wakker08,thom08b}, these clouds may or may not be representative of the Milky Way CGM as a whole. Recent evidence from absorption towards background galaxies---which have a larger effective beamsize than quasars and thus wash out small scale structure---indicate that the ``coherence scale'' of circumgalactic absorption is on the order of tens of kiloparsecs \citep{lopez18, peroux18, rubin18b}.  Therefore, to fully resolve the {\em observed} structure of the CGM, simulations must reach spatial scales of $\lesssim 1$\,kpc in order to have at least tens of resolution elements comprising such structures.
 
There are several physical processes that can give rise to multiphase structure at sub-kpc length scales. The cooling length (the sound speed times the cooling time,  see, e.g., Figure~8 of \citealp{smith17}) has scales from tens of parsecs to hundreds of kiloparsecs for typical circumgalactic densities, temperatures, and metallicities, with photoionizational cooling from the metagalactic UV background.\footnote{We note that most of the densities considered in \citet{mccourt18} are higher than the circumgalactic densities we are interested in here; these higher physical densities translate to much smaller cooling lengths than typical for the CGM.} At low densities, length scales on the smaller end of this range translate to very small mass scale---tens to hundreds of solar masses---which can be challenging to achieve in simulations evolved with fixed mass resolution (e.g., those evolved with particle-based or moving mesh codes). Resolving the cooling length in the density and temperature regimes of interest to the inner CGM, where $l_c \simeq 0.1$--$10$ kpc, is critical to resolving the thermal instability in the CGM.  This instability is potentially responsible for the formation of multiphase gas in the circumgalactic medium \citep{voit15d}.  If the simulation's physical resolution is larger than the local cooling length in the region of interest, the formation of a multiphase medium will be either dampened or suppressed entirely. 

Other physical scales that may affect the properties of the multiphase CGM include (1) the convective scale of the CGM, which is approximately equal to the pressure scale height of the stratified medium, and is the typical physical scale over which a parcel of fluid may move and bulk mixing might occur;  (2) the scale on which turbulence is driven, which also affects mixing of metals and gas of different temperatures; and (3) the Field length, which is the physical length scale where the thermal conduction of heat into a cold gas cloud is balanced by radiative cooling, and which defines the minimum scale of cold gas clouds \citep{field65,sharma10}. The pressure scale height of a virialized Milky Way-sized galaxy halo is $\sim 20$\,kpc, suggesting the convective scale is not setting the size of the small-scale structures in the CGM.  The driving scale of turbulence, and thus the largest physical objects whose properties are likely to be dominated by this phenomenon, depends on the source of the driving. In the case of cosmological accretion, that scale would be comparable to the diameter of the cosmological filament, or a significant fraction of the virial radius.  In the case of turbulence driven by galactic winds/outflows, it is comparable in size to the width of the outflows---on the order of kiloparsecs to a few tens of kpc, depending on the galaxy's recent star formation rate or AGN behavior.  Likewise, the Field length can have values spanning orders of magnitude: in $10^6$\,K diffuse gas it is tens of kiloparsecs (substantially larger than the cooling length), but in the denser $10^4$\,K gas the Field length is orders of magnitude smaller, comparable to the cooling length. 
In principle, the Jeans or Bonner-Ebert length scales---the physical scale where gravity dominates over thermal pressure---may be important as well, but given that the local gravity field is dominated by the host dark matter halo,
it is likely that gas clouds are not self-gravitating \citep[unlike in the intergalactic medium, ][]{schaye01, peeples10a, peeples10b}, and thus hydrodynamic processes will dominate \citep[see, e.g.,][]{liang18}.  The physical scale of a cloud is likely to be smaller than the convective and turbulent driving scales, and smaller than the cooling length scale, as these drive fragmentation and the onset of the thermal instability, respectively.  The cloud must also be larger than the Field length, or else it will evaporate.  If these scales are inverted (i.e., if the Field length is larger than these scales), the physical processes that drive precipitation will be suppressed.  Most simulations to date either explore the small-scale structure of circumgalactic gas in highly idealized scenarios \citep{armillotta17,fielding18} or explore regimes more akin to the disk-halo interface (i.e., much denser gas) than the diffuse CGM \citep{schneider17}, though see \citet{churchill15} and \citet{vandevoort19}. 

In this paper, we explore the consequences of resolving the CGM to sub-kpc scales, regardless of what physical processes may be responsible for generating structure on these small scales. In other words, we explore the effects of numerical resolution without changing the adopted feedback, to assess them independently.  We describe our simulations and improved resolution technique in \S\,\ref{sec:sims}; our method for extracting and analyzing synthetic spectra is described in \S\,\ref{sec:misty}. The Keck Observatory Database of Ionized Absorption toward Quasars (KODIAQ, \citealp{lehner14,omeara17}) sample we compare these synthetic spectra to is described in \S\ref{sec:kodiaq}.
In \S\,\ref{sec:physical}, we show the impact our improved resolution has on the physical properties of the CGM, and in \S\ref{sec:clouds}\ we show that we resolve the predominant structures responsible for most circumgalactic absorption at these redshifts.
In \S\,\ref{sec:comparison}, we compare the kinematic properties of the synthetic spectra to the KODIAQ absorbers, and we provide our concluding thoughts in \S\,\ref{sec:conc}.
 As the bulk of the high signal-to-noise KODIAQ absorbers with a wide range of available line transitions are at $2\leq z\leq 2.8$, we focus our simulation resolution comparison here to $z=2$ and $z=2.5$. All lengths and distances are given in physical units unless specifically stated otherwise.

\begin{figure*}[tbh]
    \centering
    \includegraphics[width=0.5\textwidth]{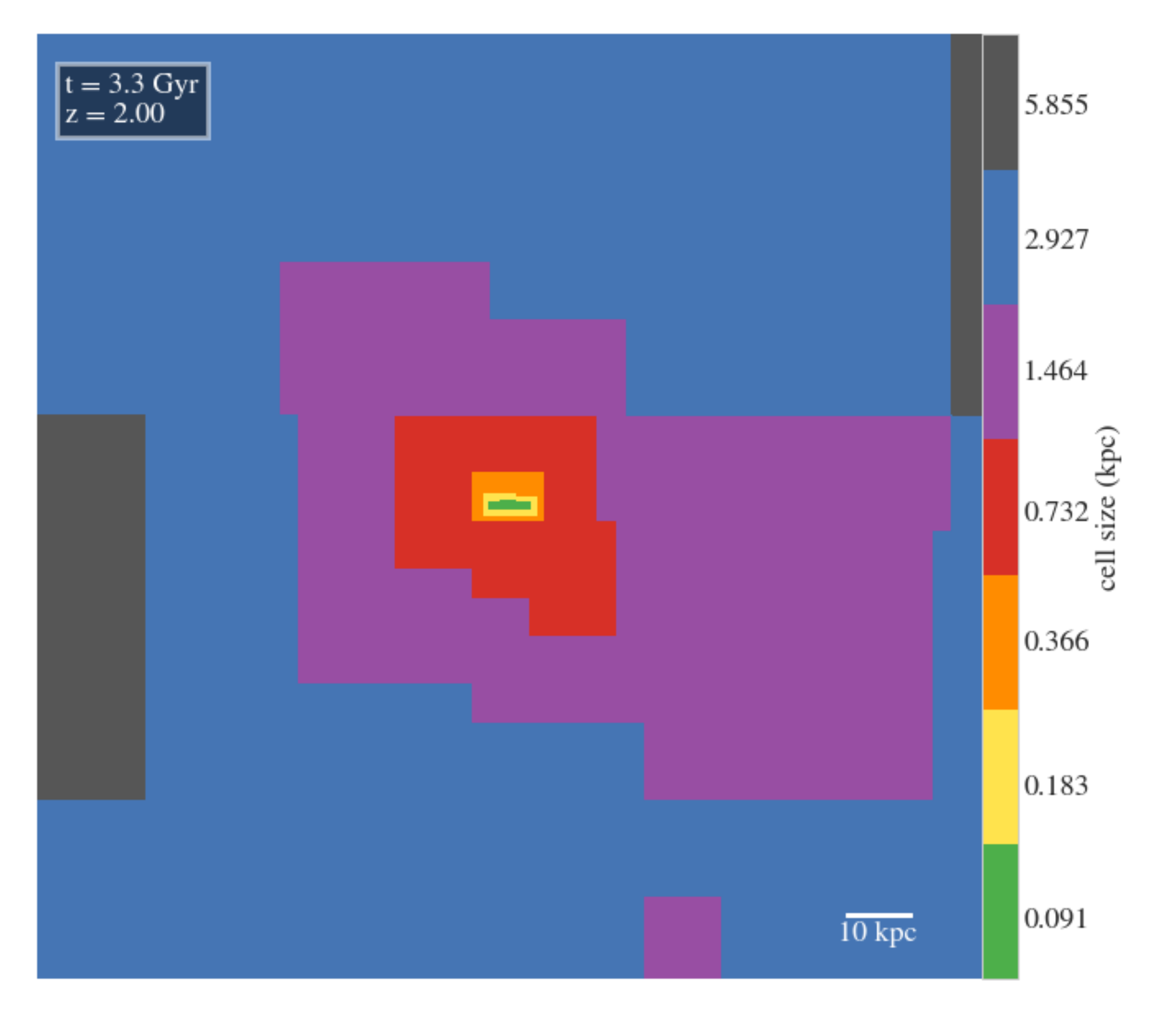}\hfill
    \includegraphics[width=0.5\textwidth]{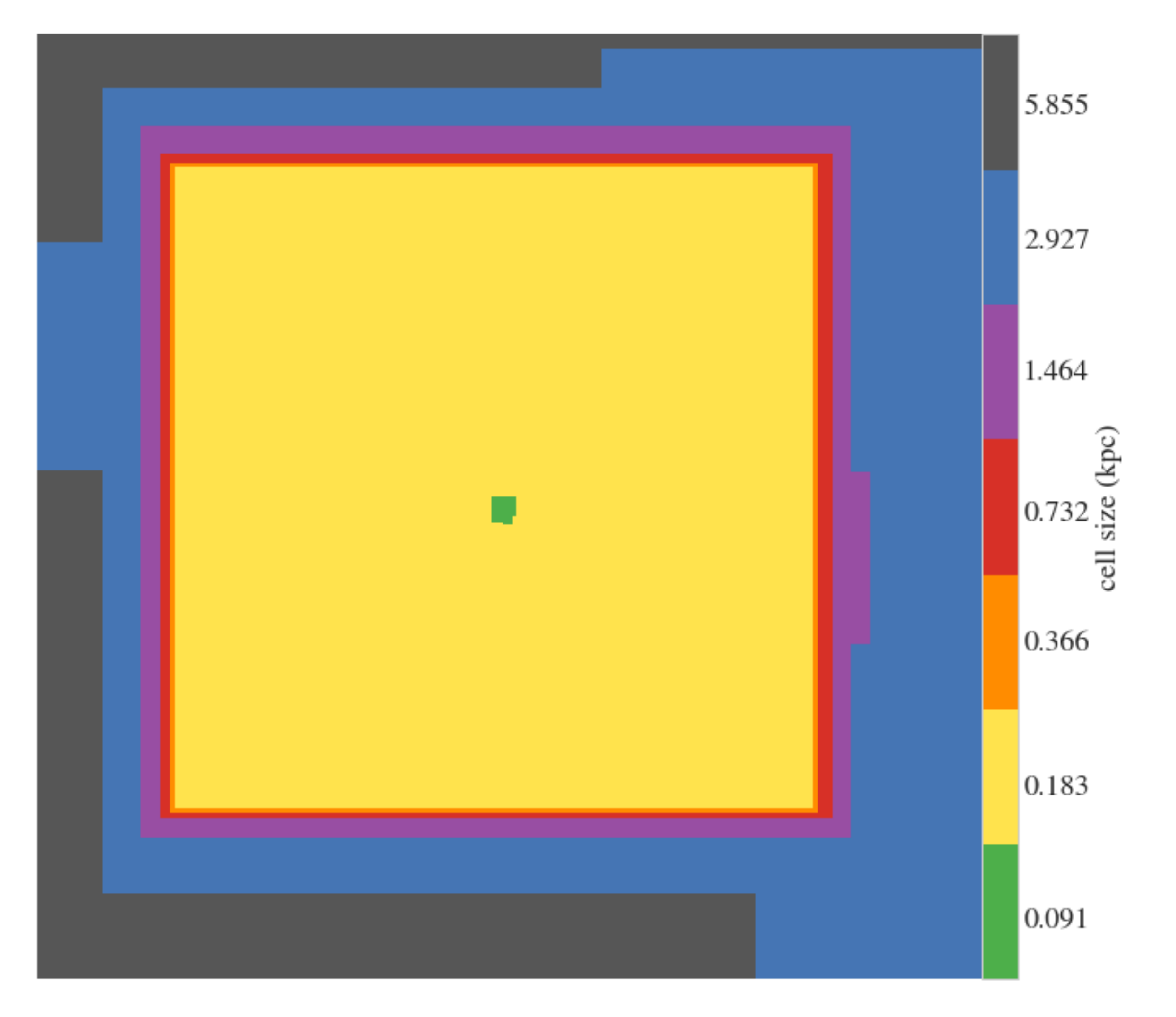}
    \caption{Slices showing the resolution of the of the standard ({\em left}) and high ({\em right}) resolution simulations at $z=2$. The cell sizes are given in physical kpc. The galaxy is located at the center of each slice; note that the ISM resolution does not change.
    \label{fig:refinement-slices}}
\end{figure*}

%%%%%%%%%%%%%%%%%%%%%%%%%%%%%%%%%%%%%%%%%%%%%%%%%%%%%%%%%%%%%%%%%%%%
%%%%%%%%%%%%%%%%%%        THE SIMULATIONS       %%%%%%%%%%%%%%%%%%%%
%%%%%%%%%%%%%%%%%%%%%%%%%%%%%%%%%%%%%%%%%%%%%%%%%%%%%%%%%%%%%%%%%%%%

\section{The Simulations} \label{sec:sims}

\subsection{Basic Properties}\label{sec:simbasics}
The Figuring Out Gas \& Galaxies In Enzo (FOGGIE) simulations are cosmological hydrodynamic simulations evolved with the block-structured adaptive mesh refinement (AMR) code Enzo \citep{bryan14} using a flat \citet{planck13} $\Lambda$CDM cosmology ($1 - \Omega_{\Lambda} = \Omega_{\rm m} = 0.285$, $\Omega_{\rm b} =0.0461$, $h = 0.695$) and a Piecewise Parabolic Method (PPM) hydro solver.  The simulation domain is a $(100\,h^{-1}{\rm cMpc})^{3}$ box. We generated the initial conditions with a simple $1{,}024^3$ particle dark matter-only simulation on the NCSA Blue Waters supercomputer. From this DM-only run we selected a halo, which we have named ``Tempest'', that has a Milky Way mass at $z = 0$ ($\mhalo \simeq 1.5 \times 10^{12}$ \Msun) and a relatively quiescent merger history at $z<1$.  The initial conditions for the Tempest halo were then regenerated using the ``cosmological zoom'' method, with a $256^3$ grid cell/particle base resolution and an effective resolution of $4{,}096^3$ particles (M$_{\rm dm} = 1.39 \times 10^6$\,\Msun) in the region of interest, which is a Lagrangian region encompassing all of the particles within two virial radii of the galaxy at $z=0$.  These ``zoom'' simulations  were then evolved on the NASA Pleiades supercomputer with a maximum of 11 levels of adaptive mesh refinement (cell sizes of $\simeq 190$ pc/h comoving). The Tempest halo has $R_{200} = 37$ kpc and $M_{200} = 3.9 \times 10^{10}$ \Msun\ at $z = 2$. 

We use Enzo's \citet{cen06} thermal supernova feedback model, forming stars in gas exceeding a comoving number density of $\simeq 0.1$~cm$^{-3}$ with a minimum star particle mass of $2 \times 10^4$~M$_\odot$. Supernova feedback is comprised of purely thermal energy that is deposited into the 27 nearest cells surrounding the star particle, after 12 gas dynamical times have elapsed since the star particle formed. The total energy imparted is $1.0 \times 10^{-5} m_{\star}c^2$, the total mass ejected is $0.25 \ m_{\star}$, and the total metal mass ejected is
\begin{equation}
    0.025 \ m_{\star} (1-Z_{\star}) + 0.25 \ Z_{\star},
\end{equation}
where all metals are tracked as a single field. As a result, particular elemental abundances throughout the paper are calculated assuming Solar relative abundances scaled to the local metallicity computed at runtime. The effects of Type Ia SNe are not included. 

The simulation includes metallicity-dependent cooling and a metagalactic background \citep{haardt12} using the Grackle\footnote{\url{https://grackle.readthedocs.io/en/latest/}} chemistry and cooling library \citep{smith17}. The code simultaneously solves a non-equilibrium six species  chemical reaction network (tracking \hi, \hii, \ion{D}{1}, \ion{D}{2}, \ion{He}{1}, \ion{He}{2}, \ion{He}{3}, and e$^-$), computing cooling directly from these species. No correction is made for self-shielding. A metal field is followed separately, with cooling calculated assuming ionization equilibrium and Solar composition.

\begin{figure*}[htb]
    \centering
    \includegraphics[width=0.48\textwidth]{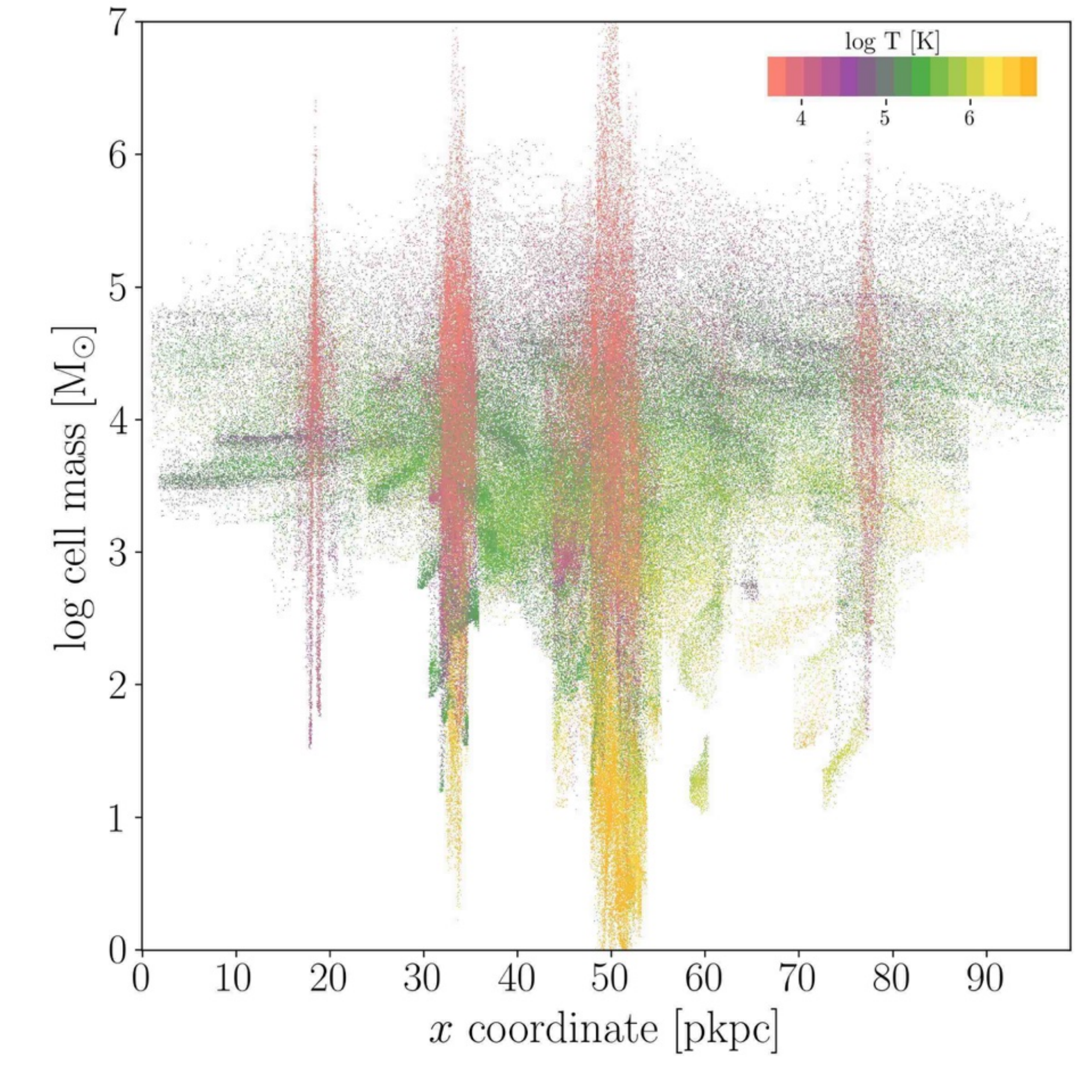}\hfill
    \includegraphics[width=0.48\textwidth]{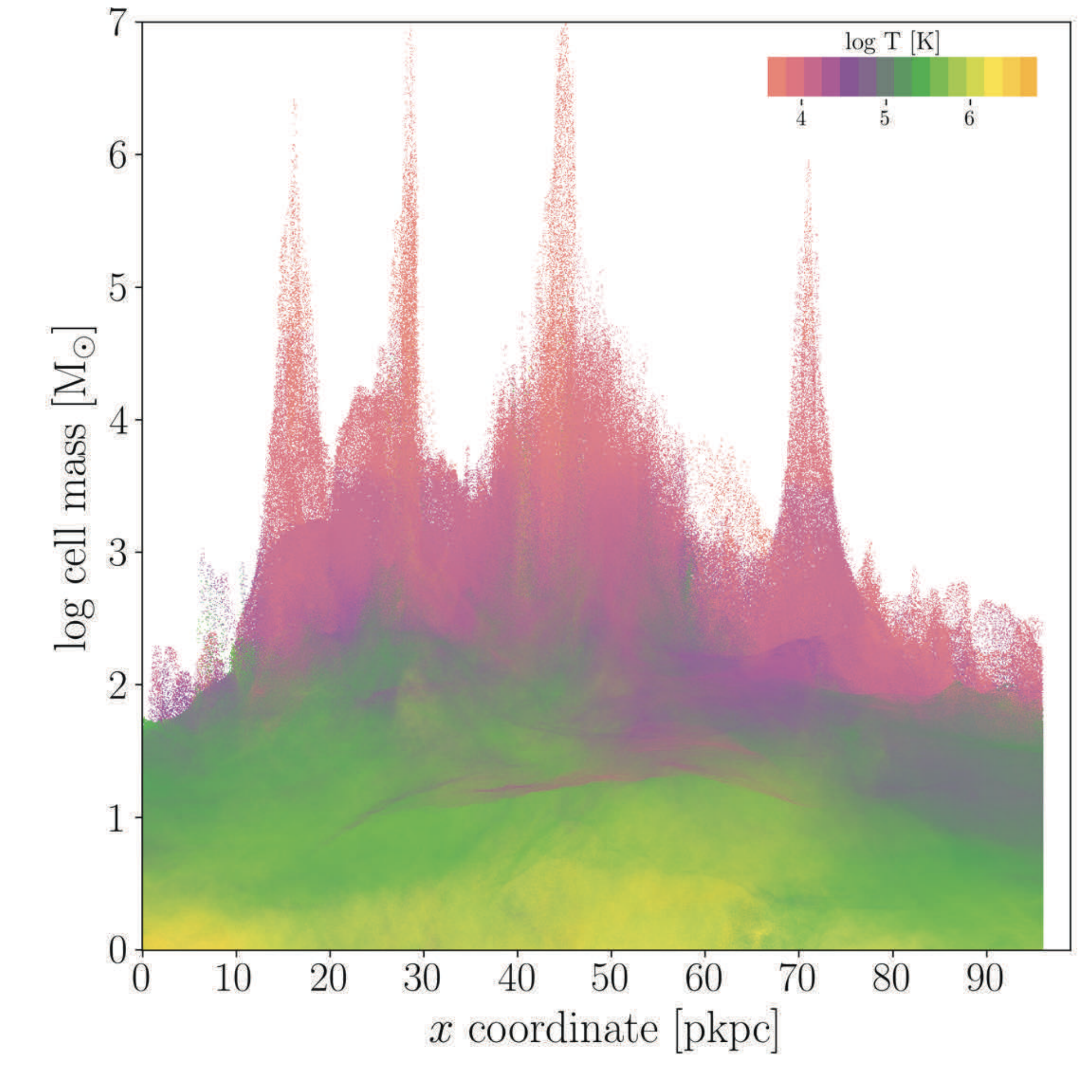} %\hfill
    \caption{
   Distribution of cell masses within the 200\ckpch\ forced refinement region  at $z = 2$, color coded by the most common temperature at that location and cell mass. In both panels, the cold, high-cell mass peaks correspond to the dense ISM of the main galaxy and its satellites, while the lower cell-mass gas at intervening positions is the CGM. The left panel shows the cell masses in the standard simulation, where cells vary in size in proportion to the local density, and most of the CGM is traced by structures $\gtrsim 10^4$\,\Msun. The right panel shows the same for the high-resolution simulation, where the uniform cell size in the CGM is 548 comoving parsec ($380/h$) and thus CGM cell masses are $\sim 1$--$100$\,\Msun. 
    \label{fig:cellmasses}}
\end{figure*}

\subsection{Forcing Resolution in the CGM}\label{sec:forcedrefinement}
In order to isolate the effects of circumgalactic resolution on observable absorption-line spectra, we consider multiple runs with the same sub-grid physics and interstellar resolution, varying only the refinement scheme that is applied in the CGM gas. To impose uniform {\em spatial} resolution on the CGM, we require a method that overrides the normal criteria for refinement on physical criteria such as gas density, dark matter overdensity, or metallicity. In its place, we use a ``forced refinement'' scheme, which imposes a fixed resolution on a specific sub-region of the computational domain that can move over time. This region is specified in a parameter file read by Enzo at runtime. For this first generation of FOGGIE simulations, we adopt a cubic forced-refinement region that runs from $-100\,h^{-1}$ to $+100\,h^{-1}$ comoving kiloparsecs from the moving center of mass of the targeted galaxy in each spatial dimension. The path of the halo of interest through the domain is obtained by tracking the halo's center of mass (including gas and dark matter) from a lower-resolution run evolved to $z = 0$ with ``standard'' refinement. This run consumes minimal resources and serves as an important controlled comparison to the forced refinement run. Once the 3D track through the domain is determined, we rerun the calculation with the forced refinement box tracking the galaxy, starting at $z = 4$, when the galaxy has formed $<5$\% of the stars it has by $z=2$. When forced refinement is on, its level of refinement is specified as a parameter and fixed. Outside the forced refinement box, standard refinement is used with a gas overdensity criterion. Within the tracked box, refinement above the fixed level can be specified to follow a density criterion. In this way, our primary production run uses a fixed refinement $\nref = 10$ levels ($380h^{-1}$pc comoving) over the (200\ckpch)$^{3}$ box, with additional refinements one level further to $\nref = 11$ ($190 h^{-1}$cpc) where the density criterion is met. We find the forced $\nref = 10$ refinement increases to 11 only in the galaxy's disk and the ISM of satellites. This uniform method of refinement is generally {\em much} more computationally efficient than a standard AMR run with a similar total number of resolution elements because the grid patches are typically larger and more uniformly sized, meaning there are fewer overall ghost zones and better cache utilization.

Figure \ref{fig:refinement-slices} illustrates this technique with slices of the cell size (i.e., effective resolution) through the center of the halo at $z=2$; note that the interstellar resolution of $190h^{-1}$ comoving pc (91 physical pc, shown in green) is the same in both simulations.  Slight differences in grid structure outside of the refinement region are due to the algorithm used for cell flagging and AMR grid creation.  We alternately refer to the standard resolution run as the ``density-refined'' run and the high-resolution simulation as the ``refined'' or ``forced refinement'' simulation. Both simulations have been evolved to $z=0$.

As a demonstration and test of convergence, we have evolved a third simulation to $z=2.5$ with a uniform 290 comoving parsec ($\nref = 11$) resolution within the 200\,ckpc volume.  We show the results from this $\nref=11$ everywhere simulation and how it compares to the $z=2.5$ in the other two simulations in Appendix \ref{app:resolution}. Broadly, we find that these two forced-resolution simulations are much more similar in the metrics we consider throughout this paper than either is to the standard-resolution simulation.

Critically, enforcing small spatial scales to be resolved translates into very low {\em masses} per resolution element in the low-density gas. Figure \ref{fig:cellmasses} shows the masses of the cells projected along a position axis through the forced-refinement region in the two simulations in temperature bins. 
The first clear difference between the two simulations is that there are {\em many} more resolution elements in the high-resolution simulation, particularly in the CGM; at $z=2$, the high-resolution run has $\sim 400\times$ as many, with $\sim 1.4\times 10^{8}$ cells within the forced refinement region, while  the standard resolution simulation has $\sim 350,000$ cells within the corresponding volume.
In the standard resolution simulation, most of the circumgalactic gas is traced by cells with masses of $\gtrsim 10^4$\,\Msun, while in the high resolution case, essentially all of the gas at temperatures $\gtrsim 10^{4.5}$\,K is resolved by cells with masses $<1000$\,\Msun, as is most of the gas at lower temperatures. As we show in \S\,\ref{sec:cloudsize}, resolving these small masses allows the FOGGIE simulations to resolve the small structures responsible for most circumgalactic absorption, even when that gas is at relatively low density.

\subsection{Synthetic Absorption Spectroscopy}\label{sec:misty}
We extract synthetic spectra from the simulations using the Trident package \citep*{hummels17}, adopting the \citet{morton03} atomic data. Trident first uses Cloudy (last described in \citealt{ferland13}) to calculate the ionization fraction of the species of interest based on the cell-by-cell density, temperature, and metallicity; as when the simulations were evolved, we assume a \citet{haardt12} metagalactic background (though the \hi\ fraction is tracked natively by Enzo). To calculate the effective redshift of absorption features, Trident assumes a smooth Hubble flow along a line of sight. This is then added to the three-dimensional velocity field of each AMR grid cell intersected by the sight line. The absorption produced by a grid cell is represented by a single Voigt profile at an instantaneous velocity $v$ and a doppler $b$ parameter specified by the temperature in the cell (i.e., the Hubble flow across the grid cell is neglected, no model is applied for turbulence on scales smaller than the cell-size, etc.). Trident returns the optical depth and normalized flux as a function of observed wavelength along the line-of-sight; in order to ease the comparison with the observations, we consider each ionic transition individually so that we can uniquely convert the observed wavelength back to the effective redshift and thus relative velocity of the system. We then resample the spectra to 2\,\kms\ per pixel, the approximate binning of Keck's High Resolution Echelle Spectrometer (HIRES, see \S\,\ref{sec:kodiaq})\footnote{Note that because Trident extracts spectra in even bins of wavelength instead of even bins of velocity, this resampling requires non-uniform interpolations. However, this does not affect our results because of the relatively small wavelength ranges considered and the extremely high resolution at which the initial spectra are extracted from the simulation.}. Finally, we convolve the spectra with a 7\,\kms\ full-width half-maximum Gaussian profile to mimic a representative HIRES line spread function (LSF).\footnote{Though it is more physical to apply the LSF and then resample the spectrum, we do this order of operations in reverse because Trident extracts the spectra in wavelength space; again, our tests show that at these spectral resolutions, the order of operations does not affect our results.}

The resampling, LSF handling, and characterization of the absorption features are part of the \spectacle\footnote{\url{https://spectacle-py.readthedocs.io/en/latest/}} package, which is still under development and which we will release publicly in the near future (Earl \& Peeples, in preparation). Here, the only analysis feature we consider is the number of minima below a given flux threshold (\S\,\ref{sec:comparison}). In practice, this threshold also serves to set the effective signal-to-noise of the synthetic spectra to be similar to the level of detected features in the real spectra without the need for developing an automated fitting routine that can also properly account for noisy data and uncertainties. As the signal-to-noise of the KODIAQ data varies from sightline to sightline (\S\,\ref{sec:kodiaq}),
we adopt a threshold in the normalized flux of 0.95 (requiring the minima absorb more than 5\% of the unattenuated normalized flux) in order to mimic the effects of noise in the real spectra.

For analyses based on these spectra (\S\S\,\ref{sec:clouds},\ref{sec:comparison}), we 
consider only simulated sightlines with $10^{16} < N_{\rm H\,I} <10^{21}$\,cm$^{-2}$, taking 100 lines of sight along each orthogonal axis at $z=2$ and $z=2.5$, for a total of 600 sightlines per simulation. Each sightline has a pathlength of $200h^{-1}$ comoving kpc, which is wholly contained within the forced refinement regions (or, for the case of the standard resolution simulation, the corresponding volume). While this pathlength is certainly smaller than the pathlengths giving rise to the LLSs seen in KODIAQ (\S\,\ref{sec:kodiaq}), this selection enables a comparison that is wholly caused by the difference in the underlying physical resolution in the two simulations.
For each metal species, we further consider only sightlines with a minimum column density, roughly corresponding to the minimum column density at which it is observed in the KODIAQ data, as given in Table~\ref{tbl:colmin}.

\begin{table}
\centering
\begin{tabular}{ cc } 
         & log (minimum\\ 
 Species & column density) [cm$^{-2}$]\\ 
 \hline
\hi\ & 16 \\
\siii\ & 11 \\ 
\siiv & 12  \\ 
\civ & 13\\
\ovi & 13\\
\end{tabular} 
\caption{Minimum considered column densities for each ion of interest.} \label{tbl:colmin}
\end{table}

\section{Observed Absorption Spectroscopy: The KODIAQ Sample}\label{sec:kodiaq}
To facilitate comparison with real data, we require a sample of high resolution, high signal-to-noise data covering \hi\ and metal line absorption across multiple ionization states. We therefore compare to spectra from the DR2 release of KODIAQ \citep{omeara17}, a large sample of Keck HIRES data on $z > 2$ QSOs.  We select those spectra that can provide an accurate measurement of \nhi\, which in turn requires that the absorption is at $z \gtrsim 2.5$ if \nhi\ is determined from the termination of the \hi\ Lyman series, or $z \gtrsim 1.6$ if the \nhi\ is determined from damping wings in the \hi\ \lya\ line.  In general, we favor absorption at $z > 2$ so that in addition to \hi, key temperature and ionization diagnostic ions such as \ovi\  are covered in the spectra.  The full KODIAQ DR2 sample includes hundreds of objects, but 
most have signal-to-noise ratios too low to provide useful comparisons to our simulated spectra (typically S/N $> 30$ per pixel at \lya).  Thus we further constrain our sample to a subset of previously studied KODIAQ absorbers in \citet{lehner14} and \citet{lehner16}.\footnote{The only exception is the absorber towards Q2126-158 at $z= 2.90731$, which was selected based on its \nhi\ from an unpublished survey \citep{burns14}.}  To facilitate a study of \ovi\ absorption in Lyman limit and damped \lyman\ systems, the spectra in \citet{lehner14} were selected to have strong ($\mlnhi > 17$) \hi\ absorption, and little to no contamination of the \ovi\ $\lambda \lambda$ 1032,\,1037\AA\ by the Lyman $\alpha$ forest. The spectra in \citep{lehner16} were selected purely on \hi\ to have $16.2 <\mlnhi <19$ and be at $z>2$.

\begin{figure}[htb]
\centering
    \includegraphics[width=0.45\textwidth]{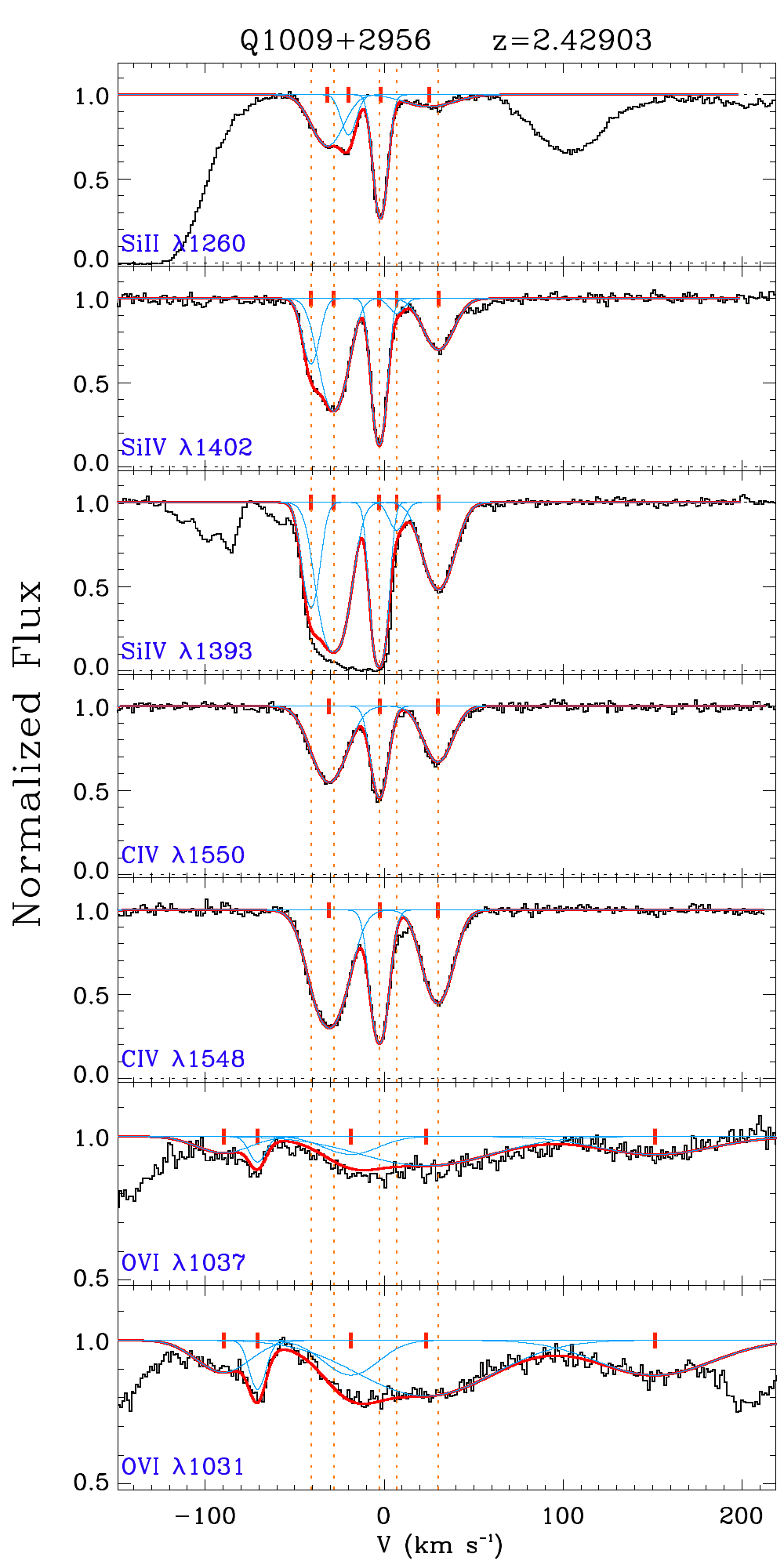}
    \caption{Example of a typical KODIAQ absorber: normalized profiles of low and high ions as a function of the rest-frame velocity for the absorber toward Q1009+2956. Each ion was fitted simultaneously (only one transition for \siii\ is shown, but three were used in the profile fitting), but each species was fitted independently. The red lines indicate the global component model and the blue lines show the individual components. The vertical orange dotted lines represent the velocity centroids for \siii\ while the red tick-marks show the velocity centroids for each species. Absorption that is not fitted is attributed to a contamination from an unrelated absorber.
    \label{fig:kodiaqexample}}
\end{figure}

We adopt the \citet{lehner14} fits for the high ions (\civ, \siiv, \ovi, \nv). However, we also want to compare FOGGIE to the observed low ions, which were not fitted in \citet{lehner14} or \citet{lehner16}. By including the low ions, we can directly determine how the kinematics of the absorption profiles change over a large range of ionization states. We select \siii\ as the main low ion because we can use several transitions (\siii\ $\lambda\lambda$1190, 1193, 1260, 1304, 1526) in our profile fitting, which allows us to assess any contamination reliably and hence accurately model the absorption profiles. To fit the individual components of \siii, we follow the overall methodology undertaken by \citet{lehner14}. In short, we use a modified version of the software described in \citet{fitzpatrick97}, which can simultaneously fit several transitions of the same ion or atom.  The best-fit values describing the gas are determined by comparing the data to composite Voigt profiles convolved with an instrumental line-spread function (LSF). The LSFs are modeled as a Gaussian with a full-width at half-maximum (FWHM) derived from the resolution of the  KODIAQ spectra (which typically vary between 4 to 10 \km\ FWHM, see \citealt{lehner14,lehner16,omeara17}).  The three parameters---the column density, $N_i$; the Doppler parameter, $b_i$; and the central velocity, $v_i$---for each component, $i$, are input as initial guesses and are subsequently freely allowed to vary to minimize the $\chi^2$ goodness of fit.  We always start each fit with the smallest number of components that reasonably model the absorption profiles of \siii, and only include additional component if it improves the reduced $\chi^2$. 

The results from these fits are summarized in Table~\ref{tbl:kodiaq} in Appendix~\ref{app:kodiaq}. In Figure~\ref{fig:kodiaqexample}, we show an example of these fits comparing low and high ions. Note that while we only show \siii\ $\lambda$1260 in this figure, the fit of the \siii\ is actually based on three transitions for this absorber ($\lambda\lambda$1260, 1304, 1526). The fitted composite profiles give us a noise- and contamination-free description of the data. We use these parameters within \spectacle\ to construct arrays of flux versus velocity, with the same 2\,\kms\ binning as the synthetic data. We subsequently use \spectacle\ to analyze both the real and simulated spectra using the same tools and assumptions. 

\begin{figure*}[htb]
    \centering
    \includegraphics[height=0.45\textwidth]{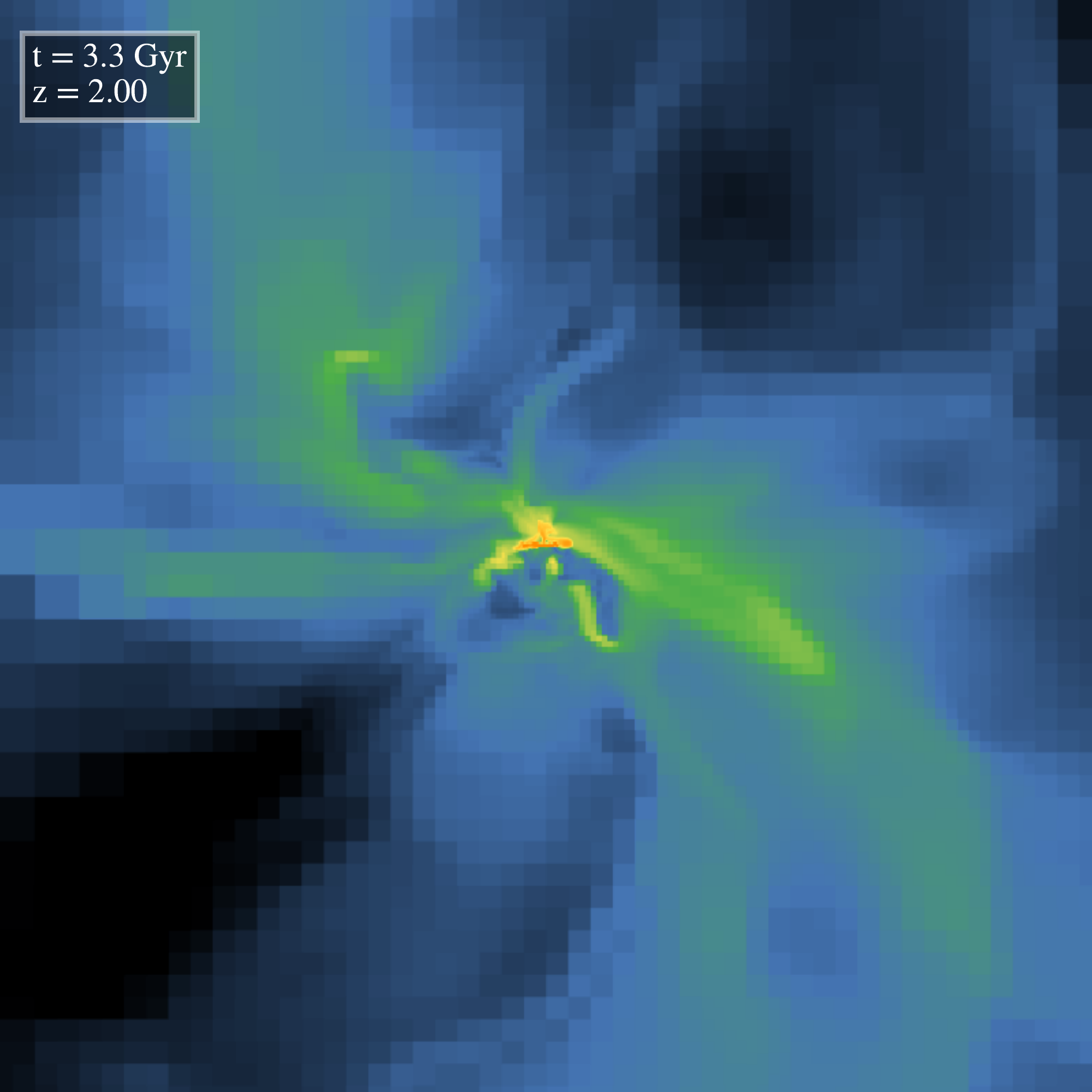}\hfill
    \includegraphics[height=0.45\textwidth]{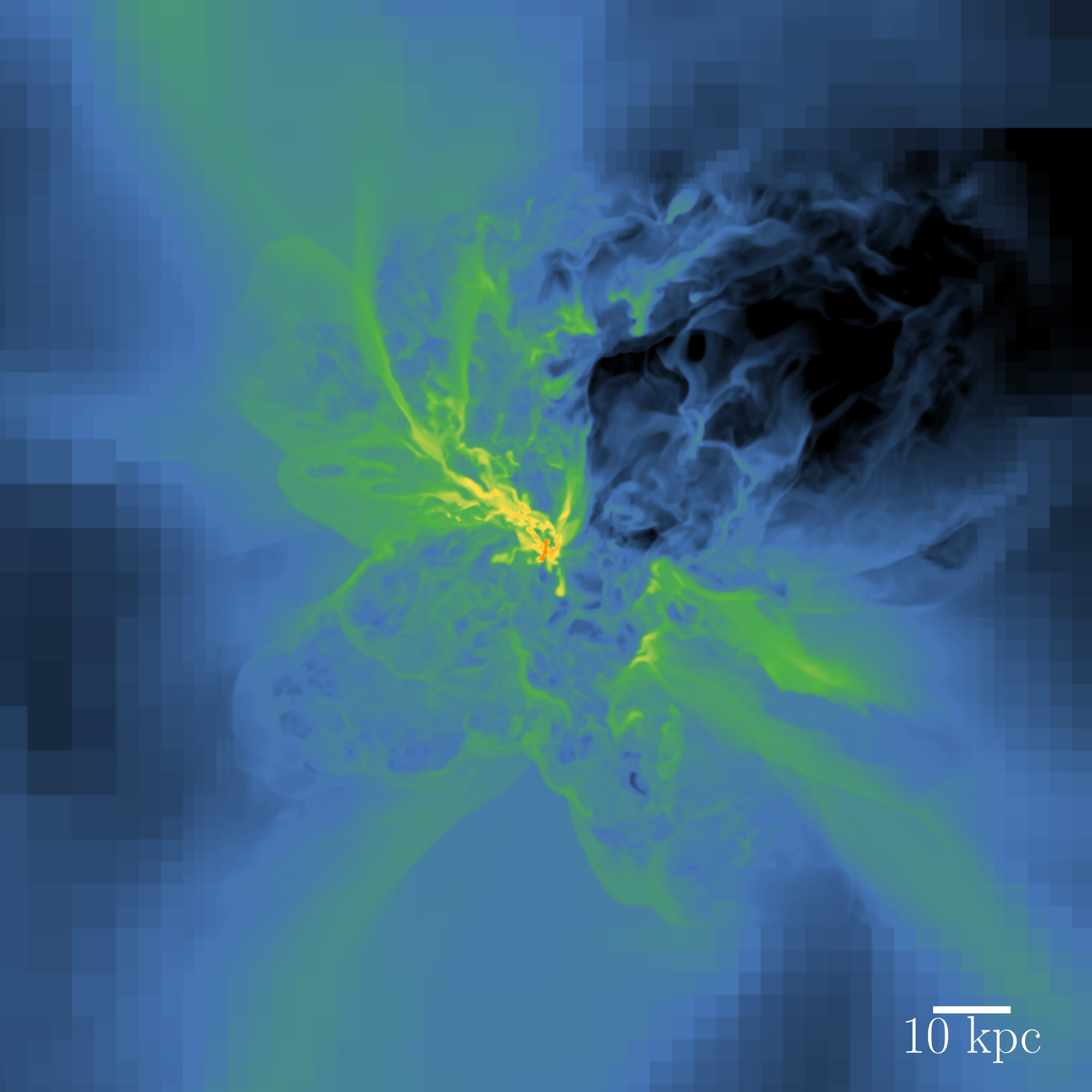}\hfill
    \includegraphics[height=0.45\textwidth]{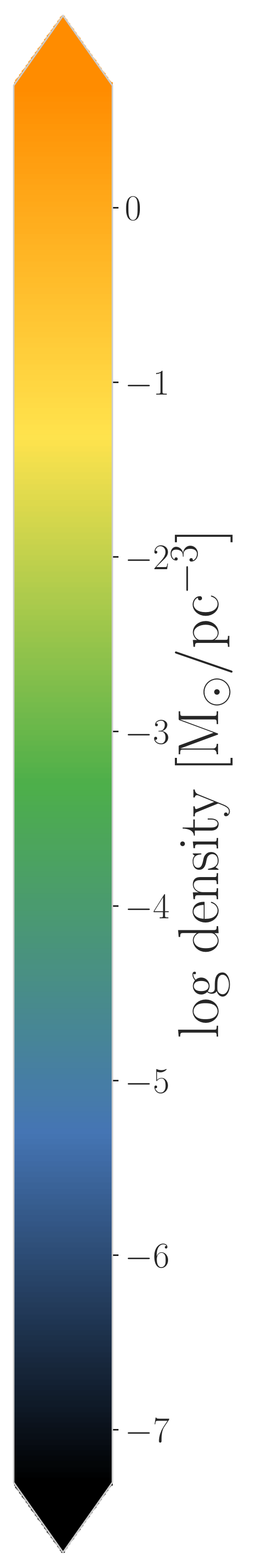}\\
    \includegraphics[height=0.45\textwidth]{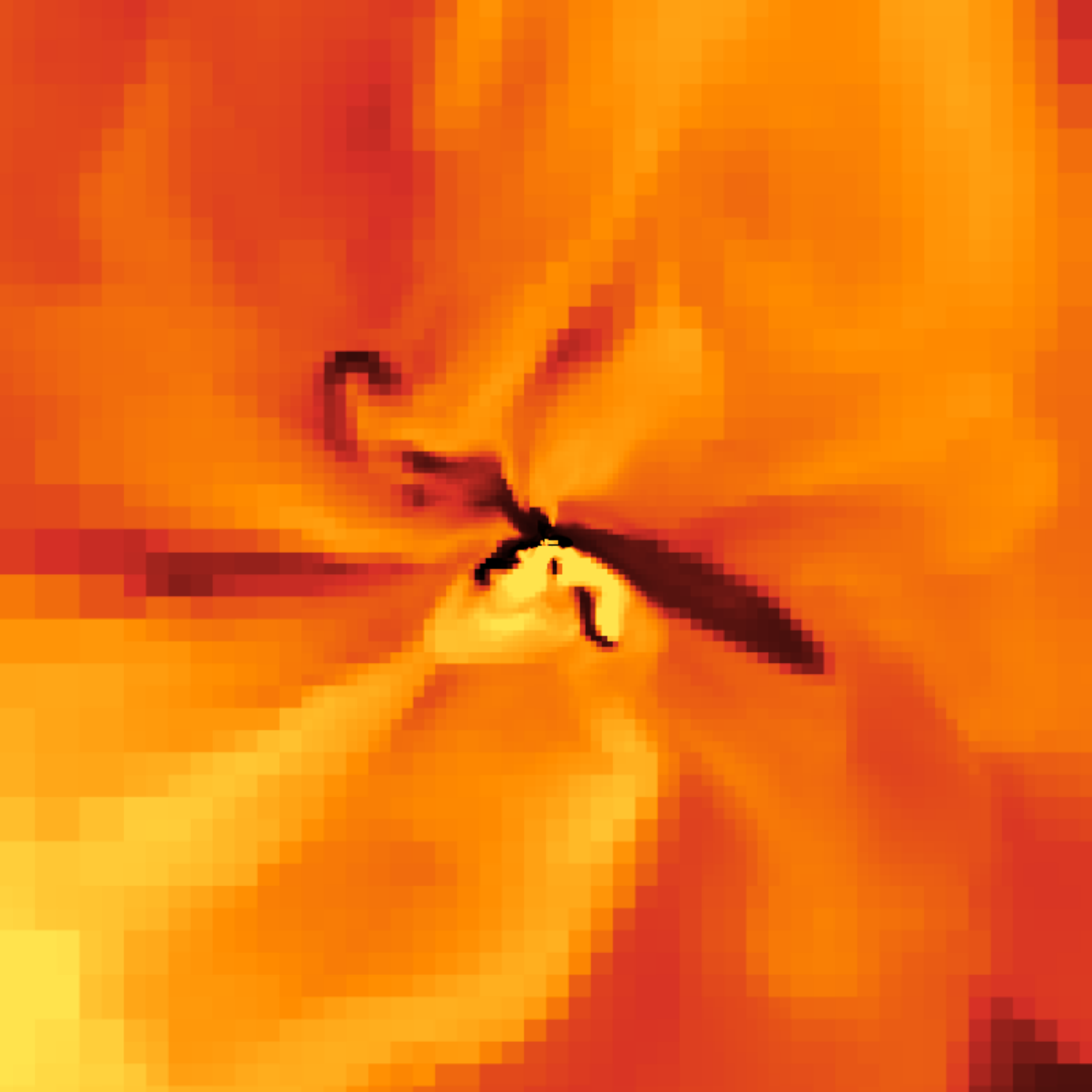}\hfill
    \includegraphics[height=0.45\textwidth]{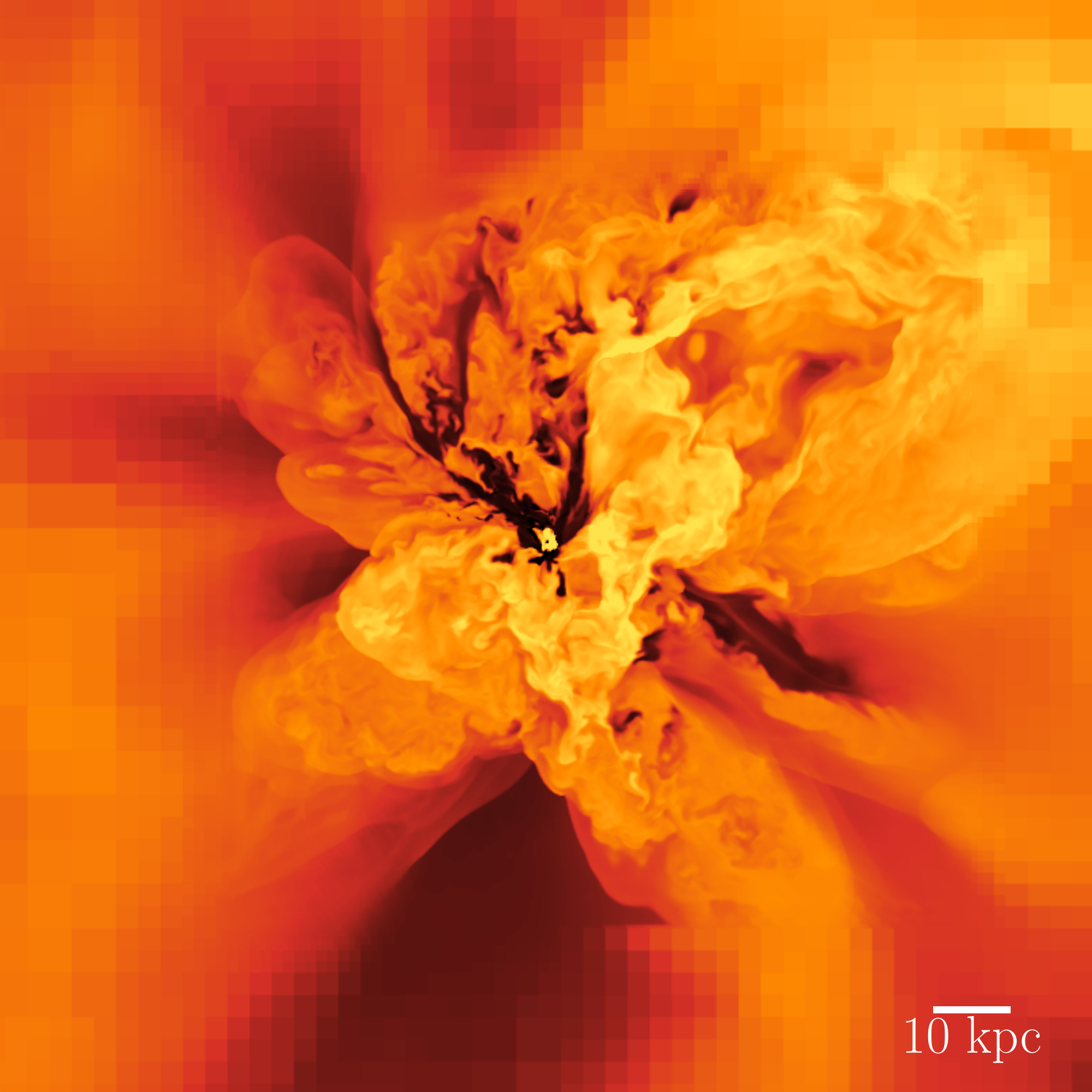}\hfill
    \includegraphics[height=0.45\textwidth]{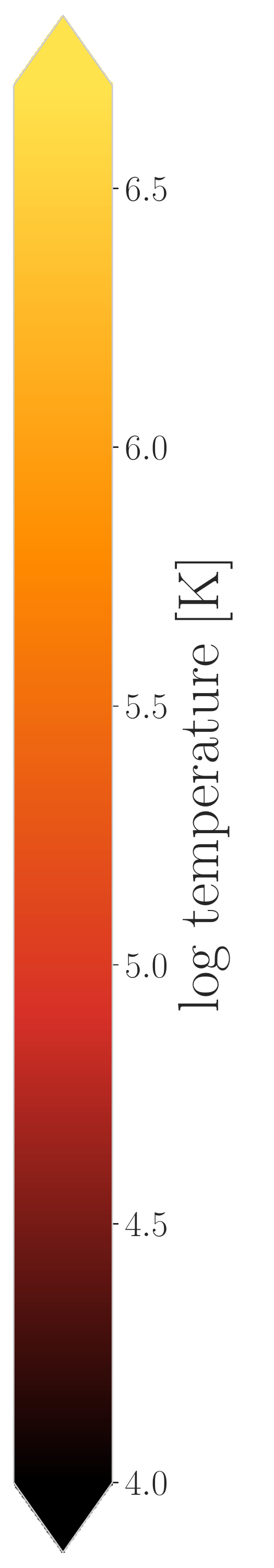}\\
    \includegraphics[height=0.45\textwidth]{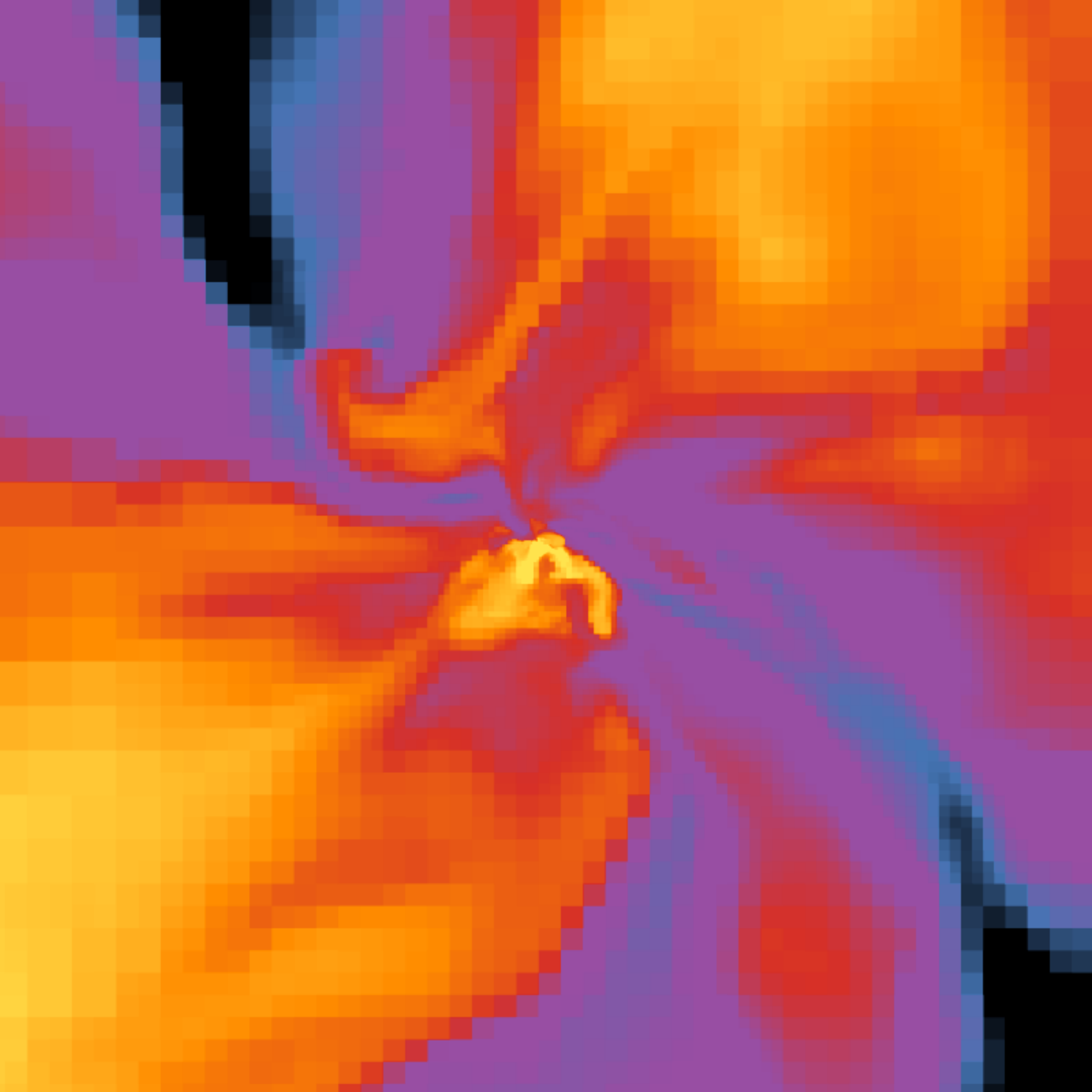}\hfill
    \includegraphics[height=0.45\textwidth]{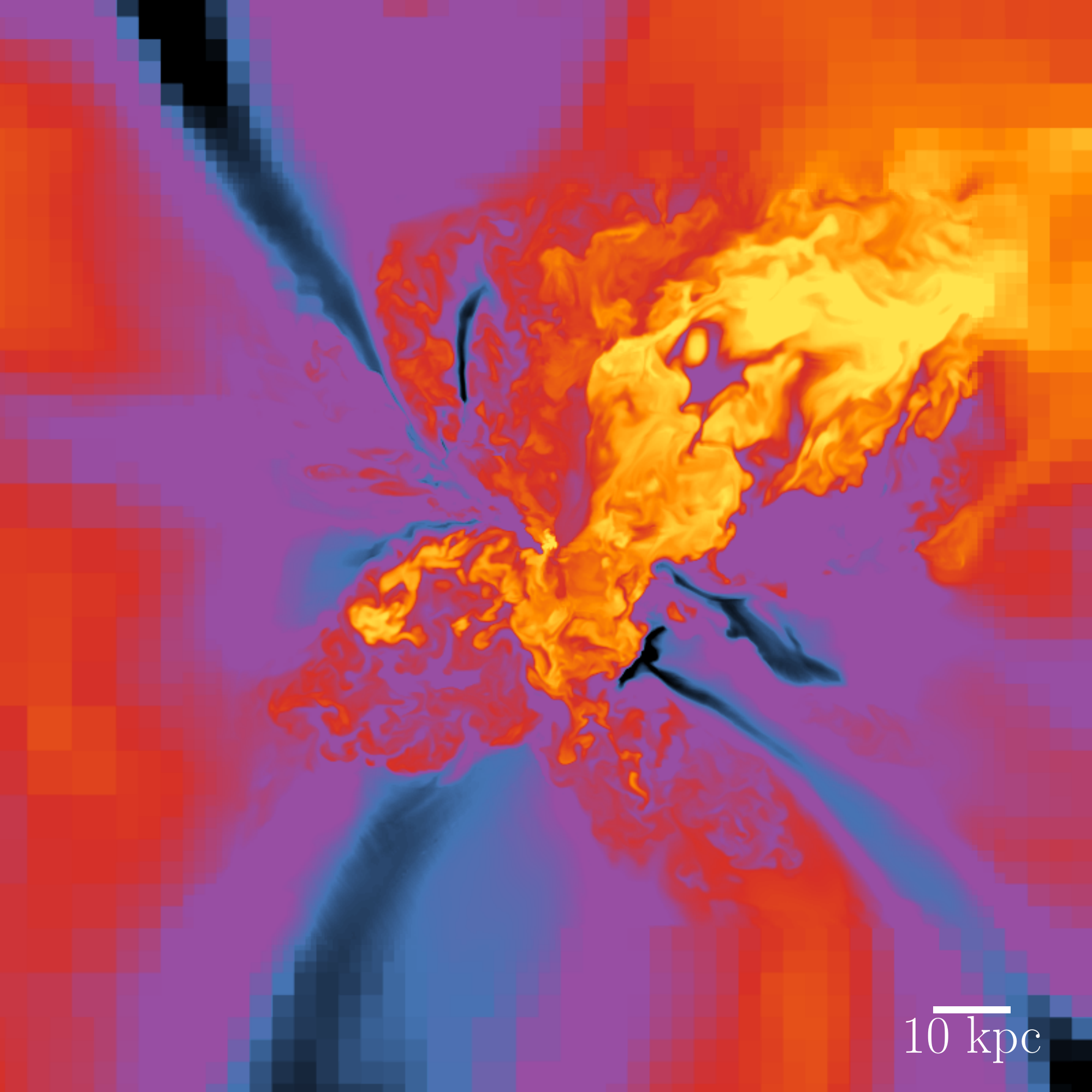}\hfill
    \includegraphics[height=0.45\textwidth]{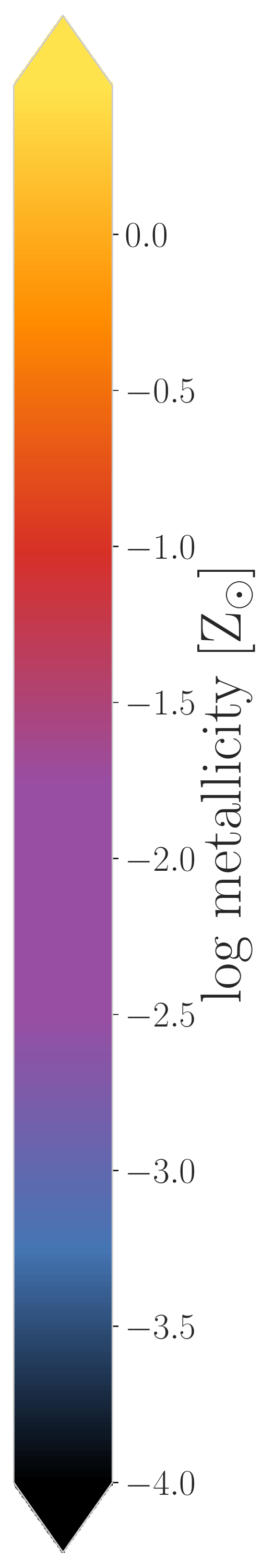}
    \caption{Slices of density ({\em top}), temperature ({\em middle}), and metallicity ({\em bottom}) through central halo in the standard ({\em left}) and high ({\em right}) resolution simulations at $z=2$.
    \label{fig:physical-slices}}
\end{figure*}

%%%%%%%%%%%%%%%%%%%%%%%%%%%%%%%%%%%%%%%%%%%%%%%%%%%%%%%%%%%%%%%%%%%%
%%%%%%%%%%%%%%%%%%%%%%   resolution and  %%%%%%%%%%%%%%%%%%%%%%%%%%%
%%%%%%%%%%%%%%%%%%%%%  physical state of gas %%%%%%%%%%%%%%%%%%%%%%%
%%%%%%%%%%%%%%%%%%%%%%%%%%%%%%%%%%%%%%%%%%%%%%%%%%%%%%%%%%%%%%%%%%%%

\section{The effects of resolution on the physical state of the gas} \label{sec:physical}
\begin{figure*}[htb]
\centering
    \includegraphics[angle=90,trim={7cm 1cm 10.5cm 1cm}, clip,height=0.142\textheight]{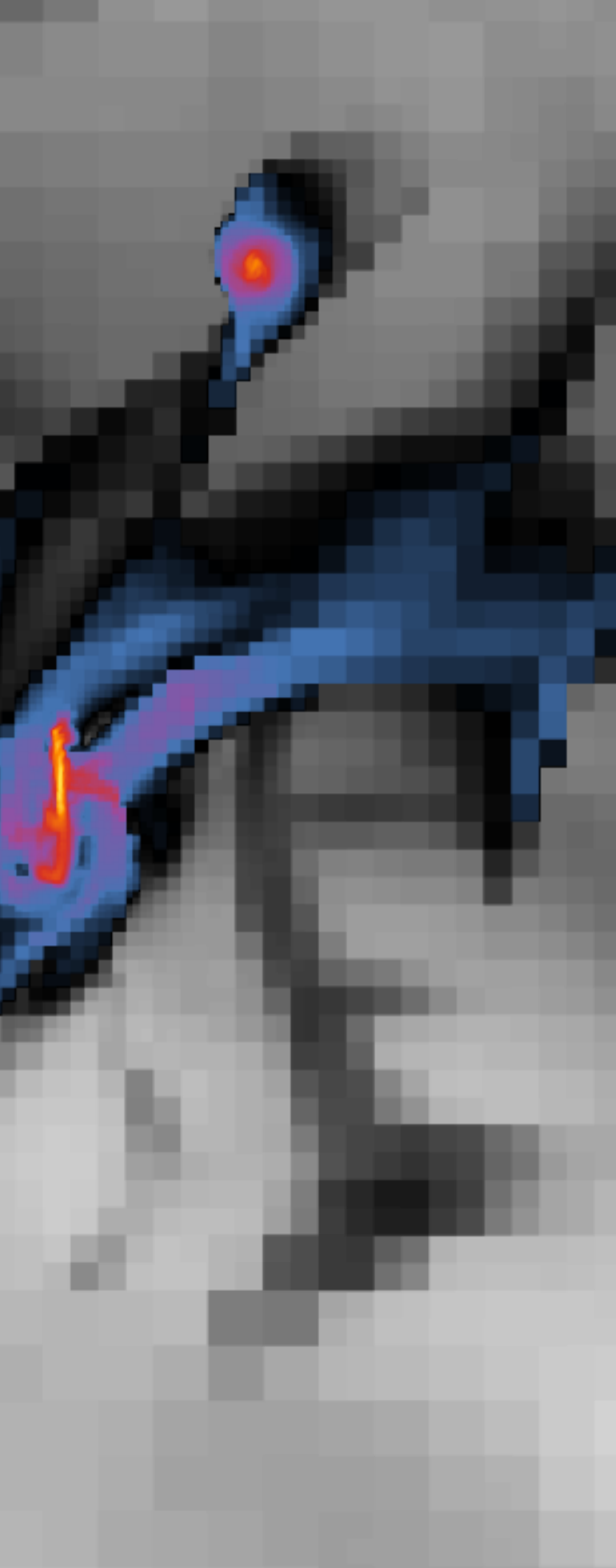}\hfill
    \includegraphics[angle=90,trim={7cm 1cm 10.5cm 1cm} ,clip,height=0.142\textheight]{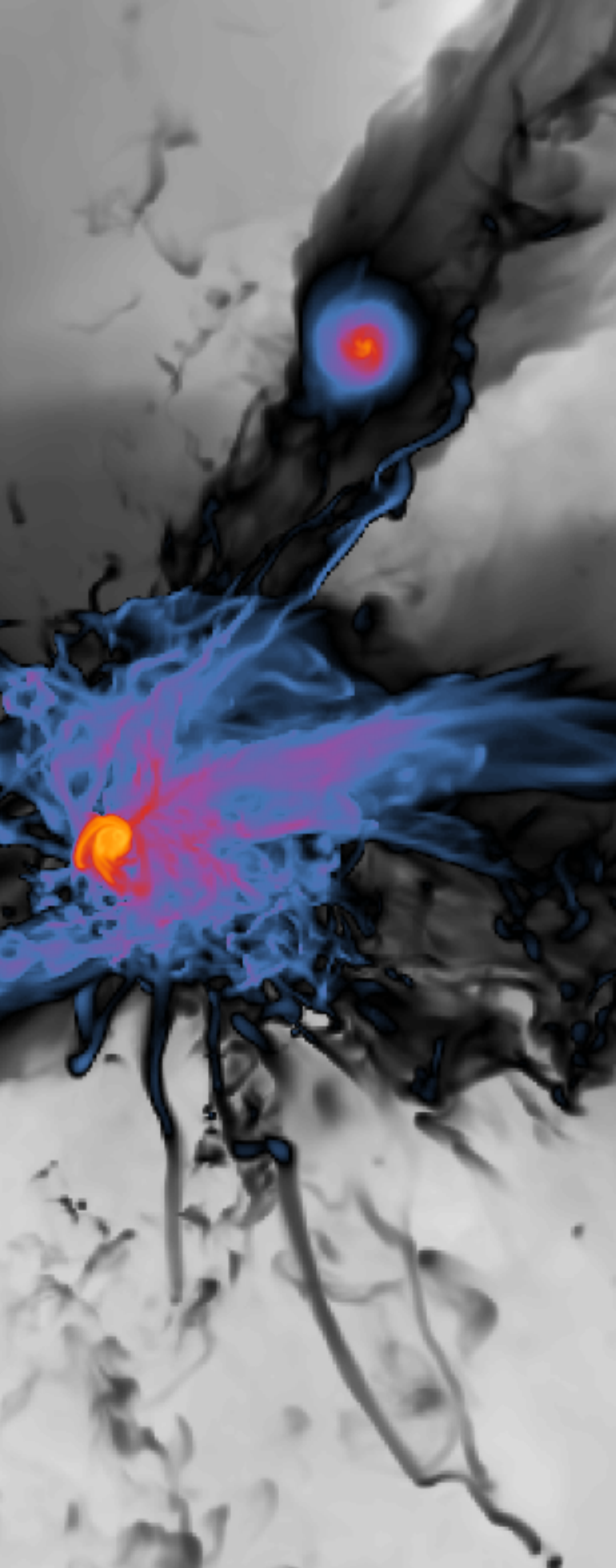}\hfill
    \includegraphics[height=0.142\textheight]{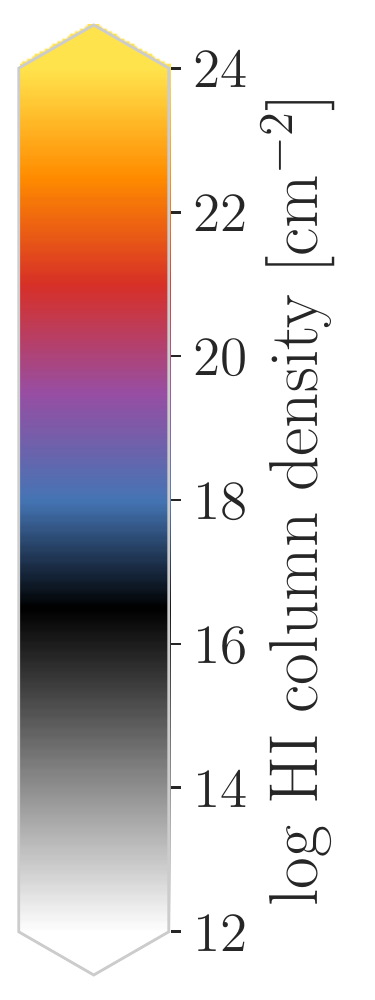}\\
    \includegraphics[angle=90,trim={7cm 1cm 10.5cm 1cm}, clip,height=0.142\textheight]{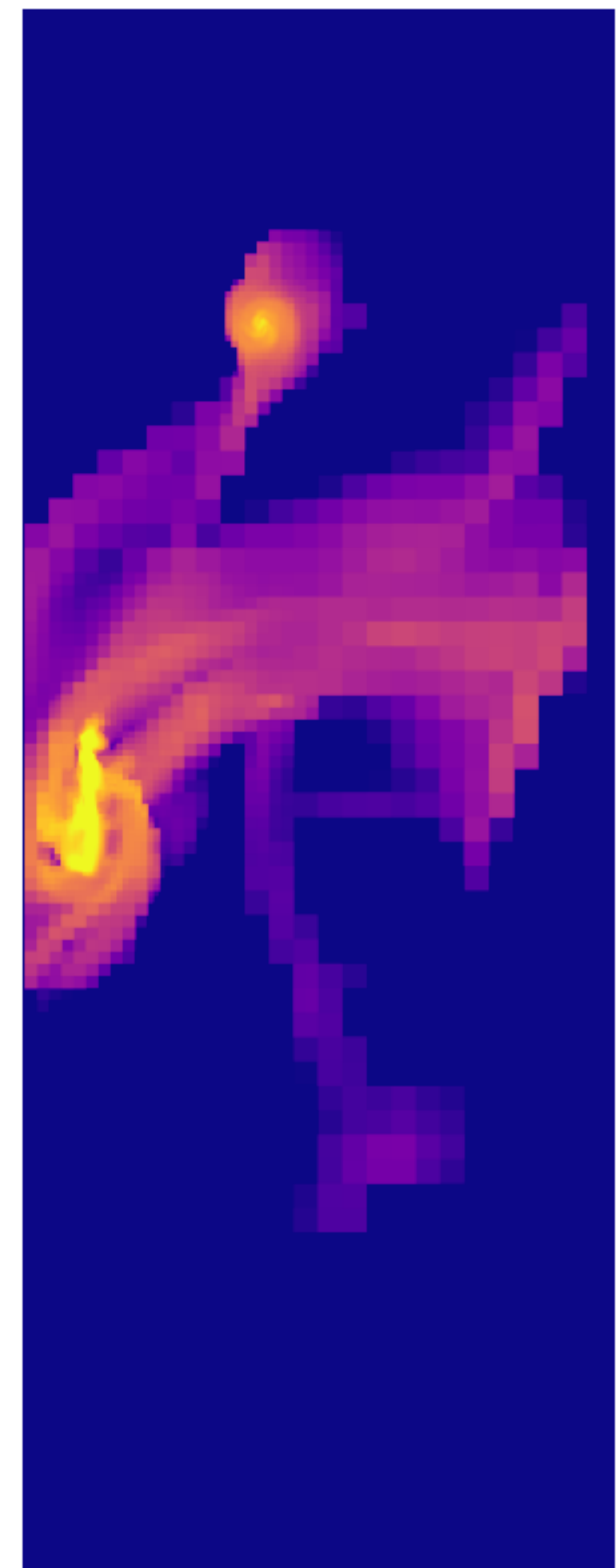}\hfill
    \includegraphics[angle=90,trim={7cm 1cm 10.5cm 1cm} ,clip,height=0.142\textheight]{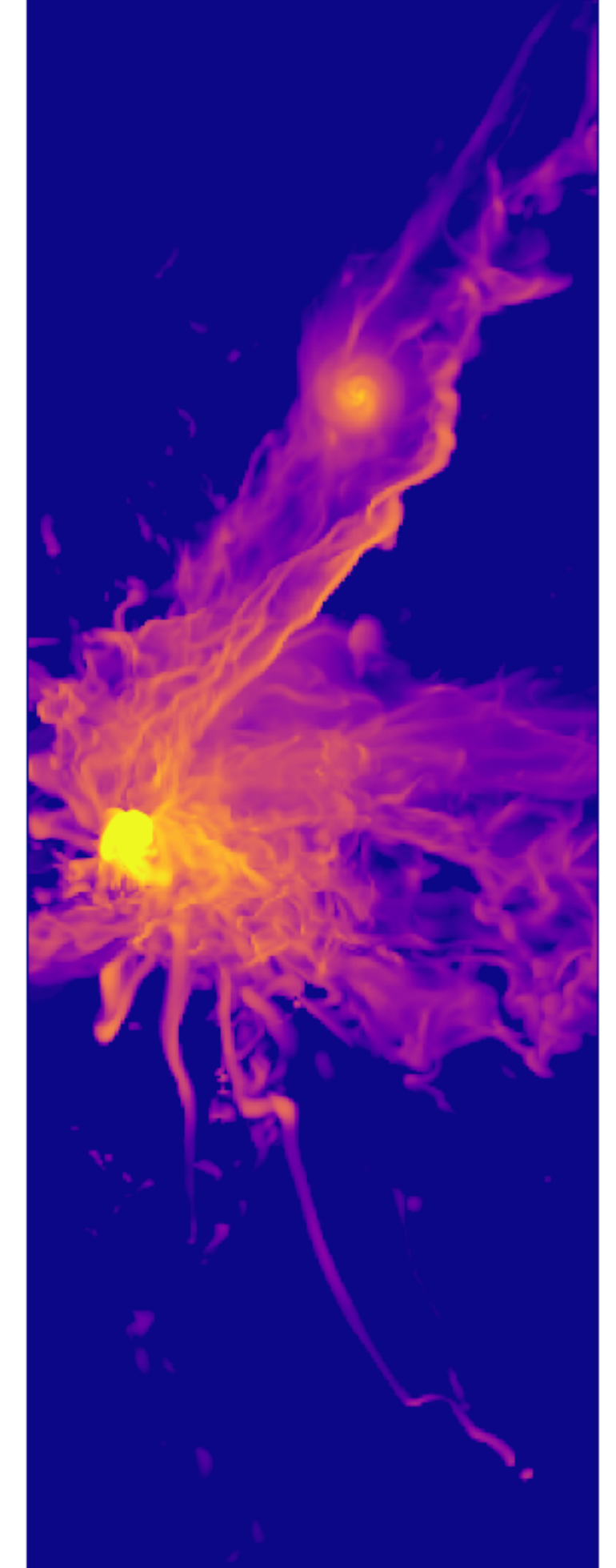}\hfill
    \includegraphics[height=0.142\textheight]{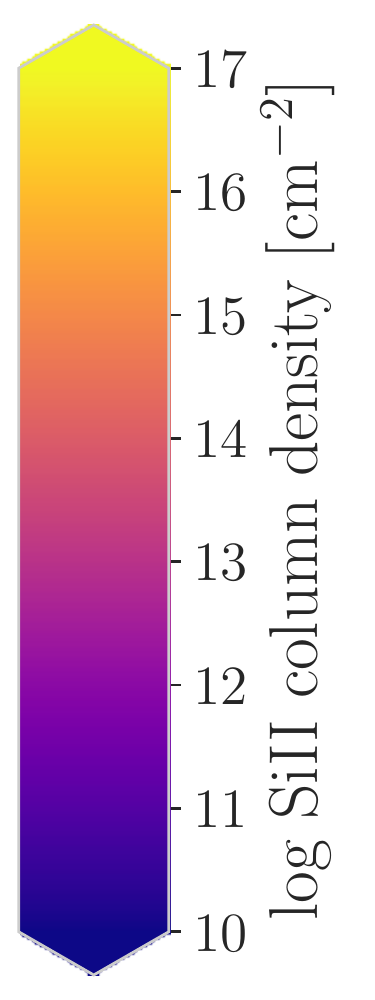}\\
    \includegraphics[angle=90,trim={7cm 1cm 10.5cm 1cm}, clip,height=0.142\textheight]{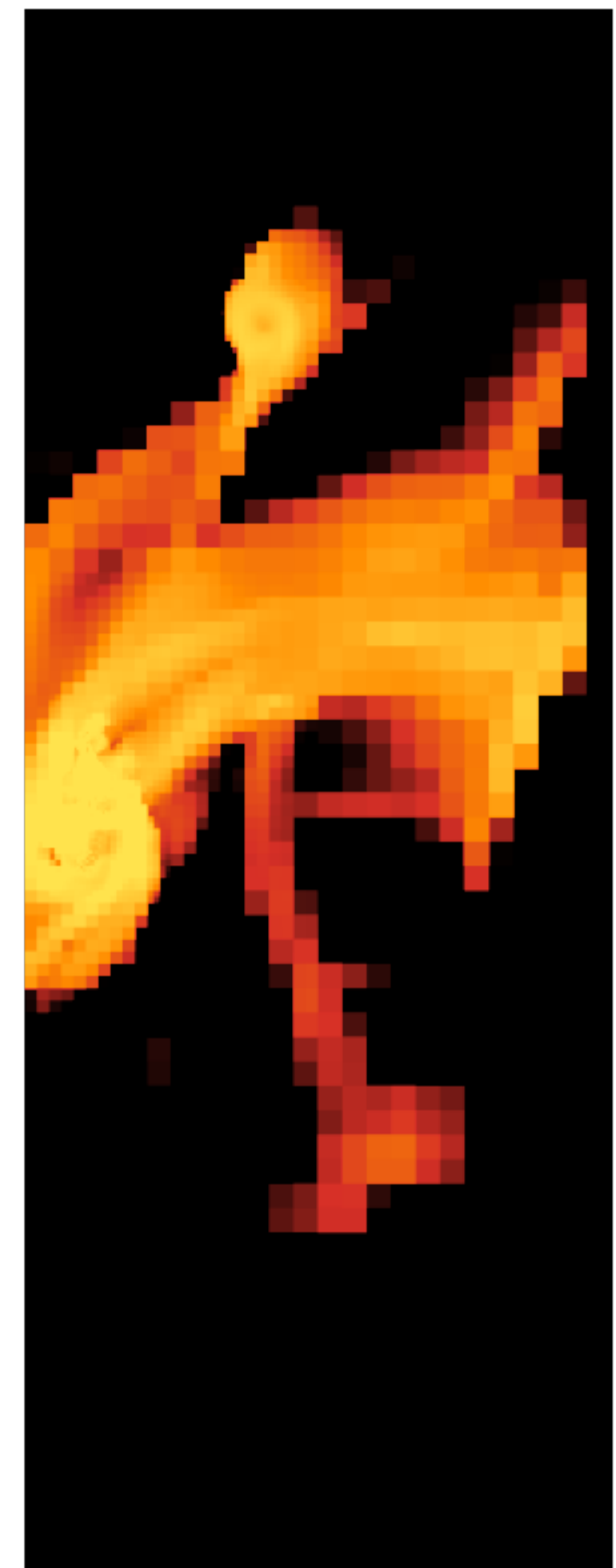}\hfill
    \includegraphics[angle=90,trim={7cm 1cm 10.5cm 1cm} ,clip,height=0.142\textheight]{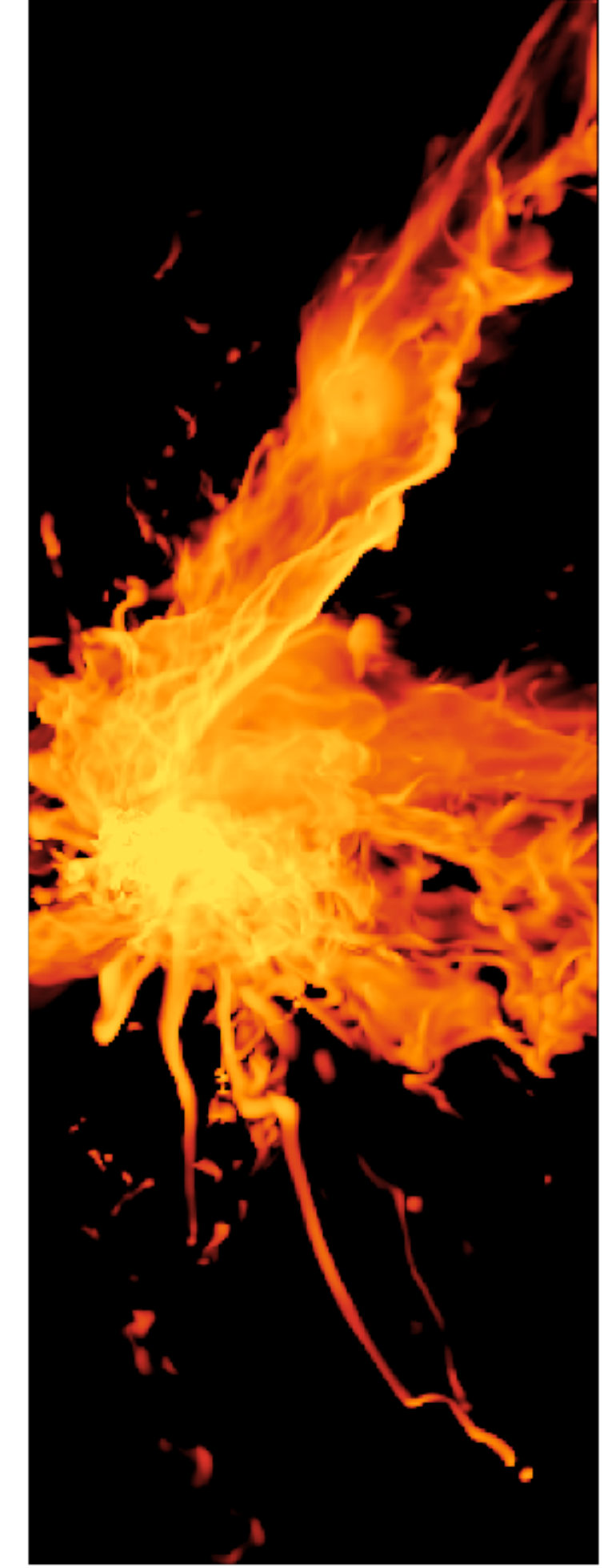}\hfill
    \includegraphics[height=0.142\textheight]{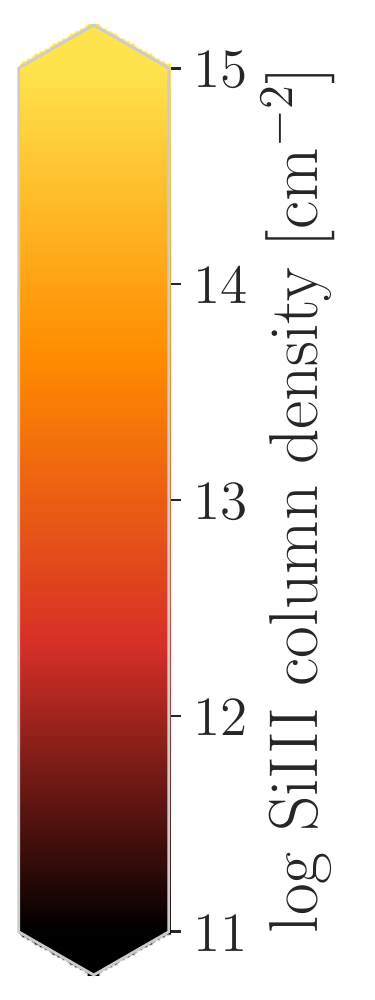}\\
    \includegraphics[angle=90,trim={7cm 1cm 10.5cm 1cm}, clip,height=0.142\textheight]{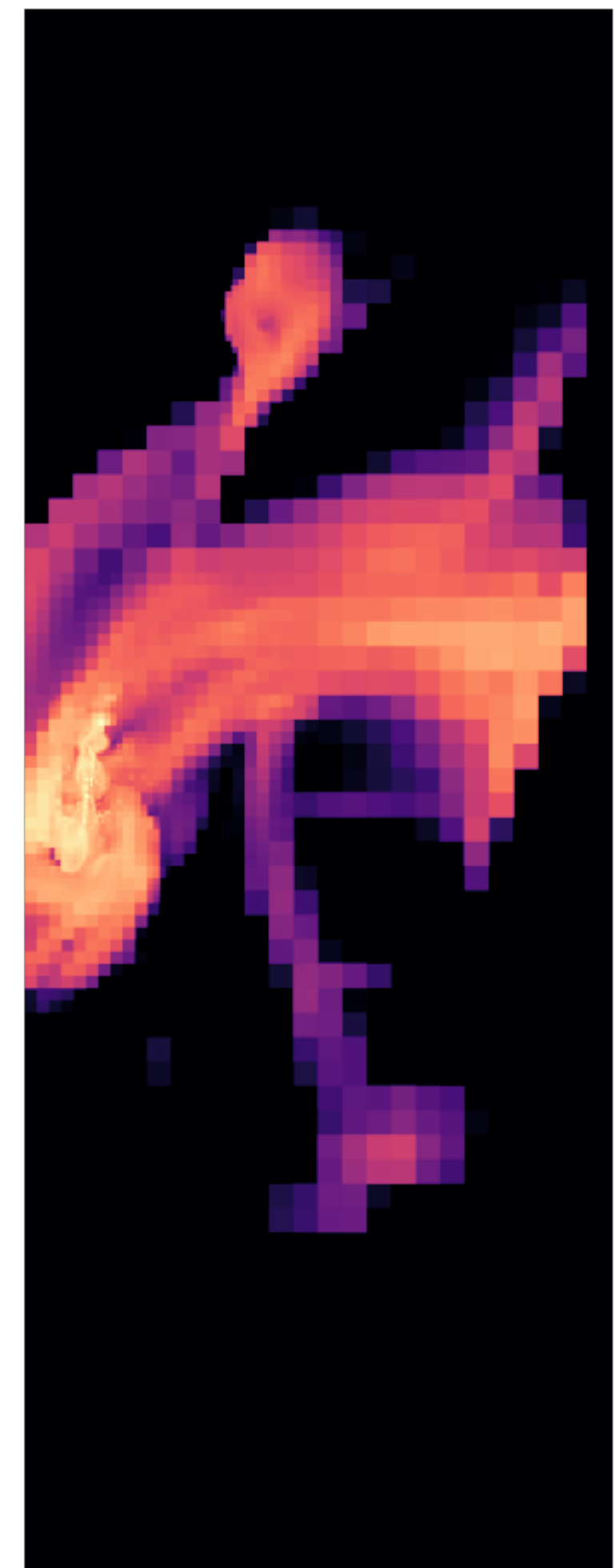}\hfill
    \includegraphics[angle=90,trim={7cm 1cm 10.5cm 1cm} ,clip,height=0.142\textheight]{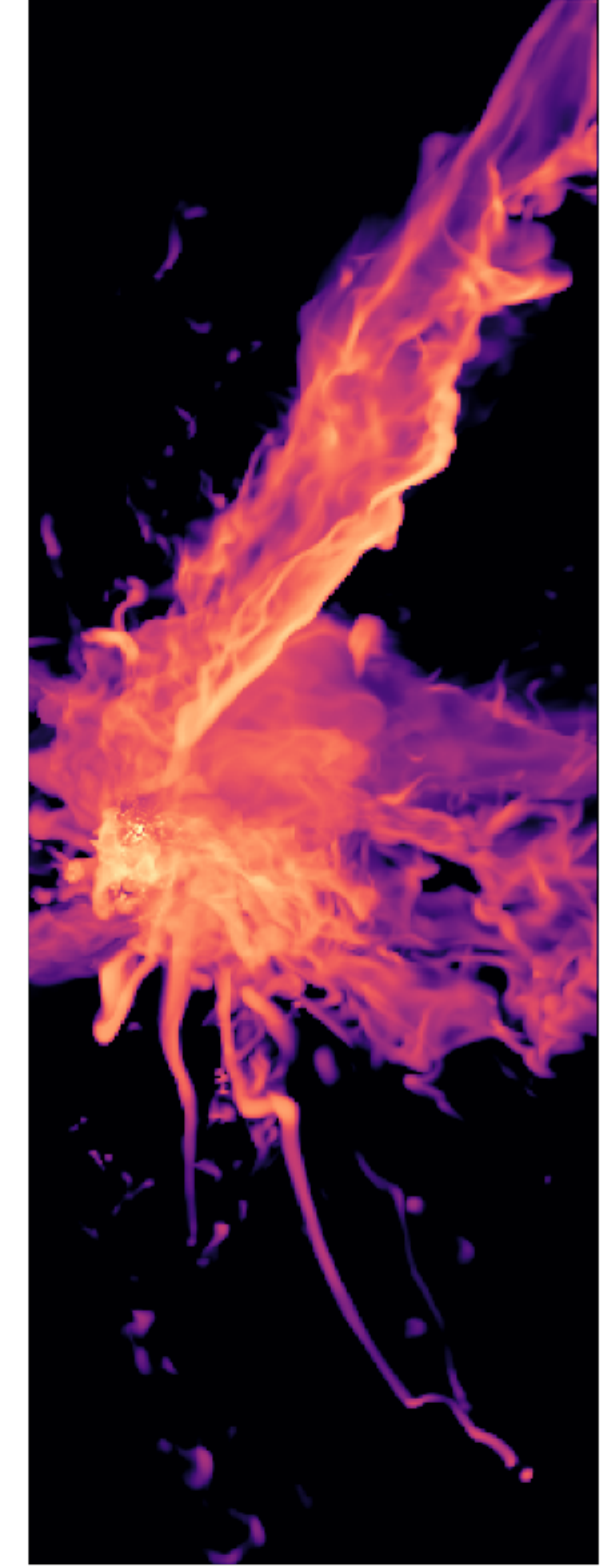}\hfill
    \includegraphics[height=0.142\textheight]{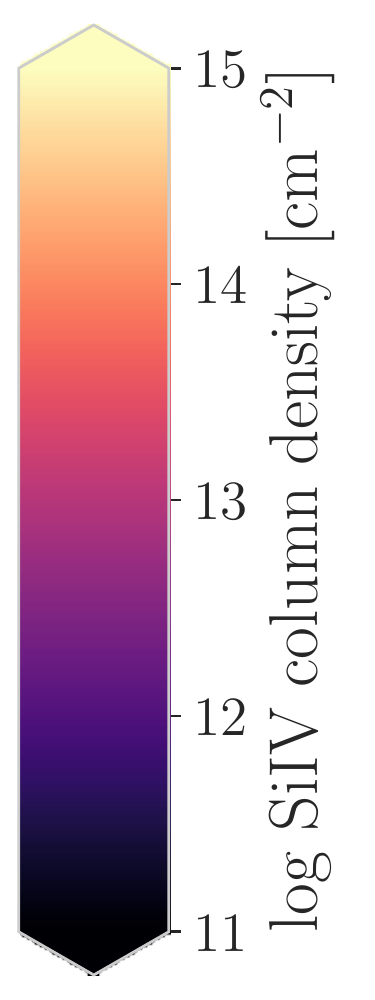}\\
    \includegraphics[angle=90,trim={7cm 1cm 10.5cm 1cm}, clip,height=0.142\textheight]{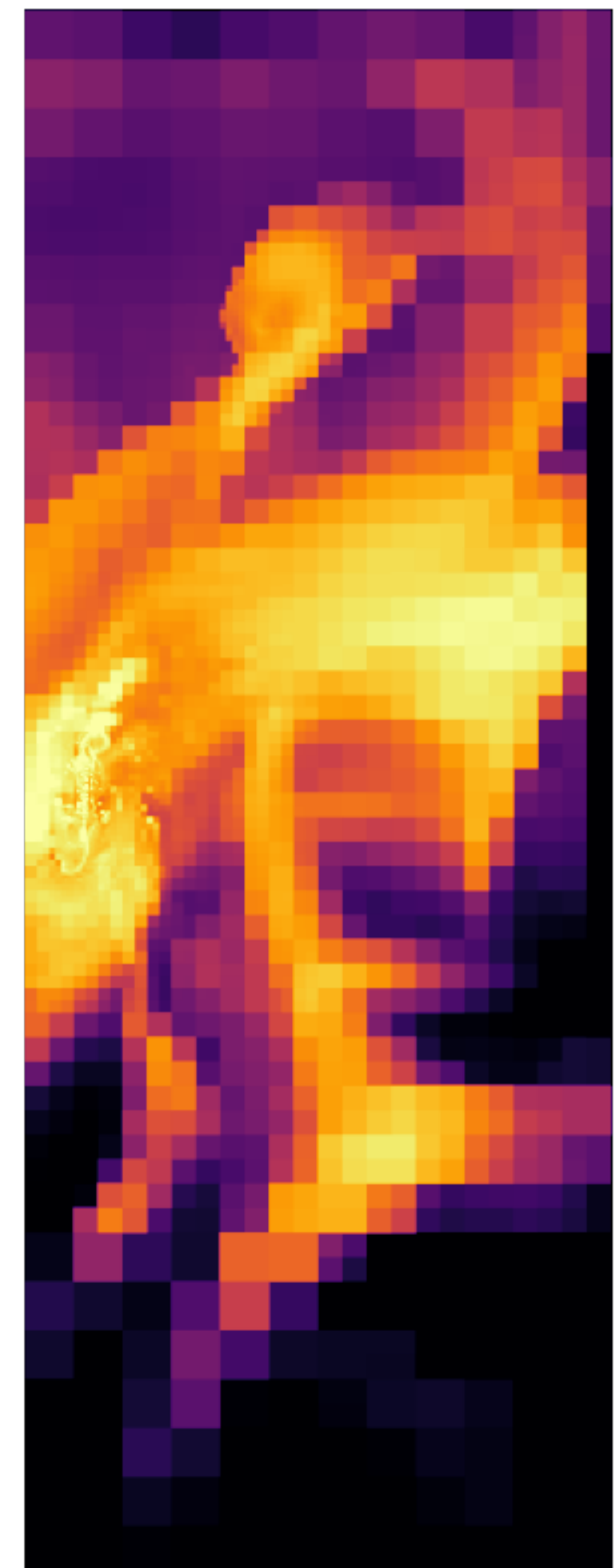}\hfill
    \includegraphics[angle=90,trim={7cm 1cm 10.5cm 1cm} ,clip,height=0.142\textheight]{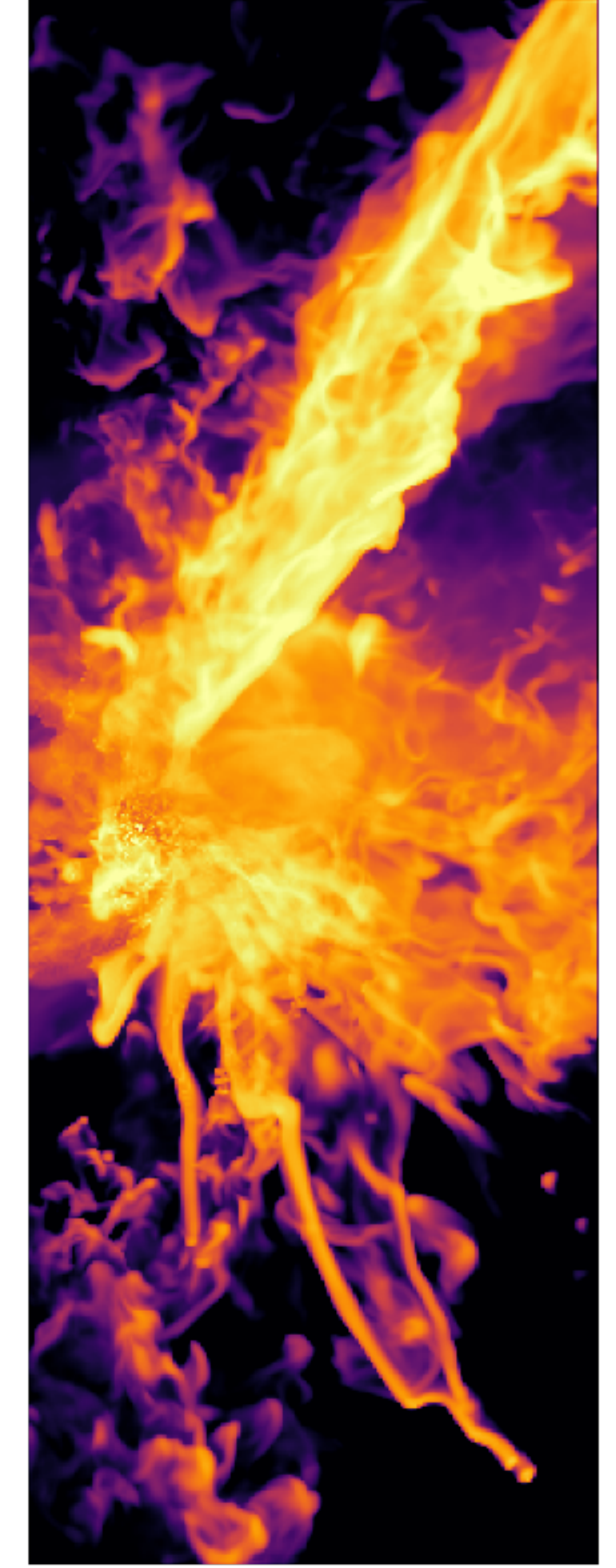}\hfill
    \includegraphics[height=0.142\textheight]{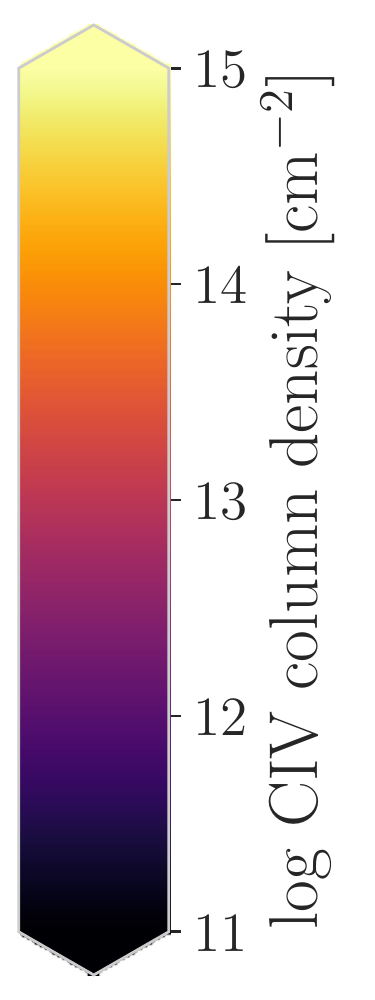}\\
    \includegraphics[angle=90,trim={7cm 1cm 10.5cm 1cm}, clip,height=0.142\textheight]{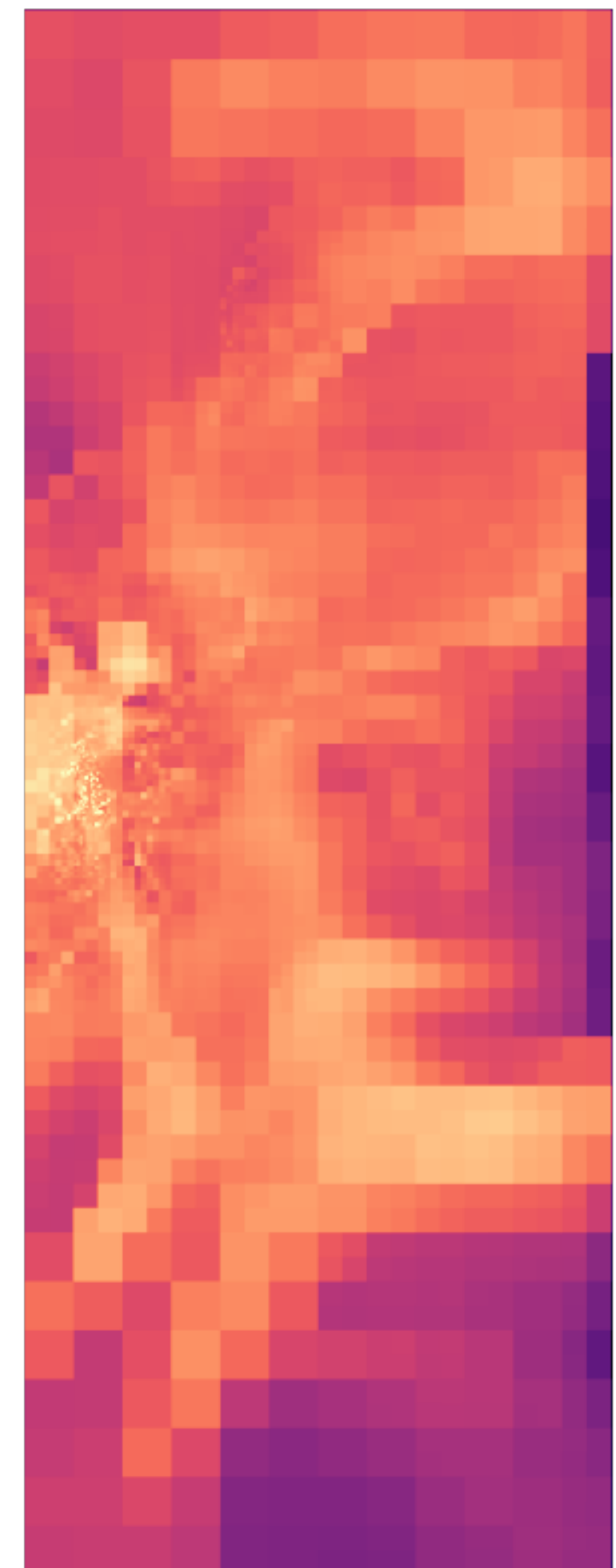}\hfill
    \includegraphics[angle=90,trim={7cm 1cm 10.5cm 1cm} ,clip,height=0.142\textheight]{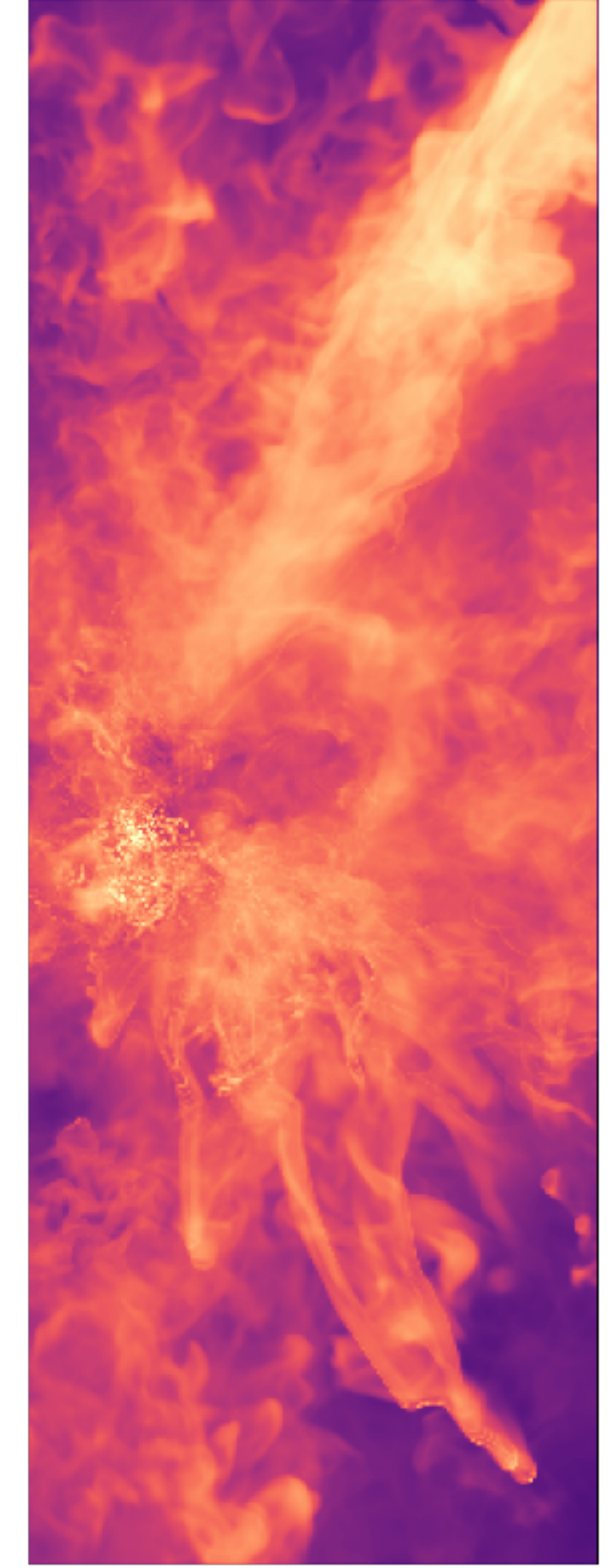}\hfill
    \includegraphics[height=0.142\textheight]{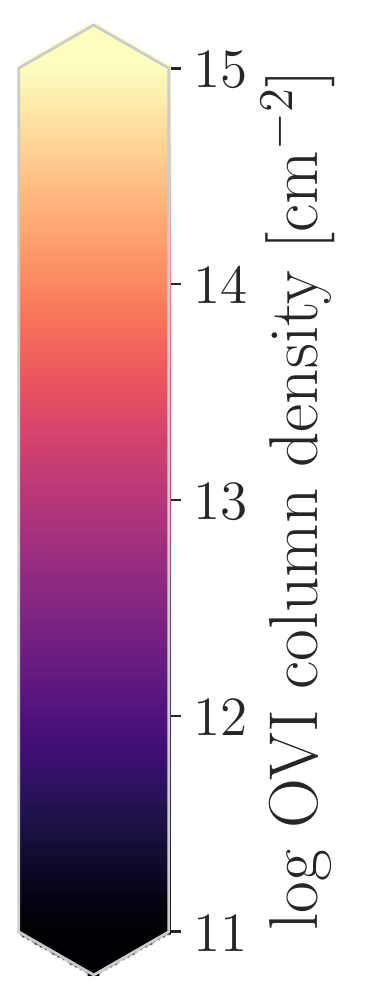}
    \caption{\hi, \siii, \siiii, \siiv, \civ, and \ovi\ projections at $z=2$ in the standard ({\em left}) and high-resolution ({\em right}) simulations. Each panel is 200\ckpch\ ($\sim 97$\,pkpc) across and deep. The colormap for \hi\ is chosen such that Lyman-limit gas is black/blue and DLA gas is red/orange, with optically-thin gas in greyscale.
    \label{fig:columns}}
\end{figure*}

\begin{figure}[htb]
\centering
    \includegraphics[width=0.45\textwidth]{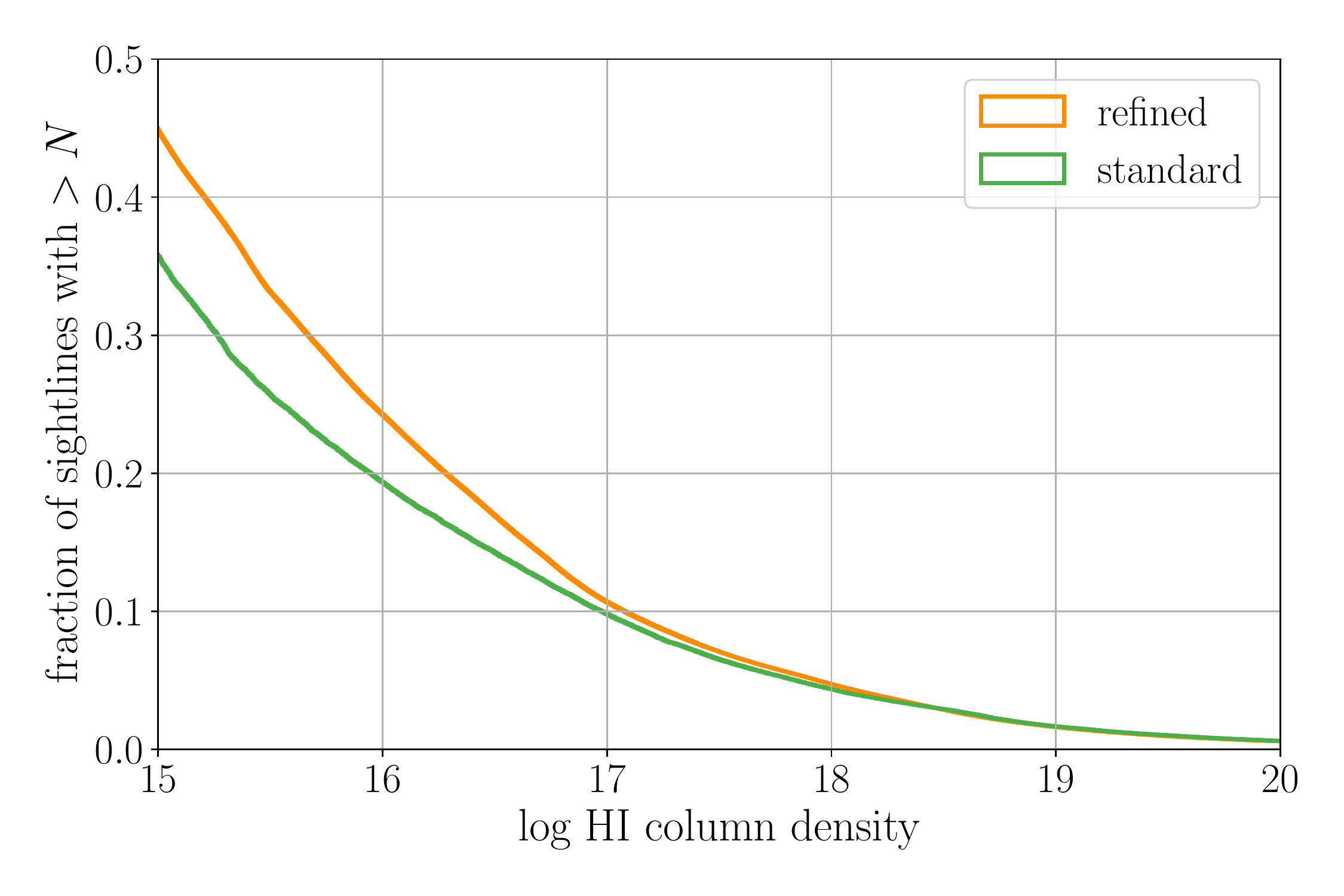}
    \caption{The cumulative \hi\ covering fraction for strong systems along each of the three cardinal axes at $z=2$ and $z=2.5$ in the two runs. The high-resolution simulation has much more strong \hi\ lines-of-sight than the standard resolution simulation.
    \label{fig:hicover}}
\end{figure}

\begin{figure*}[htb]
\centering
    \includegraphics[height=0.40\textheight]{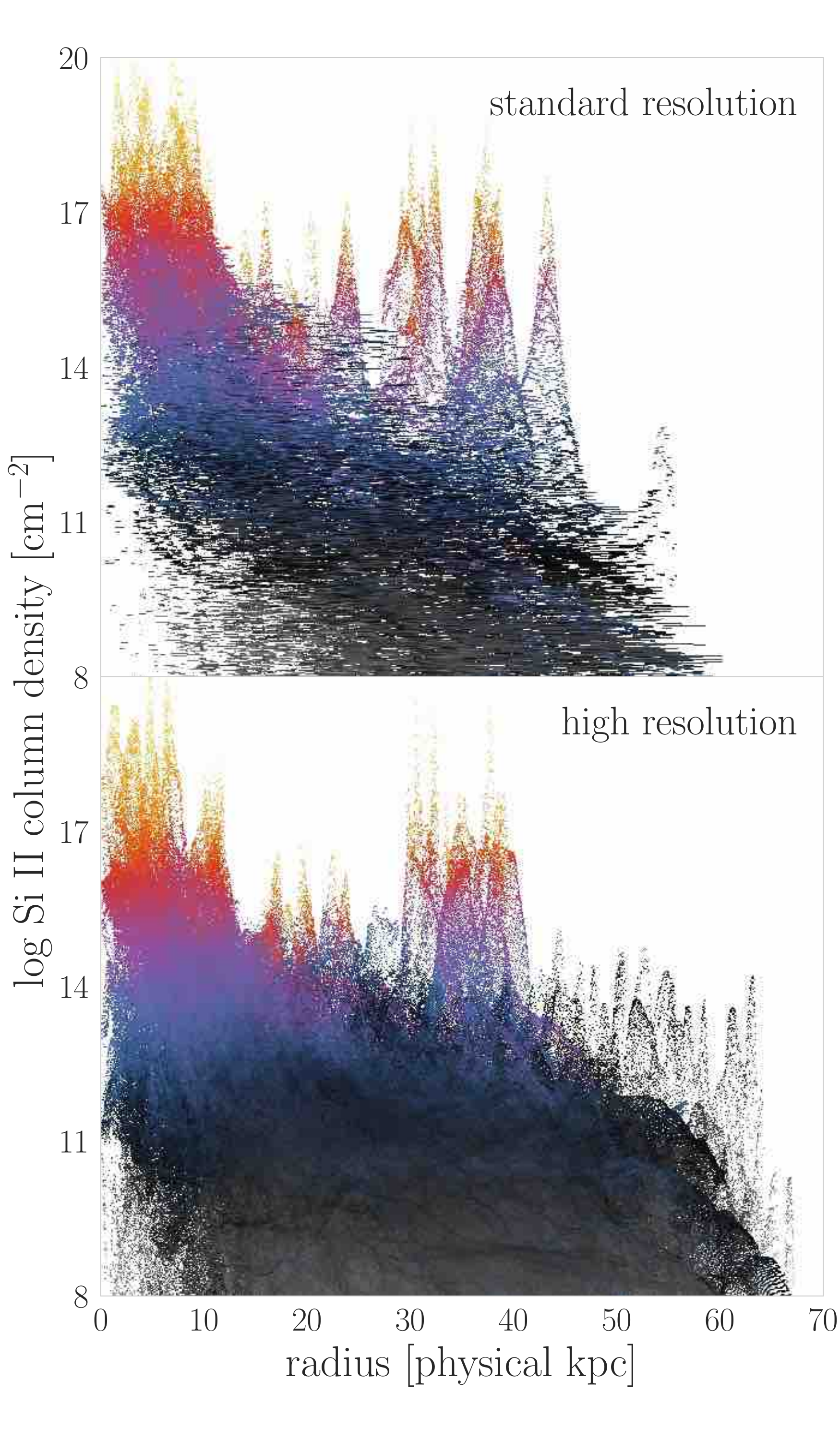}\hfill
    \includegraphics[height=0.40\textheight]{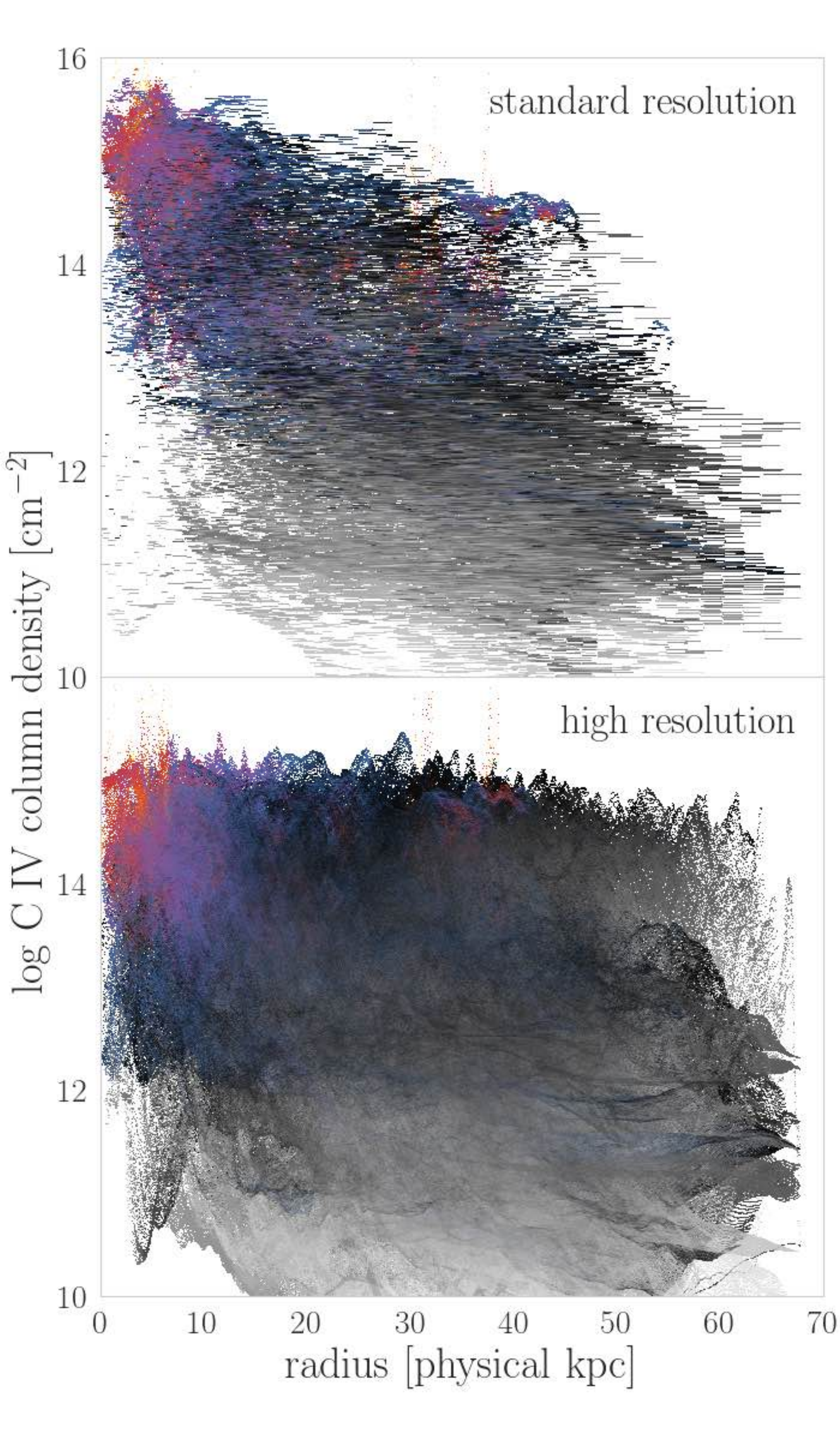}\hfill
    \includegraphics[height=0.40\textheight]{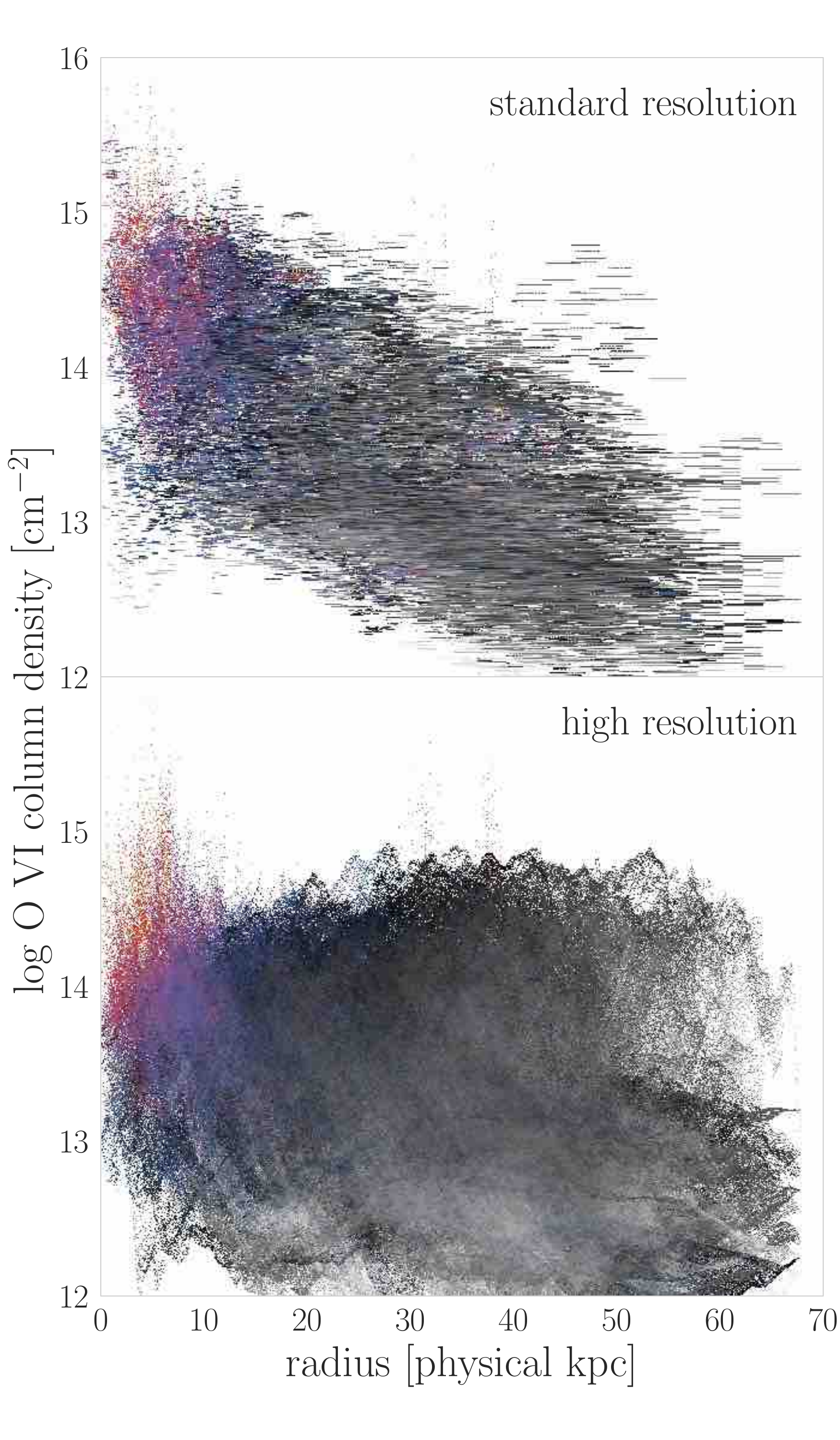}\hfill
    \includegraphics[height=0.40\textheight]{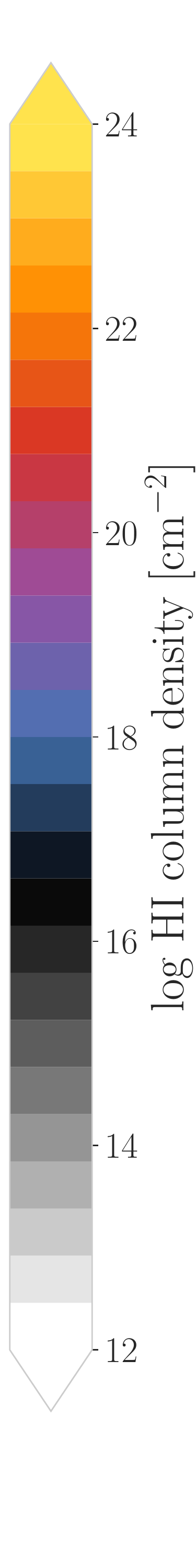}
    \caption{Radial profiles of \siii\ ({\em left}), \civ\ ({\em middle}), and \ovi\ ({\em right}) column density for the standard-resolution ({\em top}) and high-resolution ({\em bottom}) simulations along each of the three cardinal axes at $z=2$ and $z=2.5$ in the two runs, color-coded by the most frequent \hi\ column at each given radius and ionic column density. The standard-resolution simulation shows a much steeper radial trend for all species. \siii\ and \hi\ column are clearly correlated, whereas outside of the main halo, the higher ions are largely uncorrelated with \hi.
    \label{fig:columnprofiles}}
\end{figure*}

\begin{figure*}[htb]
\centering
    \includegraphics[width=0.33\textwidth]{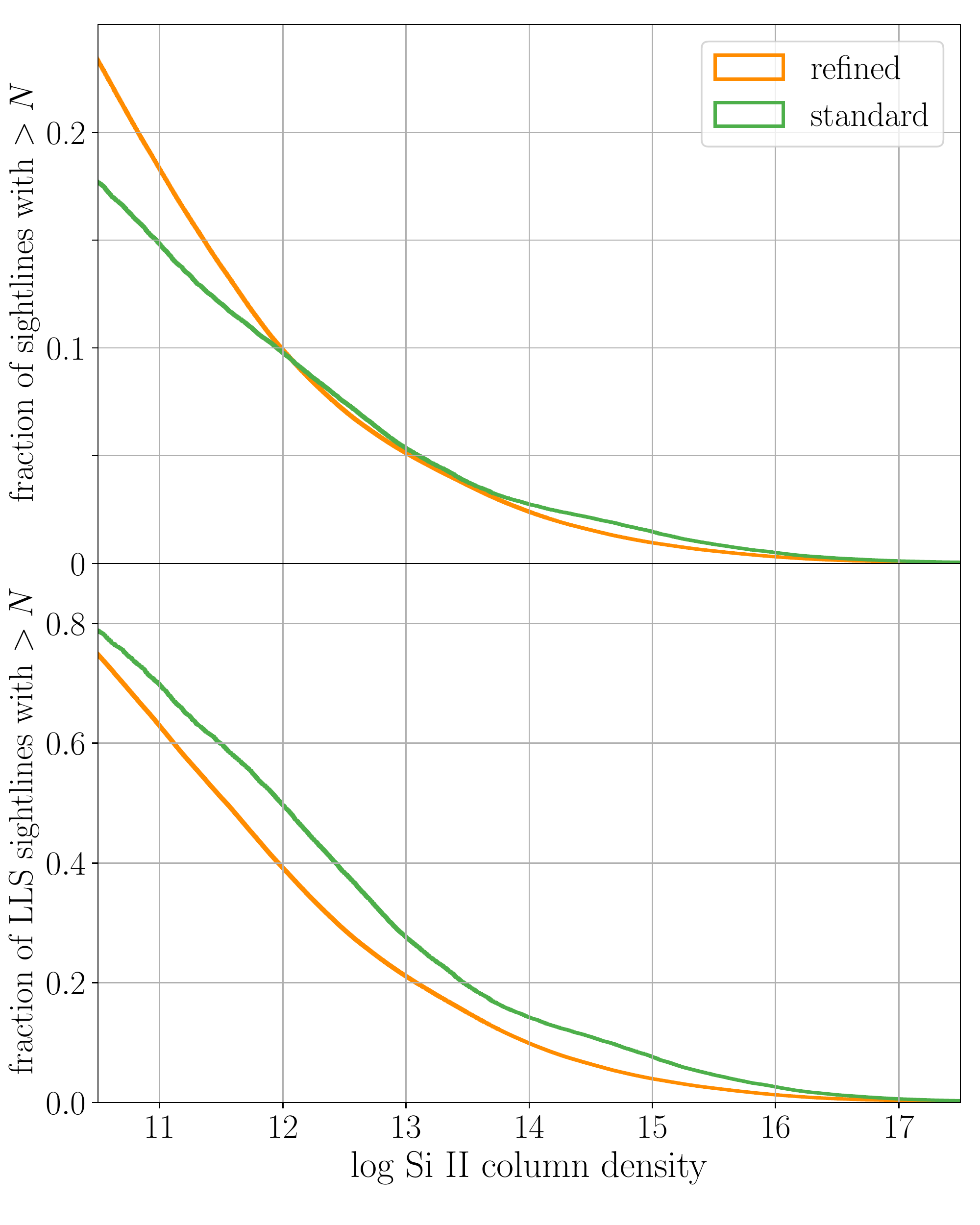}\hfill
    \includegraphics[width=0.33\textwidth]{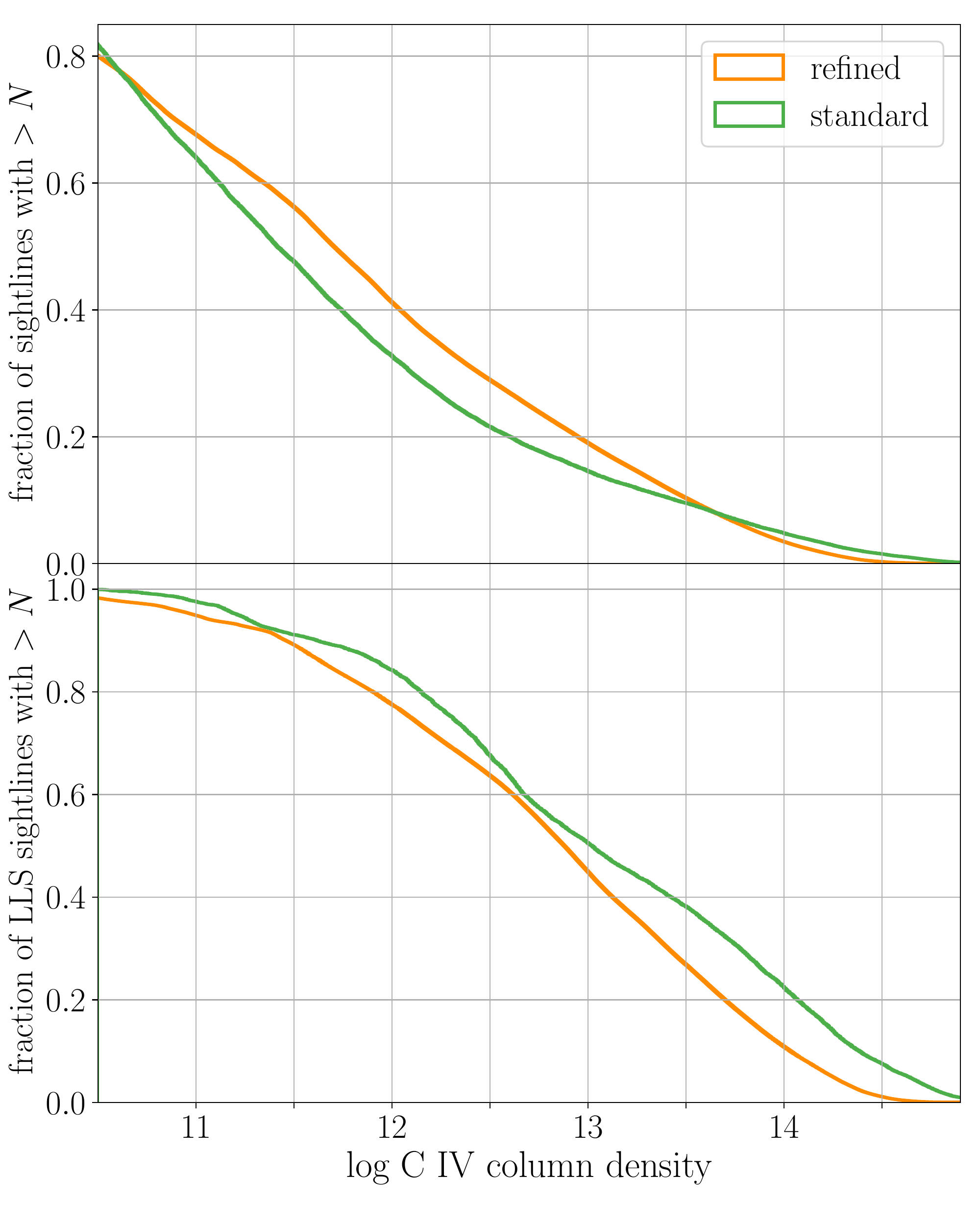}\hfill
    \includegraphics[width=0.33\textwidth]{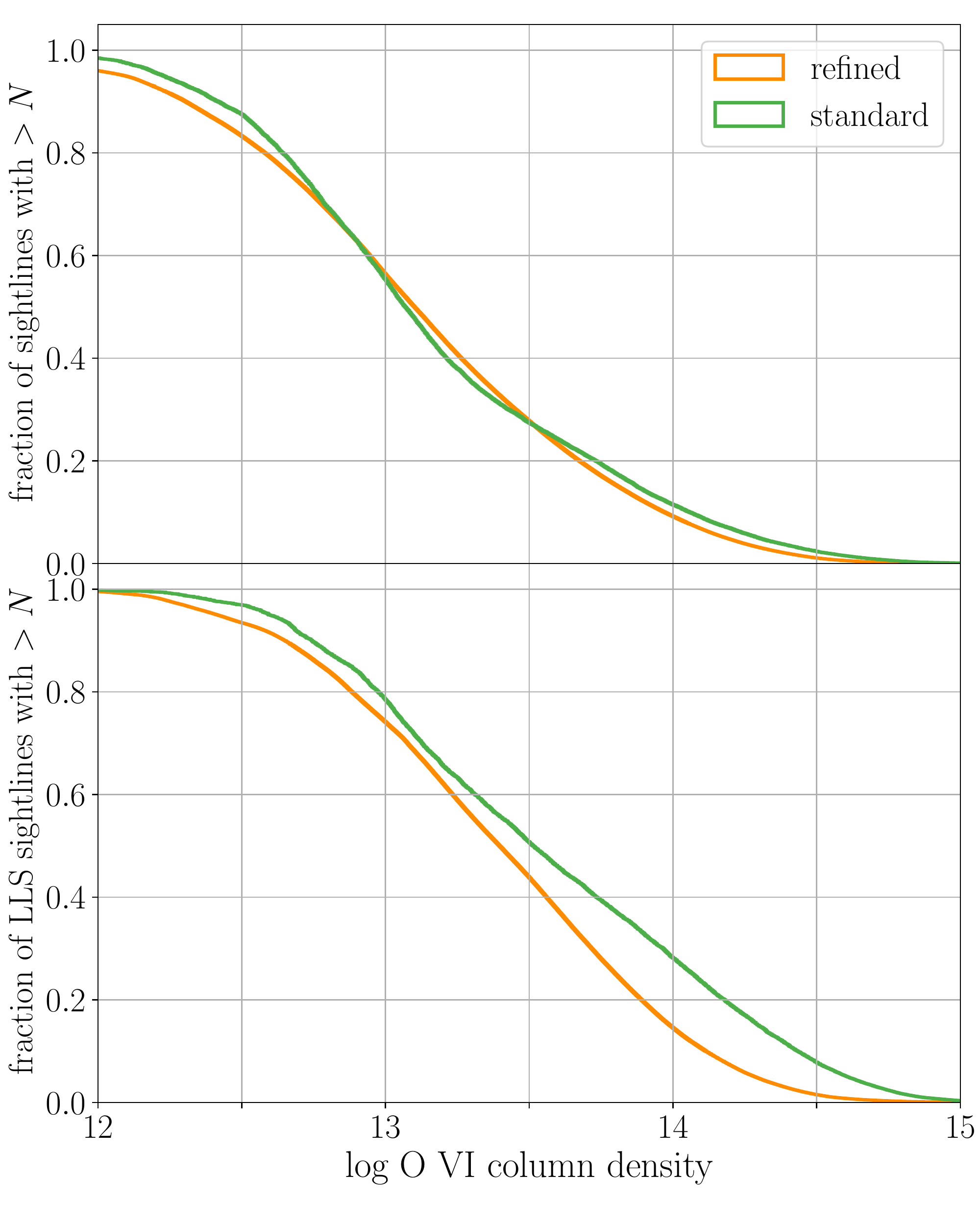}
    \caption{The \siii, \civ, and \ovi\ covering fraction for both all sightlines ({\em top}) and sightlines with $\log N_{\rm H\,I}> 16$ ({\em bottom}) along each of the three cardinal axes at $z=2$ and $z=2.5$ in the two runs. While the high-resolution simulation has a similar amount of strong metal lines as in the standard simulation, these systems are not as confined to the strong-\hi\ sightlines. While these differences are somewhat subtle, they are quite noticeable in the comparison plots we discuss in \S\,\ref{sec:comparison}.
    \label{fig:metalcover}}
\end{figure*}

We show slices of density, temperature, and metallicity through the halo at $z=2$ in both the standard and high-resolution runs in
Figure~\ref{fig:physical-slices}; as in Figure~\ref{fig:refinement-slices}, each panel is $1.5\times 200 h^{-1}$\,ckpc across to fully encompass the forced refinement region. In the standard resolution simulation it is evident that the resolution directly tracks the density of the gas, while the boundaries of the forced-refinement volume stand out clearly in the high resolution case. While the same structures are likely no longer exactly the same in the two runs owing in part to the ``butterfly effect'' \citep{genel18}, the high-resolution gas is visually much more turbulent, with more well-defined large-scale bubbles and shocks from outflowing gas than in the standard resolution simulation. 
The metallicity adds an extra dimension to these flows: in the high-resolution simulation, the high-metallicity outflowing gas is not forcibly ``over-mixed'' with its environs because it is not derefining as it expands and decreases in density. The high metallicity of the outflowing gas is maintained to much larger radii, suggesting that the difficulty of simulations to reproduce the observed pockets of high-metallicity CGM gas \citep{prochaska17} owes in part to unnatural overmixing. Likewise, the {\em low}-metallicity inflowing gas is able to penetrate to much closer to the galaxy when well-resolved than in the standard resolution simulation. We speculate that these low-metallicity flows are related to the rare low-metallicity Lyman-limit systems observed at $z>2$ \citep{fumagalli11,lehner16} and may be naturally generated in larger simulations with improved circumgalactic (or intergalactic) resolution.

The physical differences shown in Figure~\ref{fig:physical-slices} translate to quantitative differences in how the dynamic multiphase circumgalactic medium is physically structured and subsequently manifested in different ionic tracers. Though the {\em average} properties of the CGM at the two resolutions are broadly the same, the ionic absorption line tracers used to diagnose the CGM's physical structure are highly sensitive to the {\em distribution} of gas temperature, density, and metallicity (see, e.g., Figure~6 of \citealt*{tumlinson17}). The higher levels of both cold {\em and} hot gas in turn has important consequences for the column density distributions and covering fractions of commonly observed transitions, 
(as illustrated in Figure~7 and discussed in more detail in \S\,4.1 of \citetalias{corlies18}). 
This effect is evident in Figure~\ref{fig:columns}, where we show projected \hi, \siii, \siiii, \siiv, \civ, and \ovi\ column densities at $z=2$ in the two simulations; the central halo is in the bottom center of each panel. For all species, the high-resolution simulation reaches higher column densities out to higher impact parameter, and small-scale clouds and filamentary features with internally-resolved structure are clearly visible in the high-resolution projections. 

\begin{table}[hb]
\centering
\begin{tabular}{ cccr } 
 Species & High-Resolution & Standard Resolution & Ratio\\ 
 \hline
stars & $1.289\times 10^{10}$ & $1.118\times 10^{10}$ & $1.15\phantom{0}$ \\
 all gas & $1.082\times 10^{10}$ & $1.215\times 10^{10}$ & $0.890$\\
\hi\ & $3.412\times 10^{9}$ &  $4.003\times 10^{9}$ & $0.852$\\
\siii\ & $4.686\times 10^{6}$ & $5.727\times 10^{6}$ & $0.818$ \\ 
\siiv\ & $1.421\times 10^{4}$ &  $5.048\times 10^{2}$  & $28.2\phantom{00}$ \\ 
\civ & $4.119\times 10^{4}$ & $3.529\times 10^{4}$ & $1.17\phantom{0}$\\
\ovi & $4.400\times 10^{4}$  & $4.429\times 10^{4}$ & $0.993$\\
\end{tabular} 
\caption{Mass of all gas and different ionic species within the $(200\,{\rm ckpc}/h)^3$ forced refinement volume in the two simulations at $z=2$; masses are given in \Msun. The rightmost column gives the ratio of the mass in the high-resolution simulation to that in the standard-resolution simulation. The total baryonic masses within the refined region (stars $+$ gas) differ by $<2$\%.} \label{tbl:mass}
\end{table}

We plot the cumulative $N_{\rm H\,I}$ covering fractions in Figure~\ref{fig:hicover}, considering the three cardinal axes at both $z=2$ and 2.5.
We find that the high-resolution simulation produces a higher covering fraction of strong \hi\ absorbers than the standard-resolution simulation, though at $\log N_{\rm H\,I}>18$ this effect is subtle. We summarize the masses of different species of interest within the forced refinement region at $z=2$ in Table~\ref{tbl:mass}; despite the higher prevalence of strong \hi\ absorbers, the  total \hi\ mass within the forced-refinement region in the high-resolution simulation is $\sim 15$\% {\em lower} than in the standard-resolution simulation. This difference, however, can be mostly explained by the difference in the total amount of {\em gas} in the two boxes, which in turn owes to some combination of more star formation and (perhaps) more efficient outflows in the high-resolution simulation. 

Using the AREPO code, \citet{vandevoort19} similarly found that improving circumgalactic resolution increased the amount of \hi\ and the incidence of strong \hi\ absorbers, bringing their simulated galaxies more in line with the COS-Halos $z\sim 0.25$ \hi\ observations \citep{thom12, tumlinson13, prochaska17}.  This similarity of result given the different hydrodynamic codes, initial conditions, and redshifts suggests that resolving smaller circumgalactic structures generically increases incidence of strong \hi\ absorbers.

\begin{figure*}[htb]
\centering
    \includegraphics[width=0.33\textwidth]{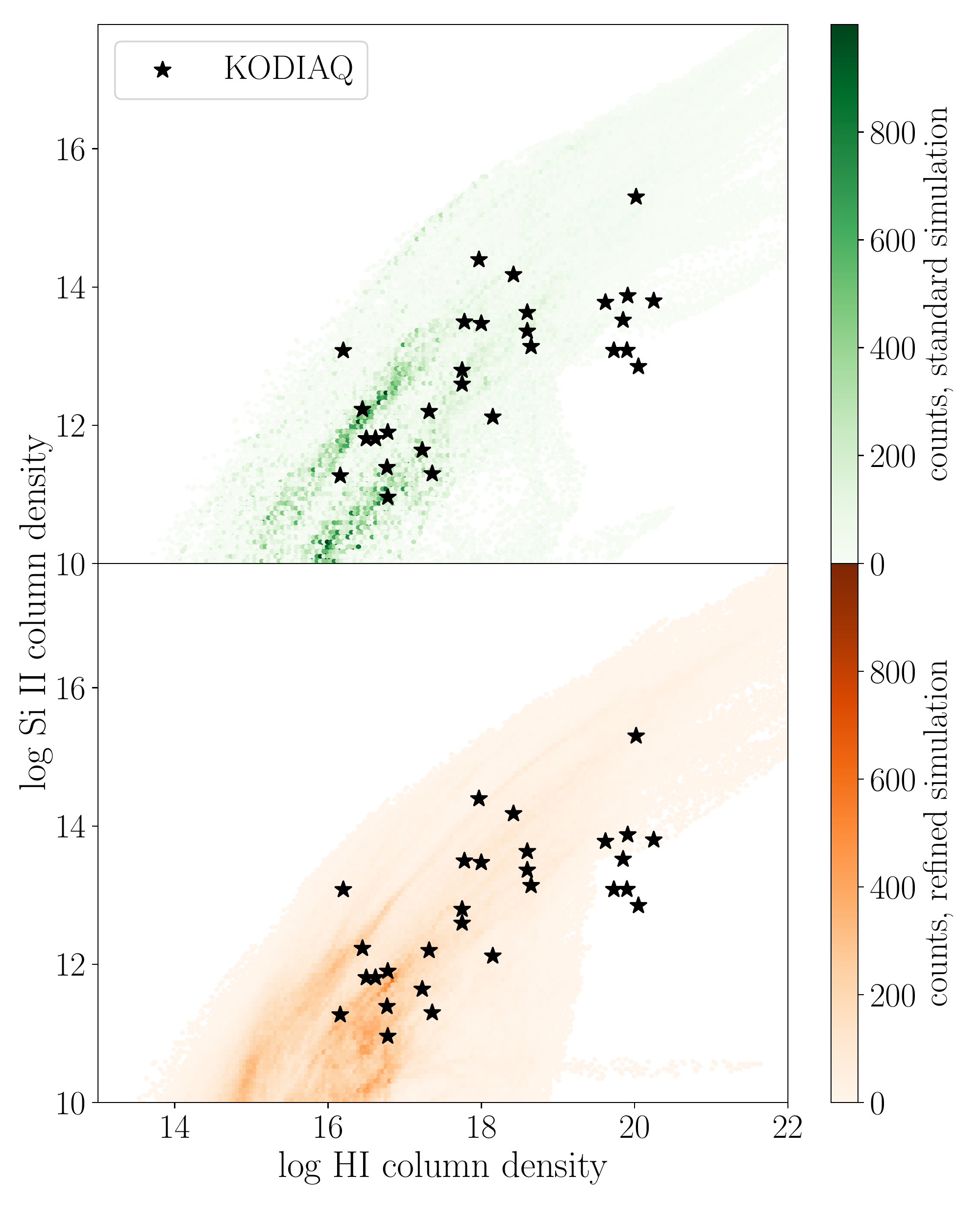}\hfill 
    \includegraphics[width=0.33\textwidth]{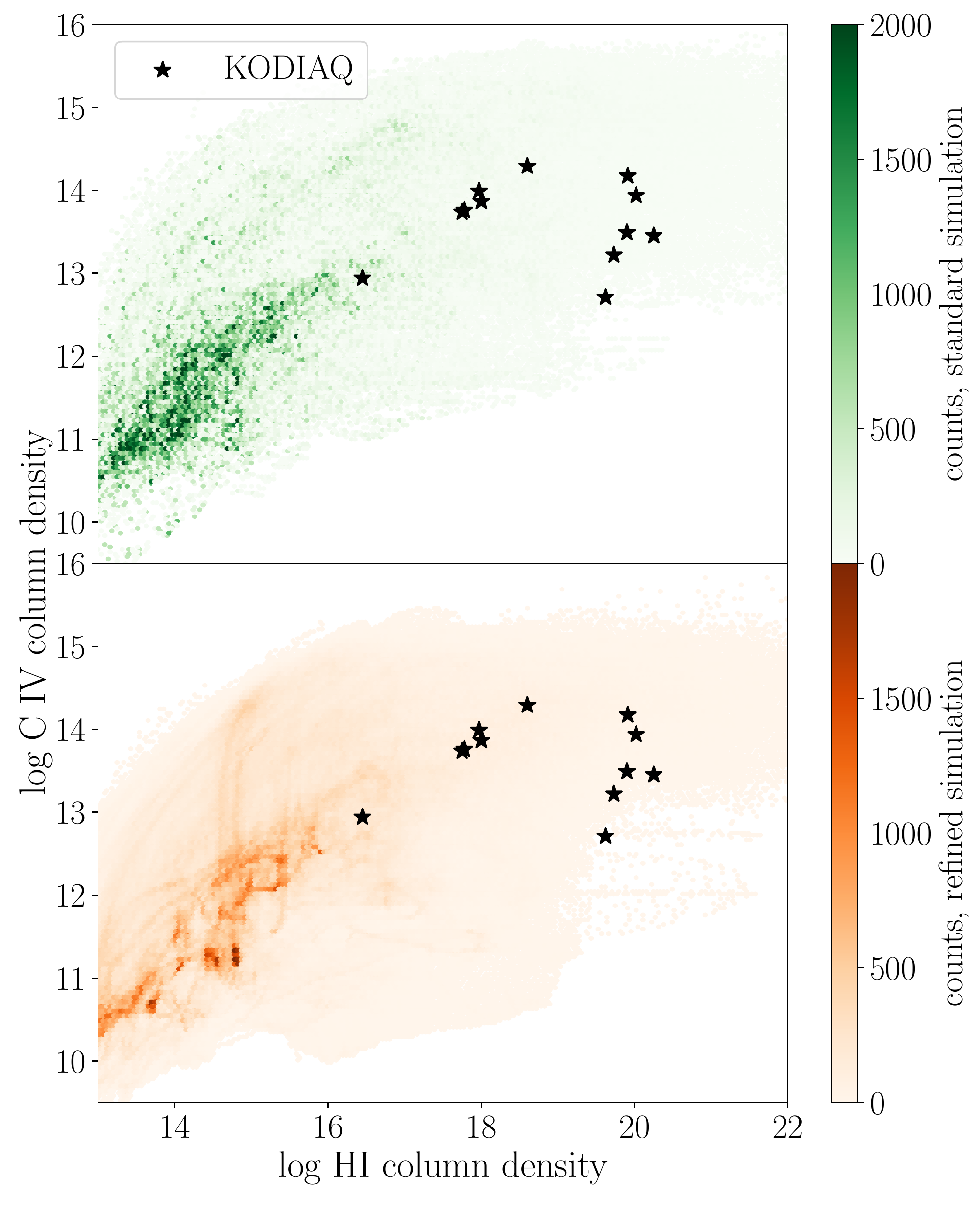}\hfill
    \includegraphics[width=0.33\textwidth]{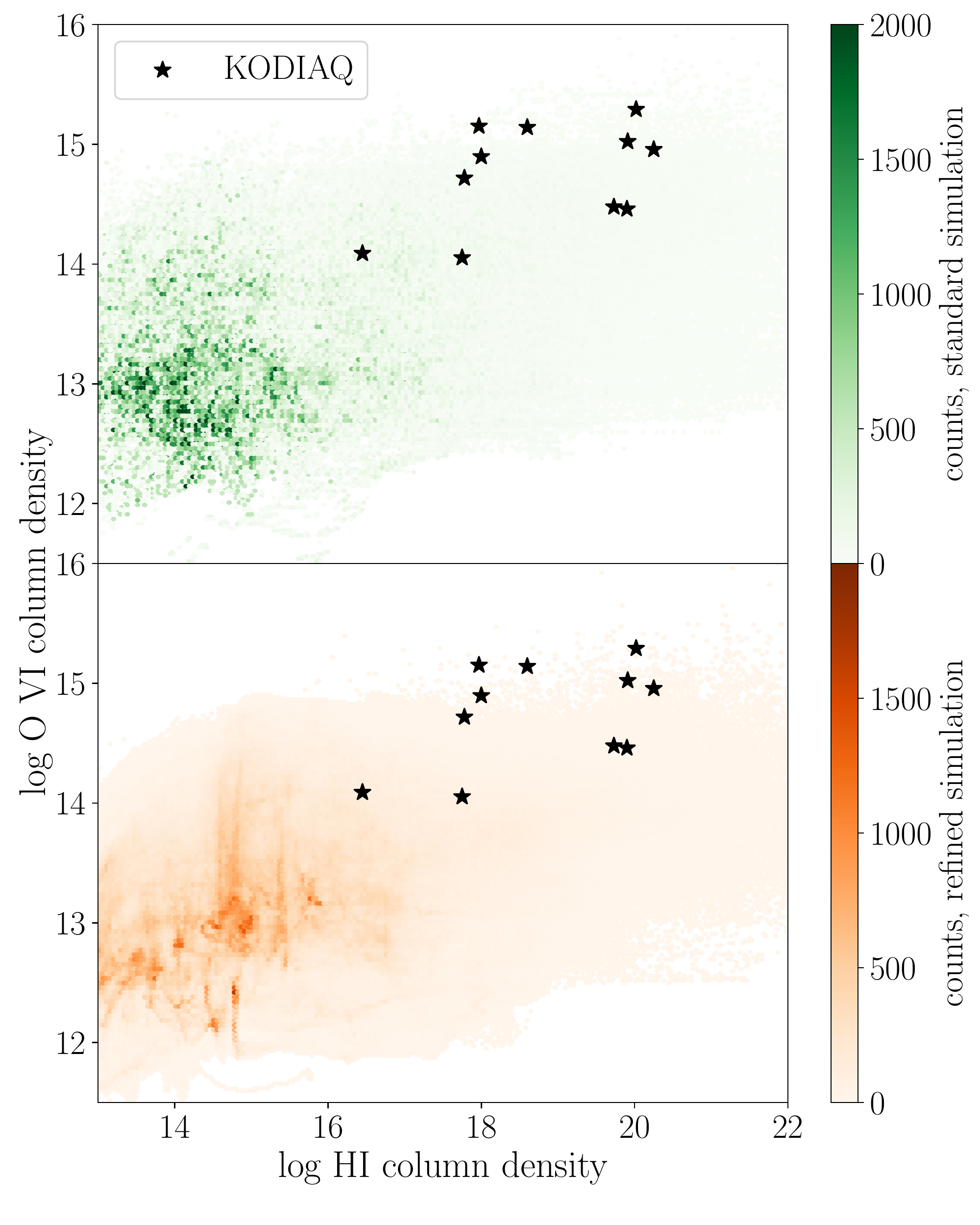}\hfill
    \caption{\siii\ ({\em left}), \civ\ ({\em center}), and \ovi\ ({\em right}) versus \hi\ column densities along each of the three cardinal axes at $z=2$ and $z=2.5$ in the high-resolution ({\em top}) and standard ({\em bottom}) simulations, as compared to the KODIAQ data (\S\ref{sec:kodiaq}). While broadly speaking the relationship between strong \hi\ and the metal ions do not change, the high-resolution simulation more smoothly samples the available parameter space.
    \label{fig:cddf}}
\end{figure*}

While we do not find a significant difference in the column density distribution functions of many of the metal ions, the spatial distribution of the metal ions and their association with strong \hi\ is significantly different in the two simulations. We show in Figure~\ref{fig:columnprofiles} column density versus radius for the two simulations, again considering projections along each of the three cardinal axes at both $z=2$ and 2.5. The median radial profiles of metallic ion column densities are steeper in the standard resolution simulation, especially for the higher ions. 
The horizontal features in the standard-resolution simulation panels are a manifestation of its coarse resolution: its column densities do not change over relatively large spans of radius. Figure~\ref{fig:metalcover} shows the cumulative covering fractions of \siii, \civ, and \ovi\ for both all sightlines and sightlines with $\log N_{\rm H\,I}>16$. Likely owing to the differences in the radial profiles and in the higher prevalence of strong \hi\ in the high-resolution simulation,  LLS-selected sightlines in the high-resolution simulation counter-intuitively have slightly {\em weaker} metal-line absorption than in the standard-resolution simulation, but these differences are well within the typical uncertainties in covering fraction measurements.

Finally, we plot in Figure~\ref{fig:cddf} the \siii, \civ, and \ovi\ column densities versus \hi\ column, independent of radius, for the two simulations.
We hypothesize that the more detailed structures seen in the high-resolution panels of Figure~\ref{fig:cddf} are caused by multiple sightlines probing the same underlying physical structure, whereas the patchiness seen in the standard resolution panels is caused by small numbers of resolution elements being responsible for most of the line-of-sight column density. 
Unfortunately, much of the parameter space traced by KODIAQ (\S\ref{sec:kodiaq}) are not probed by this simulation, so our comparison to the data will be somewhat qualitative.\footnote{Our preliminary tests using the new self-shielding method in Grackle indicate that, as expected, the \hi\ fraction and column densities are higher when self-shielding is taken into account.}

%%%%%%%%%%%%%%%%%%%%%%%%%%%%%%%%%%%%%%%%%%%%%%%%%%%%%%%%%%%%%%%%%%%%
%%%%%%%%%%%%%%%%%%%%%%   resolution and  %%%%%%%%%%%%%%%%%%%%%%%%%%%
%%%%%%%%%%%%%%%%%%%%%     cloud sizes   %%%%%%%%%%%%%%%%%%%%%%%%%%%%
%%%%%%%%%%%%%%%%%%%%%%%%%%%%%%%%%%%%%%%%%%%%%%%%%%%%%%%%%%%%%%%%%%%%
\section{The Effects of Resolution on Circumgalactic Absorption} \label{sec:clouds}
Circumgalactic observations, like those in most of astronomy, suffer from the projection of the third spatial dimension onto a velocity scale. In a medium like the CGM that is highly structured in density, ionization, and metallicity, this projection effect causes a significant loss of diagnostic information. Clouds with large physical separations along the line of sight can appear close to each other in velocity space. Weak components can be concealed by stronger ones. Low metallicity gas can be screened out by high metallicity gas. These kinds of issues greatly complicate the interpretation of observed absorption line profiles and are a large source of systematic uncertainty in conclusions about the CGM. 
A major advantage of using a simulated universe to help interpret real data is that in simulations we can de-project the observable spectra and investigate in detail how complex, multiphase absorption profiles are produced.
We show how  dynamic multiphase circumgalactic gas translates to absorption-line systems in \S\,\ref{sec:phasespace}, arguing in \S\,\ref{sec:cloudsize} and \ref{sec:cloudvel} that the bulk of the absorption is resolved in our high-resolution simulation but not in the standard resolution simulation.

\begin{figure*}[htb]
\centering
    \includegraphics[trim={1.4cm 0.3cm 1.7cm 0.3cm},clip, width=\textwidth]{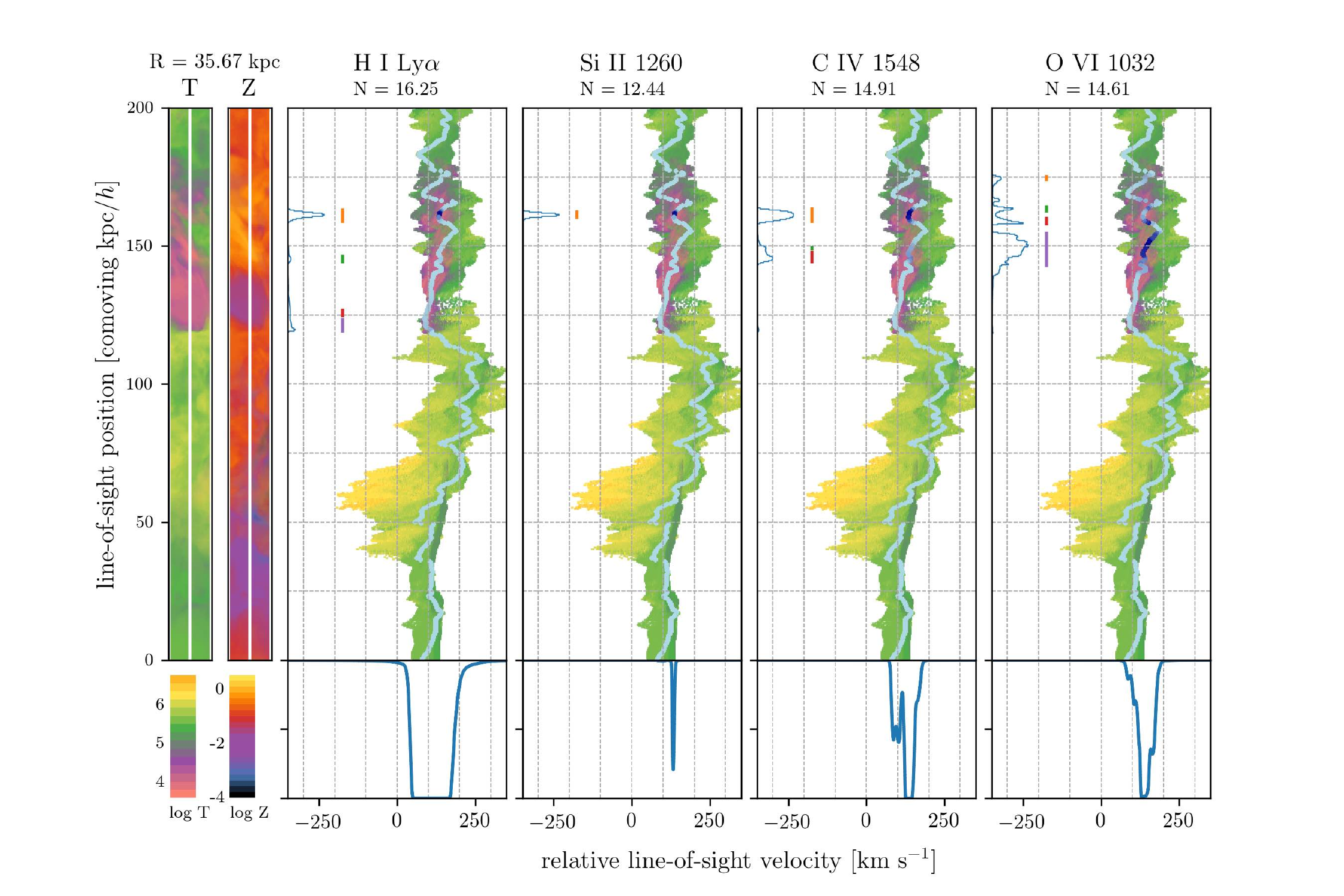}
    \caption{How multiphase gas becomes multi-species absorption, here for a sightline through the high-resolution simulation at $z=2$. 
    The two thin windows at left render the temperature and metallicity in a rectangular prism $\pm 10$\,\ckpch\ to either side of a 200\,\ckpch\ sightline; the impact parameter with respect to the central halo is given in the upper-left corner in physical kpc. As in Figure~\ref{fig:cellmasses}, the color at each position in these panels is given by the most common temperature of a cell at that location. In the four main panels, the same cells are rendered again with the temperature shading but with the line-of-sight velocity on the lower axis. The blue traces denote the path of the ray through the velocity--line-of-sight phase space, with darker blue regions denoting higher ionization fractions in each ion; the ``N = '' at the top of these panels gives the total column density for that ion along the sightline in $\log$\,cm$^{-2}$. The ionization fractions are also shown as the interior left histograms (arbitrarily scaled) of each of these panels, with identified ``clouds'' marked with colored bars. The spectra in the bottom panels are aligned with the same velocity axis and have had no rebinning or LSF applied; the dashed grey lines are at fixed 100\,\kms\ intervals to guide the eye.  See
    \S\,\ref{sec:phasespace} for a more thorough description. 
    \label{fig:velphaseref}}
\end{figure*}

\begin{figure*}[htb]
\centering
    \includegraphics[trim={1.4cm 0.3cm 1.7cm 0.3cm},clip,width=\textwidth]{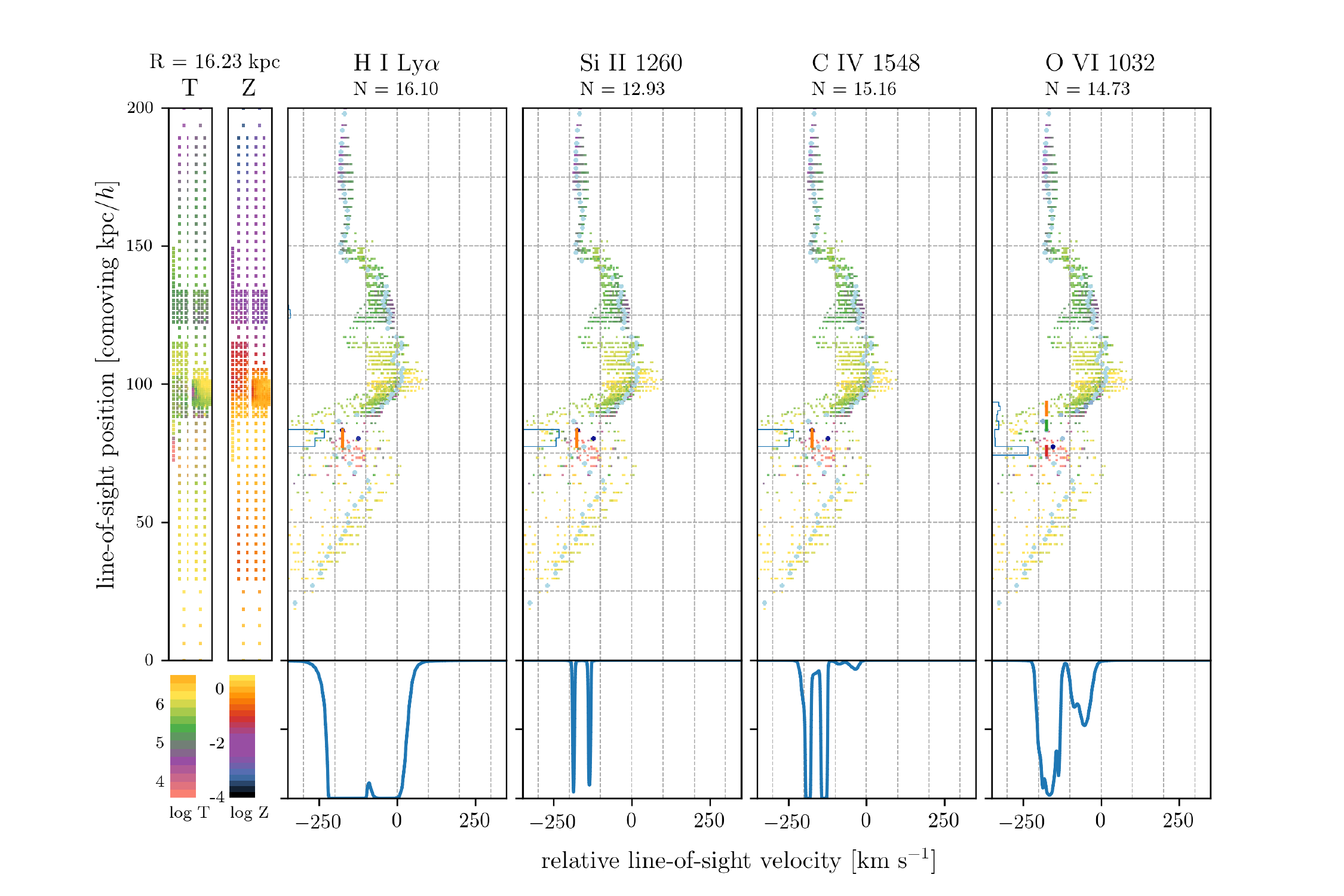}
    \caption{The same as Figure~\ref{fig:velphaseref}, but for a sightline through the standard resolution simulation at $z=2.5$. The contribution of individual cells is clearly evident in the temperature and metallicity renderings.
    \label{fig:velphasenat}}
\end{figure*}

%%%%%%%%%%%%%%%%%%%%%%%%%%%%%%%%%%%%%%%%%%%%%%%%%%%%%%%%%%%%%%%%%%%%
\subsection{Velocity--Line-of-Sight Phase Space}\label{sec:phasespace}
Figures~\ref{fig:velphaseref} and \ref{fig:velphasenat} show diagrams  mapping the 3D spatial fields of temperature and metallicity to the line-of-sight velocity phase space and resulting absorption-line systems for example sightlines in the high- and standard-resolution simulations, respectively. We show further illustrative examples in Appendix~\ref{app:velphase}.
In each diagram, the shaded panels at left show a ``core sample'' extending 200\ckpch\ along the line-of-sight and $\pm 10$\ckpch\ to either side of the ray in two orthogonal axes. These core sample place the infinitesimally narrow sightline (shown by the white lines) into the context of the structures it intercepts.

The four main panels render the same cells from the core sample, color-coded by temperature, 
with the horizontal axis now denoting the line-of-sight velocity along the ray, $v_{\rm los}$. 
The most obvious behavior seen here is that structures at many densities, temperatures, and metallicities extending over the 200\ckpch\ of the ray are still quite tightly constrained in velocity space, with most of the gas ranging over $\pm 200$\,\kms.
Each of these phase-space panels corresponds to an individual ion (here, \hi, \siii, \civ, and \ovi).
The thin trace marked in shades of blue (a color absent from the temperature scale, to provide contrast) shows the locations in phase space occupied by the cells directly along the line of sight, enabling one to immediately read off the interplay between the physical position and line-of-sight velocity for a given sightline. The darkness of blue shades along these traces is in proportion to the number density of the ion species of interest in that cell; the darkly shaded cells therefore generally line up with the absorption peaks in the simulated spectra shown in the bottom panels. In the left of each phase-space panel, there is a thin blue histogram showing the distribution of each ion's number density (arbitrarily scaled for visibility). Overlaid on these are the colored bars marking the extent of the identified ``clouds'';\footnote{Though these absorbing regions are clearly continuously fluctuating regions, we will for simplicity refer to them as ``clouds''.} we explore the properties of these clouds in detail in \S\S\,\ref{sec:cloudsize},\ref{sec:cloudvel}. Finally, we mark the locations of the spectral minima analyzed in \S\,\ref{sec:comparison} in the bottom panels. These panels are arranged such that reading over from the core samples to the phase space panels and down to the spectra ties together the temperature, metallicity, ionic number density, and absorption signature of each cell along the line of sight. 

These diagrams enable us to immediately draw several conclusions about the simulated CGM. First, the strongest absorption is mostly confined to within $\pm 200$\,\kms\ of the systemic velocity, with the hot gas having the highest velocity dispersion and the cold gas generally the lowest. We can see also that the strongest absorption---those components that make detectable impressions on the spectrum---are often confined to only a few physical locations, with the higher ions arising from gas that is more spatially extended than the lower ions. Generally, these locations correspond to where the cells have significant metal enrichment (with the notable exception of the \hi-traced gas) and a temperature that is suitable for the species to have a significant ionization fraction. 

Contrasting Figures~\ref{fig:velphaseref} and \ref{fig:velphasenat}, we immediately notice how sparsely-sampled both the underlying physical {\em and} velocity spaces are in the standard-resolution simulation.  As expected from \S\ref{sec:physical}, the high-resolution simulation shows much more variation along the line of sight than in the standard-resolution simulation. As we show in \S\,\ref{sec:cloudsize}, the bulk of the gas giving rise to the ``observed'' absorption in the standard-resolution simulation is often from only a few resolution elements, while these structures are more fully resolved in the high-resolution simulation.

%%%%%%%%%%%%%%%%%%%%%%%%%%%%%%%%%%%%%%%%%%%%%%%%%%%%%%%%%%%%%%%%%%%%
\subsection{Resolving the Absorbers Spatially}\label{sec:cloudsize}
\begin{figure*}[htb]
\centering
    \includegraphics[width=0.49\textwidth]{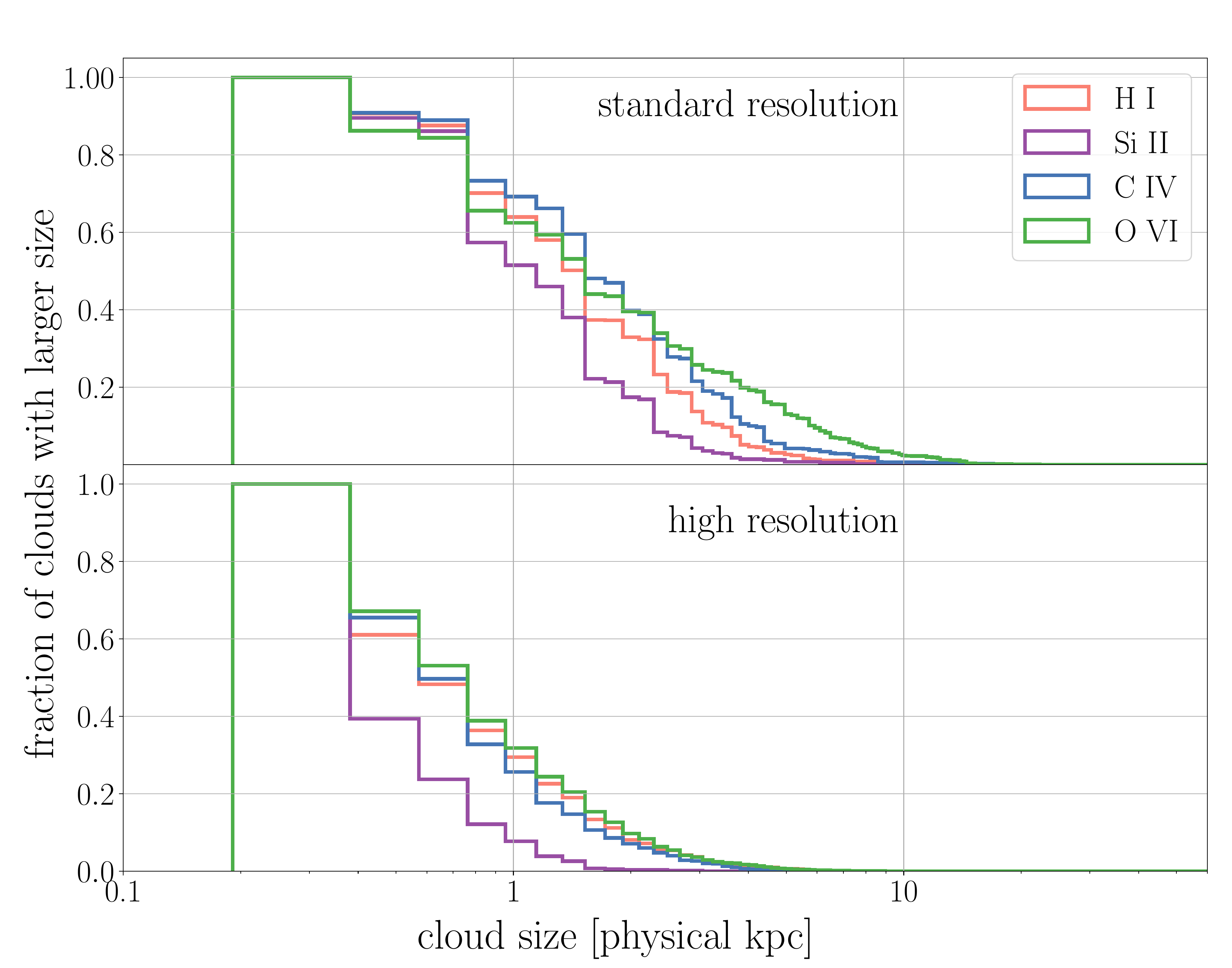}\hfill
    \includegraphics[angle=90,width=0.49\textwidth]{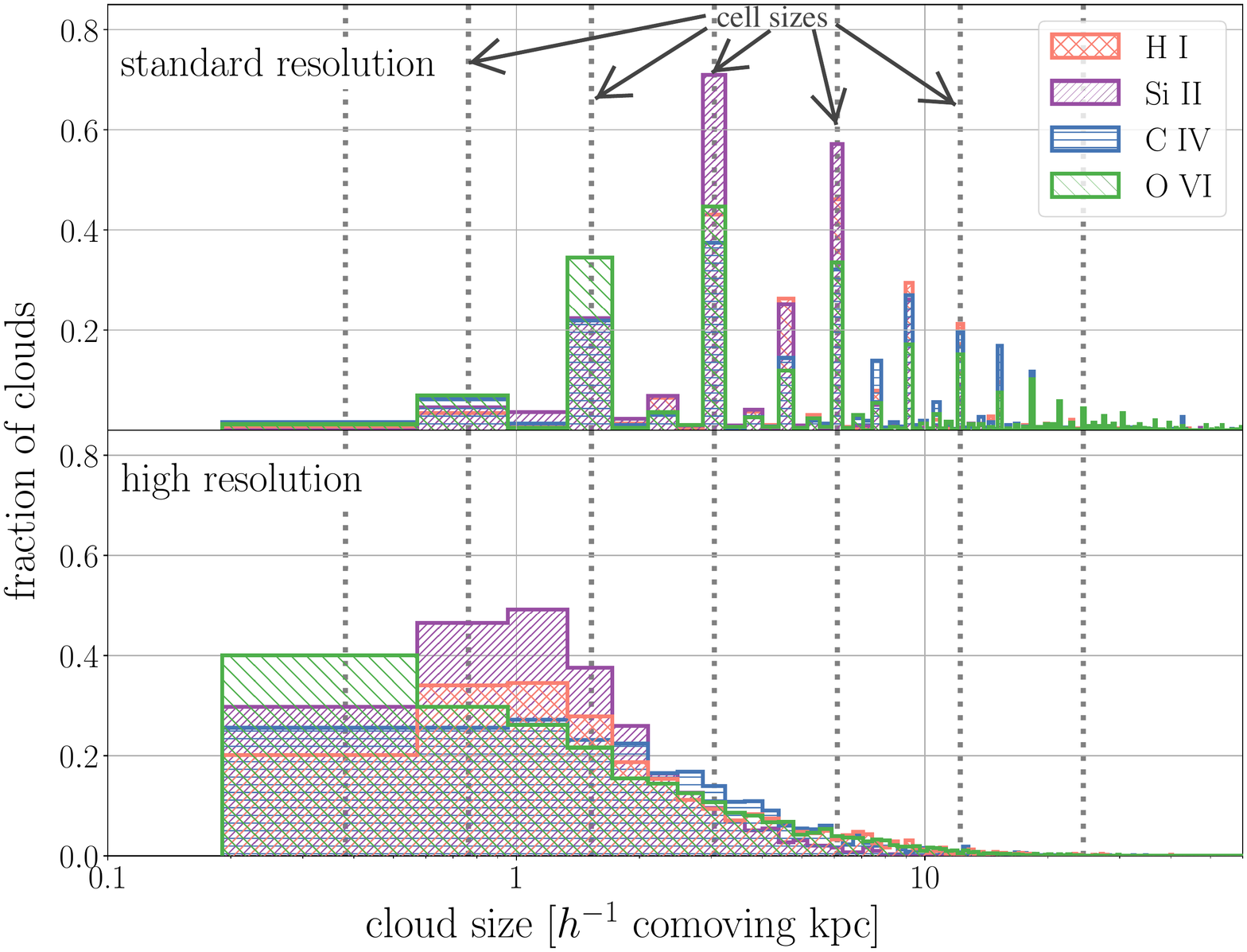}
    \caption{Cumulative ({\em left}) and differential ({\em right}) distributions of \hi\ (pink), \siii\ (purple), \civ\ (blue) and \ovi\ (green) cloud lengths in the standard- ({\em top}) and high-resolution ({\em bottom}) simulations. We plot the differential distributions in comoving coordinates to show that in the standard-resolution simulation, cloud sizes tend to cluster around the cell sizes (vertical grey dashed lines in the right-hand panels), indicating that the clouds are unresolved.
    \label{fig:cloudsizehist}}
\end{figure*}

\begin{figure}[htb]
\centering
    \includegraphics[width=0.49\textwidth]{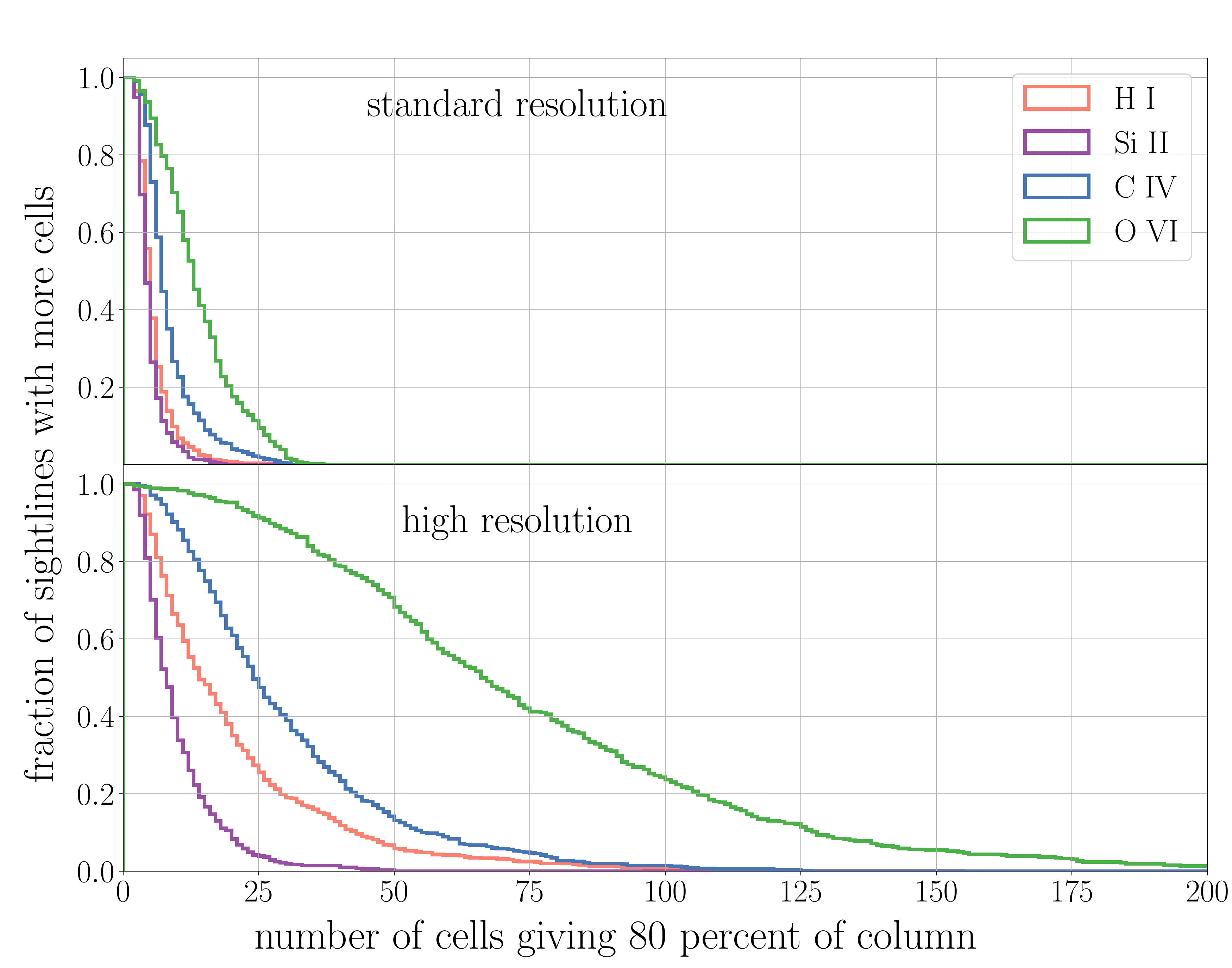}
    \caption{Cumulative distribution of the number of resolution elements giving 80\% of the absorption along the line-of-sight per absorber for \hi\ (pink), \siii\ (purple), \civ\ (blue) and \ovi\ (green).  
    \label{fig:cloudcellshist}}
\end{figure}

To quantify our comparisons of absorption structures generated from simulations at different resolutions, we define a relatively unbiased, empirical definition of what constitutes a ``cloud'' that is independent of the simulated absorption. For each ion, we find the threshold in the ionic number density $n_{\rm ion}$ above which 80\% of the total sightline column density is produced. Using all cells with number densities above this threshold value, we find the positions along the line of sight where $n_{\rm ion}$ crosses this threshold on its way up and then down again to define a single cloud. Between up and down threshold crossings, $n_{\rm ion}$ can take on any value, and so the clouds can be complex in shape and size. 
While our choice of 80\% of the total sightline column density is somewhat arbitrary, we find that much larger percentages often lead to the near-entirety of the sightline contributing to single clouds, which, while interesting, is less useful as a diagnostic tool.
The identified clouds and their extents are denoted with the vertical colored bars next to the line-of-sight histograms in the velocity phase-space plots (Figures~\ref{fig:velphaseref} and \ref{fig:velphasenat}). 

The immediate conclusion from this analysis is that only a few clouds dominate the simulated column density of each sightline, and that the number of distinct clouds responsible for 80\% of the column density increases as we go up the scale of ionization potential from \hi\ to \ovi. For example, though the ions shown in Figure~\ref{fig:velphaseref} are mostly at the same line-of-sight velocity, the \hi\ gas is largely confined to a single $\sim 5$\,kpc cloud, while the \ovi-traced gas is spread out over $\sim 35$\,kpc, most of which is not directly associated with the \hi-bearing gas. Moreover, the low-ionization \siii\ largely (though not entirely) traces the \hi, while the intermediate \civ\ ion has one cloud largely associated with the \hi\ and \siii\ and one cloud overlapping with the most predominant \ovi-bearing cloud.
This finding by itself calls into question the general tendency to treat observed absorbers as single objects with homogeneous conditions. 

In Figure~\ref{fig:cloudsizehist}, we show the distribution of the cloud sizes in the standard and forced refinement runs for \hi, \siii, \civ, and \ovi. 
The dashed grey lines in the differential histogram denote the cell sizes at different refinement levels found in the CGM in the standard resolution simulation: the gas contributing to the bulk of the absorption in this simulation is clearly often arising from single resolution elements and is therefore unresolved. In contrast, the highly-resolved CGM has a more power-law distribution of cloud sizes, as would be more generally expected if the cloud size is set by turbulence (in analogy to the mass distribution of star-forming clouds driven by turbulence, \citealp{padoan02}).   Figure~\ref{fig:cloudcellshist} further quantifies the differences in cloud sizes by showing the distribution of the numbers of cells along the line-of-sight per cloud in the two simulations. 
In the standard resolution simulation, $\sim 75$\% of the sightlines with $\mlnhi>16$ 
have only  $\leq 5$ cells giving rise to $>80$\% of their \hi\ column density, whereas this fraction is $<20$\% in the high-resolution simulation. This difference is still stark for the highly-ionized \ovi: $\sim 17$\% of the sightlines with $\log N_{\rm OVI}>13$ have  $>80$\% of this column density in only $\leq 5$ cells, whereas this fraction is $\sim 1$\% in the high-resolution simulation.

\begin{figure}[thb]
\centering
    \includegraphics[width=0.49\textwidth]{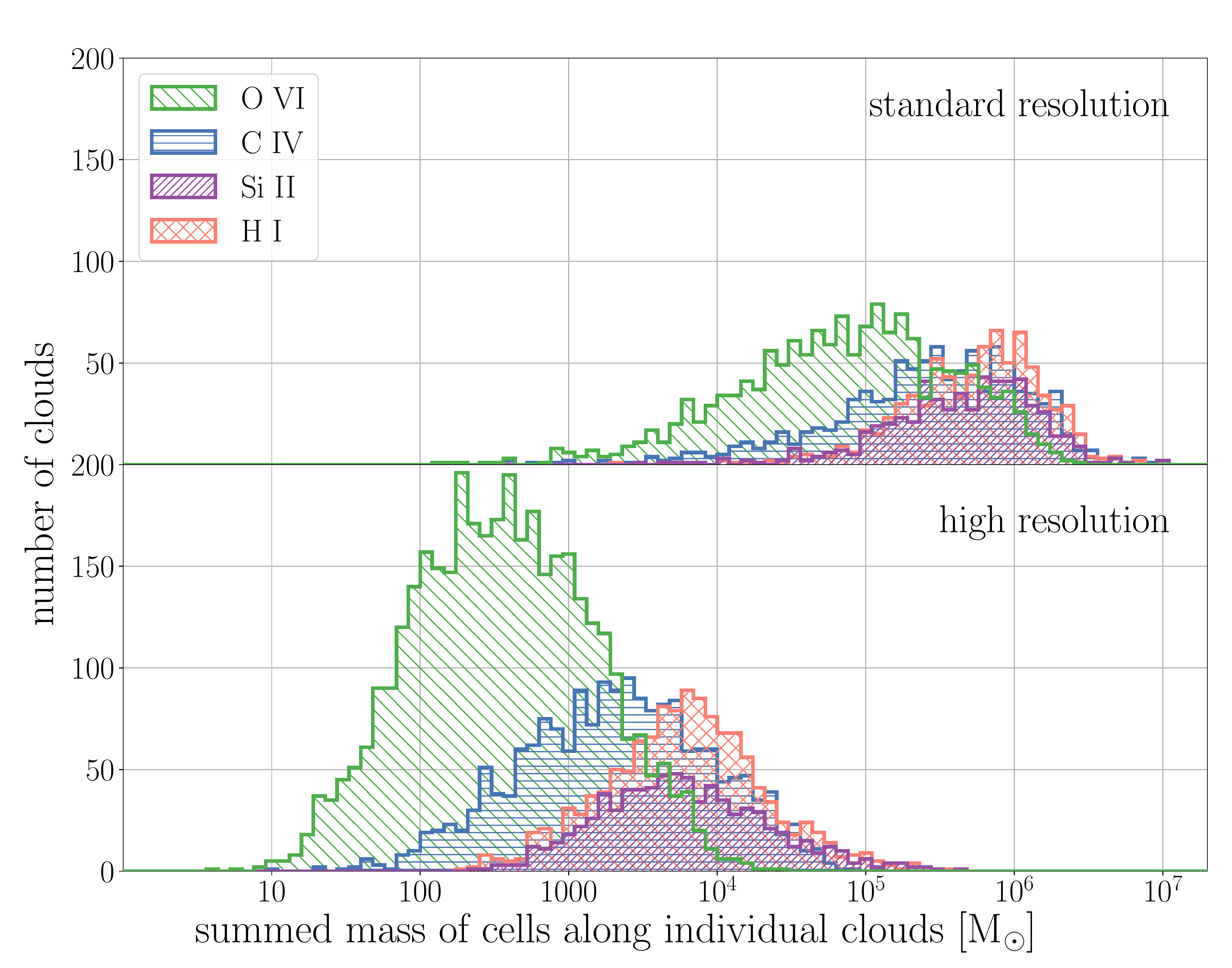}
    \caption{Distributions of 1D cloud masses for \hi\ (pink), \siii\ (purple), \civ\ (blue) and \ovi\ (green) in the standard- ({\em top}) and high-resolution ({\em bottom}) simulations.
    \label{fig:cloudmasses}}
\end{figure}

In Figure~\ref{fig:cloudmasses}, we show how the improvement in {\em spatially} resolving the absorbing clouds translates back into the {\em mass} resolution commonly enforced by Lagrangian codes. In the standard resolution simulation, we find that most of the absorbing clouds have masses of $\gtrsim 10^4$\,\Msun, consistent with the masses of resolution elements in FIRE \citep{hopkins14,hopkins18}, Illustris \citep{vogelsberger14}, or EAGLE \citep{schaye15}. In the high-resolution simulation, however, the absorbing clouds typically have masses two to three orders of magnitude lower, suggesting that the shift to lower cell masses seen in Figure~\ref{fig:cellmasses} results in lower mass structures being resolved.
The more highly ionized \ovi\ tends to arise in larger clouds, more clouds along the line-of-sight, and much less mass per cloud than the lower ionization metal species, which is roughly what is expected from the more highly ionized species tracing lower density gas. This combination of results highlights the fact that to adequately resolve low-density gas---even over large spatial scales---low {\em mass} scales must also be achievable: in the high-resolution simulation, $>75$\%  of the clouds responsible for $>80$\% of the absorption in the $>10^{13}$\,cm$^{-2}$ \ovi\ sightlines have 1D cloud masses of $<1000$\,\Msun (and $\sim 20$\% of the clouds have masses $<100$\,\Msun).

\begin{figure}[htb]
\centering
    \includegraphics[width=0.49\textwidth]{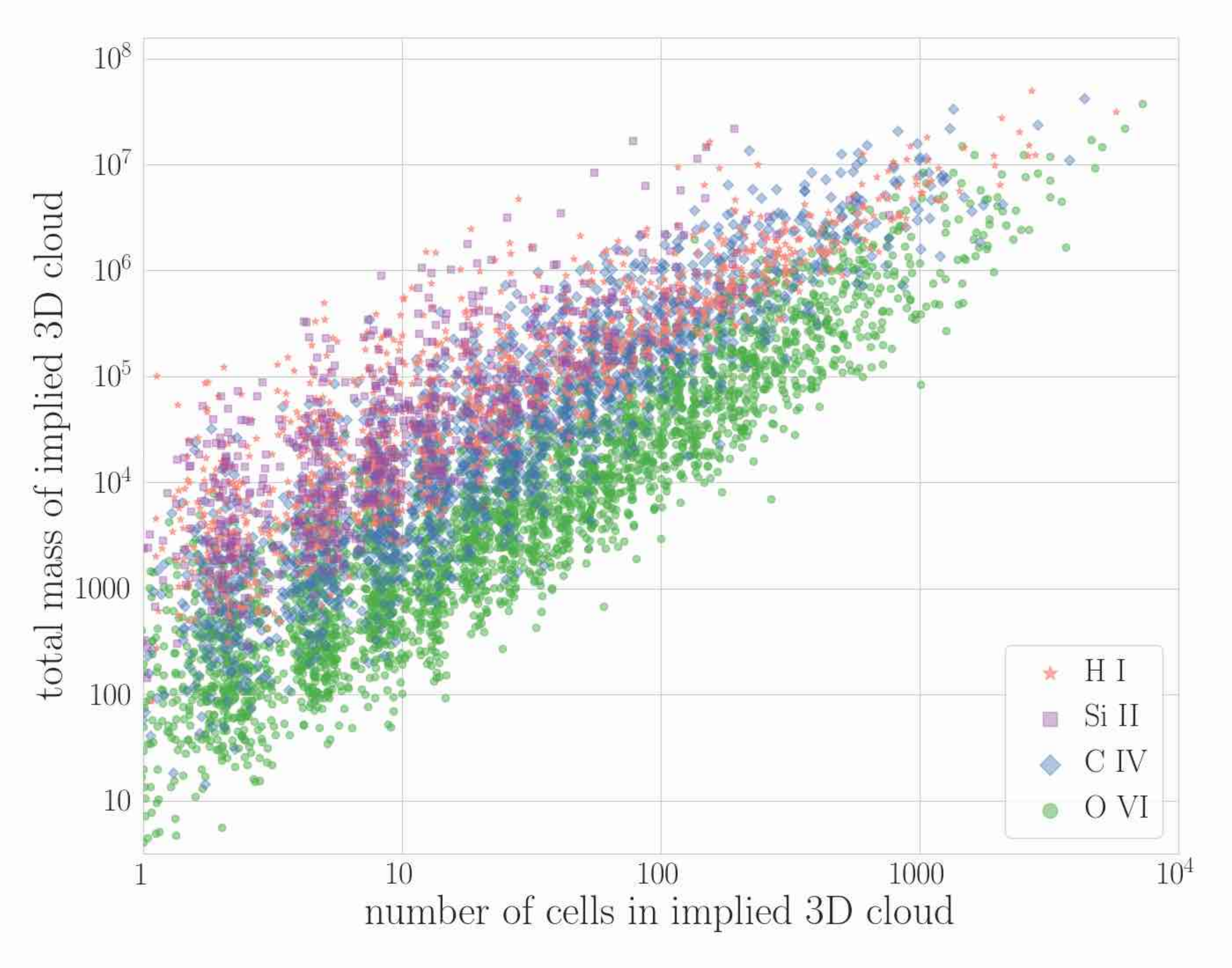}
    \caption{ The implied 3D cloud masses (see Equation~\ref{eqn:mcloud}) of \hi\ (pink), \siii\ (purple), \civ\ (blue) and \ovi\ (green) versus the implied 3D number of cells in the high-resolution simulation. Small random offsets have been applied to the cell numbers to make the trends easier to see. At the typical mass resolution of most current Lagrangian simulations ($\sim 10^4$--$10^6$\Msun), FOGGIE resolves most absorbers with tens to hundreds of resolution elements.
    \label{fig:3dclouds}}
\end{figure}

We calculate an estimated 3D cloud mass, $\mcloud$, by calculating the average density along the sightline and assuming spherical geometry; that is,
\begin{equation}\label{eqn:mcloud}
\mcloud = \frac{4}{3}\pi\left(\frac{\mbox{cloud length}}{2}\right)^3 \times \sum\limits_{\rm{cells\,in\,cloud}}\frac{\mbox{cell mass}}{\mbox{cell volume}}\\
\end{equation}
Likewise, for a uniform grid, the estimated number of cells in a 3D cloud is $(\pi/6)\times$(the number of cells in the 1D cloud)$^2$.
We plot the 3D cloud masses versus estimated number of cells in the 3D cloud for the high-resolution simulation in Figure~\ref{fig:3dclouds}. As the cell sizes are effectively uniform within the forced refinement region, the trend in cloud mass for a fixed number of cells between the different ions is a tracer of the underlying density, with the lower-ionization species being traced by higher-density gas, as expected. We also find that for 3D cloud masses of $\gtrsim 10^5$\Msun, FOGGIE resolves the structures giving rise to most circumgalactic absorption with $\gtrsim 10$--1000 resolution elements.

%%%%%%%%%%%%%%%%%%%%%%%%%%%%%%%%%%%%%%%%%%%%%%%%%%%%%%%%%%%%%%%%%%%%
\subsection{Kinematically Resolving Circumgalactic Absorbers}\label{sec:cloudvel}
In real absorption-line observations, we are unable to measure distance along the line of sight; instead we attempt to disentangle the physical positions of absorbing clouds from their line-of-sight velocity. While coarsely-resolved simulations can be used to understand observed large-scale kinematic trends \citep{turner17}, we are interested here in what sets the kinematics within individual absorption complexes.
As can be seen from Figures~\ref{fig:velphaseref} and \ref{fig:velphasenat} (and the other examples in Appendix~\ref{app:velphase}),  absorbing gas that is many tens of kiloparsecs apart may be coincident in line-of-sight velocities. This complicates the interpretation of single absorption line ``components'' and the comparison of absorption seen at the same velocity in species with different ionization levels. Likewise, a flow along the line-of-sight can imprint itself on an absorption feature by spreading nearly co-spatial gas out over a broad range in velocity.  Observationally, this means that bulk motions contribute significantly to the non-thermal widths of broad absorption features, and it is generally the hotter, more thermally broadened gas, that has higher relative velocities. 

\begin{figure*}
    \centering
    \includegraphics[width=0.48\textwidth]{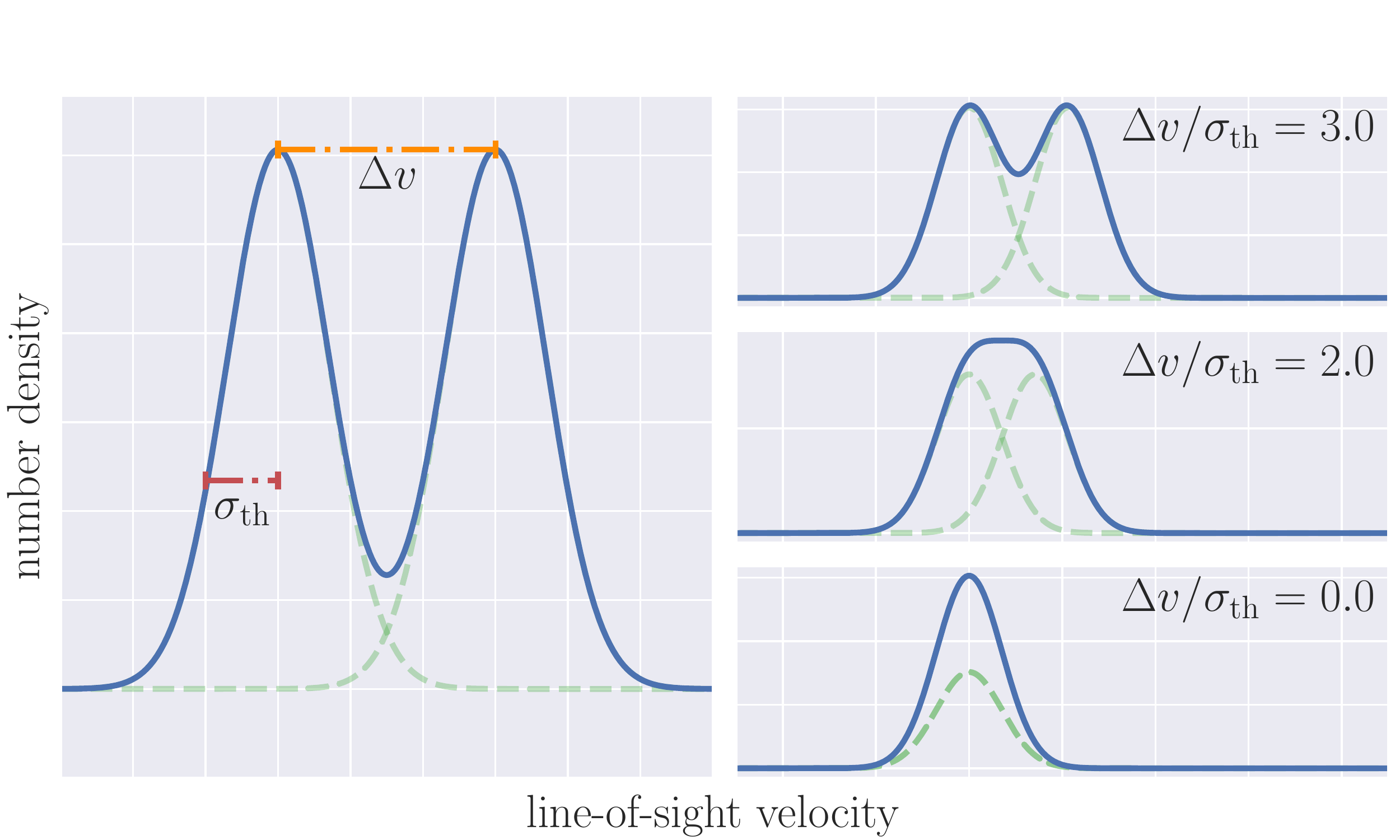}\hfill
    \includegraphics[width=0.48\textwidth]{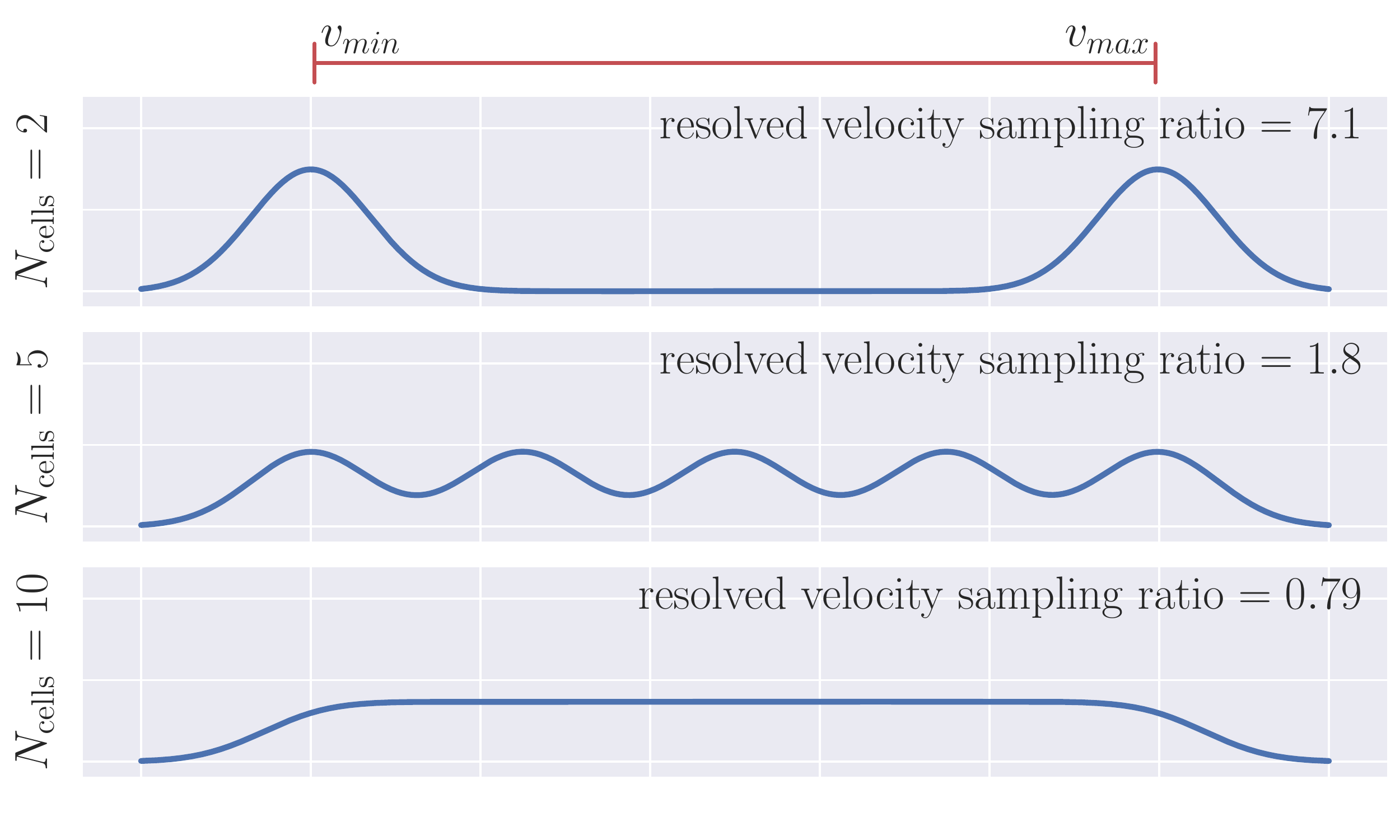}
    \caption{{\em Left}: How the cloud velocity width $\dvcloud$ and average cell thermal broadening $\sigth$ correspond to the 
    cloud velocity sampling $\dvcloud/\sigth$ (Equation~\ref{eqn:sampling}). In the case of two components, systems with $\dvcloud/\sigth\leq 2$ show only one extremum, while less well-sampled systems are less blended. {\rm Right}: How better resolved velocity flows translate into smaller resolved velocity sampling ratios  (Equation~\ref{eqn:velratio}) for a range of $\Ncells$; here, the total area under the curves in each panel is the same. \label{fig:cartoon}}
\end{figure*}

It is this latter case that is particularly tricky to simulate: the underlying velocity field is discretized by the sampling of the individual resolution elements but then smeared out by thermal broadening when creating the mock spectrum. While a high-resolution simulation may better resolve the dynamic structure of the CGM, a low-resolution simulation may have more kinematic components in the {\em spectra} because of single cells giving rise to individual absorption-line components (see Figure~\ref{fig:cartoon}), whereas the same feature when fully resolved may be blended out into one larger, more complicated feature. The \siii\ and \civ\ panels of Figure~\ref{fig:velphasenat} show good examples of this phenomenon: though the identified ``clouds'' in these panels comprise multiple cells, it is clear that the sharp absorption features are arising from individual cells---and that if the underlying flow was better resolved, the resulting spectrum would be more blended. We explore the impacts of resolution on observable kinematic structure and how they compare to the KODIAQ data in \S\,\ref{sec:comparison}, but we first examine here the interplay between the number of kinematic components in an ``observed'' spectrum and how well resolved the the underlying clouds are. 

Consider an individual 1D cloud comprised of some number of cells, $\Ncells$. For a given ion, the relevant average temperature of the cloud is the ion-density weighted temperature
\begin{equation}
    \bar{T}_{\rm ion,\,cloud} = \frac{1}{N_{\rm ion,\,cloud}} \times \sum\limits_{\rm cells}T_{\rm cell}\Ncells,
\end{equation}
where $N_{\rm ion,\,cell}$ is the column density of the ion in that cell and $N_{\rm ion,\,cloud}$ is the column density of the ion in the entire cloud. From here, we define the average velocity dispersion from thermal motions per cell in the cloud as
\begin{equation}\label{eqn:sigth}
    \sigth = \sqrt{\frac{k\bar{T}_{\rm ion,\,cloud}}{m_{\rm ion}}},
\end{equation}
where $k$ is the Boltzmann constant and $m_{\rm ion}$ is the mass of the ion.
Taking the velocity width of the cloud $\dvcloud$ to be the difference between the maximum and minimum line-of-sight velocities of cells in the cloud, we can define a cloud velocity sampling as
\begin{equation}\label{eqn:sampling}
    \mbox{cloud velocity sampling} = \frac{\dvcloud}{\sigth}.
\end{equation}

This cloud velocity sampling then gives us an estimate of how many {\em kinematic} resolution elements are required to resolve a cloud. 
In order to meaningfully compare the cloud velocity sampling to $\Ncells$, we define a flow to be ``resolved'' if the average spacing between Gaussian profiles results in a fully blended profile, i.e., if there is only one resulting peak with no troughs in the combined profile. For identical Gaussian profiles, this occurs when the spacing between centroids is $\lesssim 2\sigma$, as we illustrate in Figure~\ref{fig:cartoon}. Extrapolating to many cells, we define the ``resolved velocity sampling ratio'' as
\begin{equation}\label{eqn:velratio}
    \mbox{resolved velocity sampling ratio}\equiv
    \frac{\mbox{cloud velocity sampling}}{2(\Ncells-1)}.
\end{equation}

Here, a cloud that is better sampled (and thus better kinematically resolved) has a lower resolved velocity sampling ratio, whereas clouds with ratios $>1$ likely have individual cells as the predominant contributor to individual absorption-line components, as illustrated in the right-hand panels of Figure~\ref{fig:cartoon}. (In cases with $\Ncells=1$ (and thus $\Delta v\equiv 0$), we set the resolved velocity sampling ratio to $10^6$, as these clouds are clearly unresolved.)
As an example, consider the \siii\ absorbers in Figures~\ref{fig:velphaseref} and \ref{fig:velphasenat}. In the high-resolution case, there is a single \siii\ cloud comprised of 5 cells with $\Delta v=3.17$\,\kms\ (and length 1.9\,\ckpch) and $\sigth=1.8$\,\kms, yielding a cloud velocity sampling of 1.7 and a resolved velocity sampling ratio of 0.21. The velocity structure of this cloud is clearly resolved. In contrast, the example standard-resolution \siii\ absorber is visibly unresolved. Quantitatively, this absorption arises from a single ``cloud'' with only two cells separated by 51\,\kms\ (and 6.1\,\ckpch) with $\sigth=2.2$\,\kms, yielding a resolved velocity sampling ratio 11.7.

\begin{figure}
    \centering
    \includegraphics[width=0.48\textwidth]{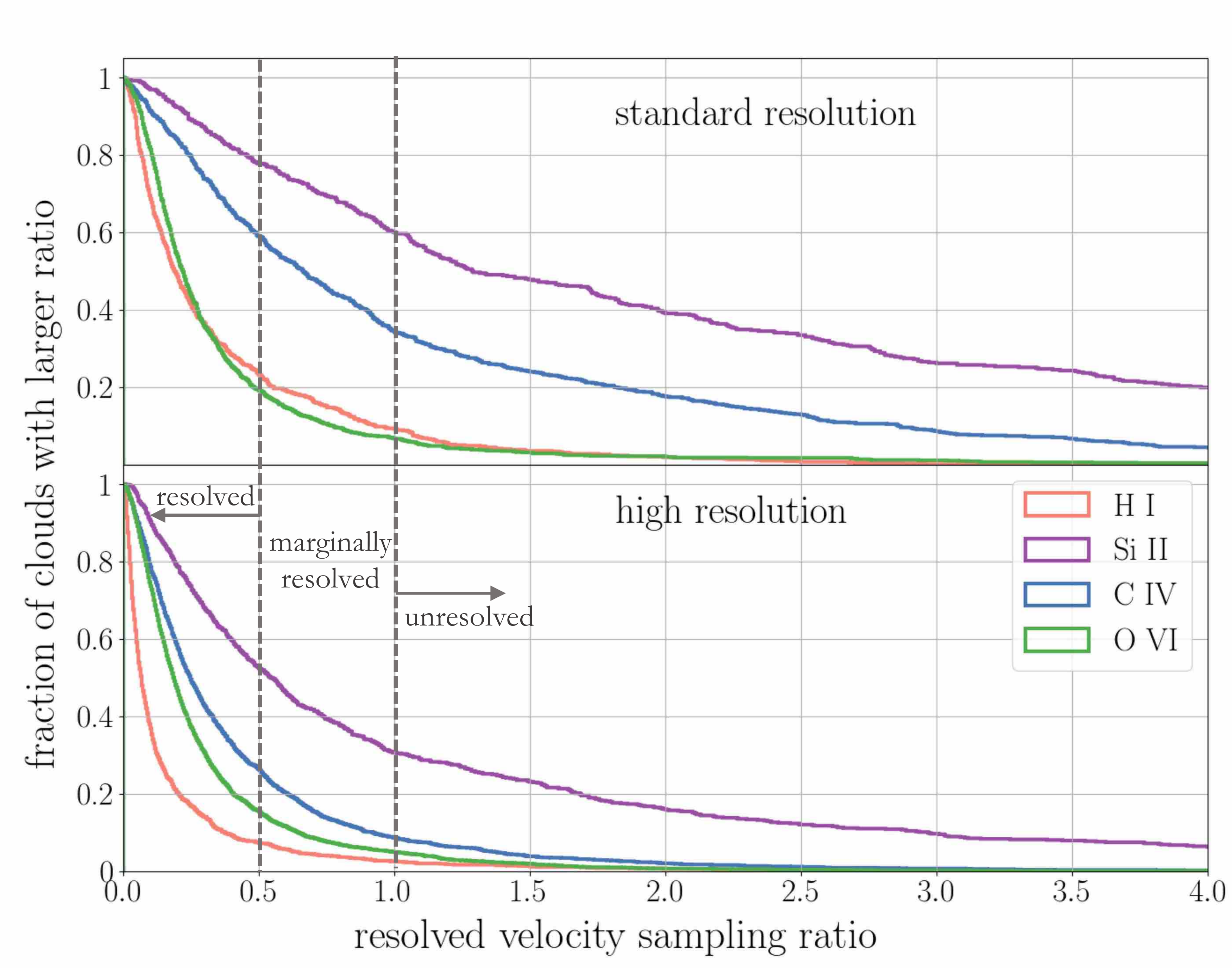}
    \caption{Cumulative distributions of the resolved velocity spacing ratio (Equation~\ref{eqn:velratio}) in the standard- ({\em top}) and high-resolution ({\em bottom}) simulations. \label{fig:velratio}}
\end{figure}

Figure~\ref{fig:velratio} shows the cumulative distributions of the resolved velocity sampling ratio for the two simulations. 
We draw several conclusions from this analysis. First, for the metallic ions in both simulations, the kinematics of the lower-ionization gas are less well-resolved than for the more highly-ionized gas, an effect that which here is likely further exaggerated by the relatively high atomic mass of Silicon (which enters via $\sigth$ as defined in Equation~\ref{eqn:sigth}). Conversely, the low atomic mass of Hydrogen (and our high $\mlnhi$ selection) leads to relatively well-resolved \hi\ kinematics.
Second, while the internal kinematics of absorbing clouds is better resolved in the high-resolution simulation, there is still a significant fraction of clouds that likely have individual cells dominating individual absorption-line components, especially for \ion{Si}{2}: 39\% of \ion{Si}{2} clouds in the high-resolution simulation and 66\% of sightlines in the standard-resolution simulation have resolved velocity spacing ratios $>1$. We find a large improvement in the resolution of \civ\ kinematics, with 55\% of \civ\ clouds with resolved velocity spacing ratios $<1$ in the standard-resolution simulation increasing to 82\% in the high-resolution case.

\begin{figure*}
    \centering
    \includegraphics[width=0.24\textwidth]{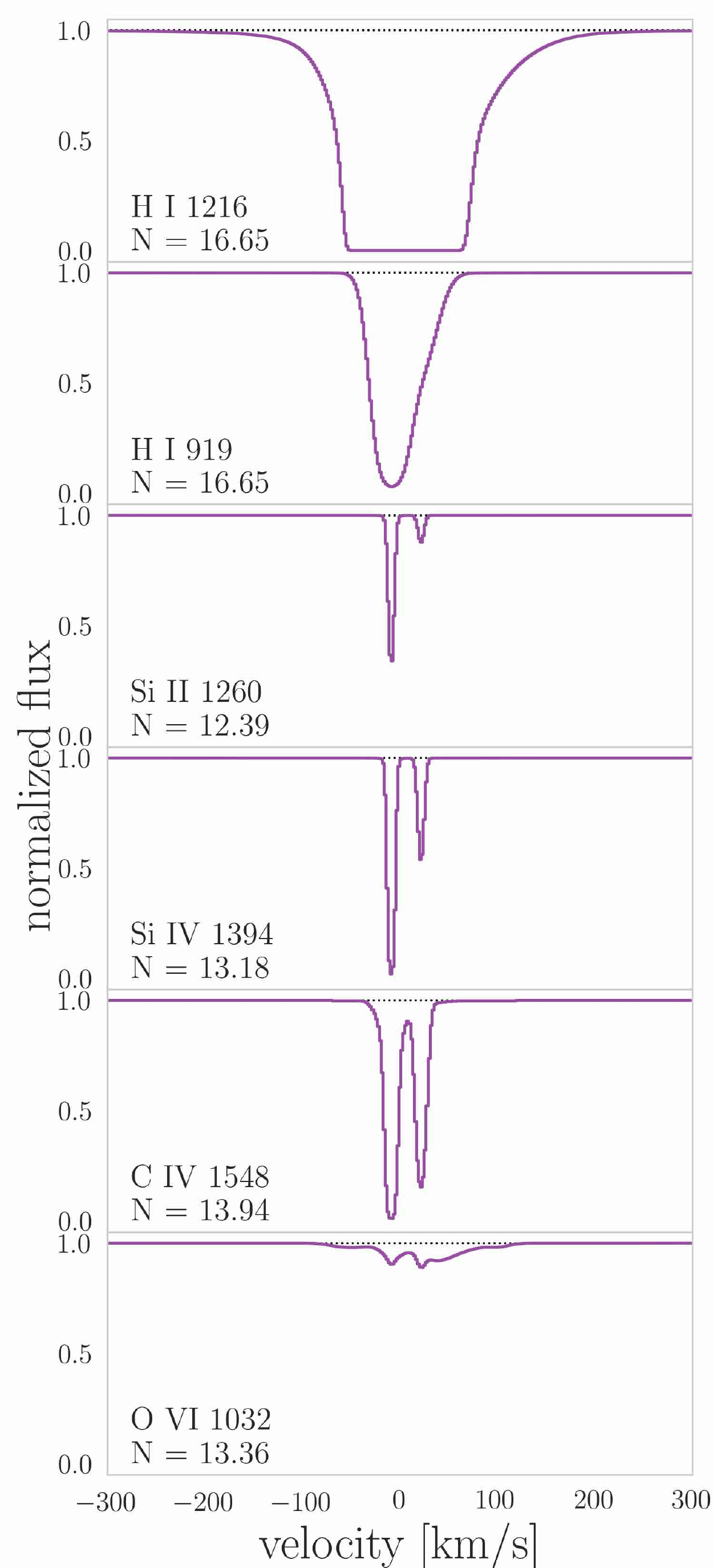}\hfill
    \includegraphics[width=0.24\textwidth]{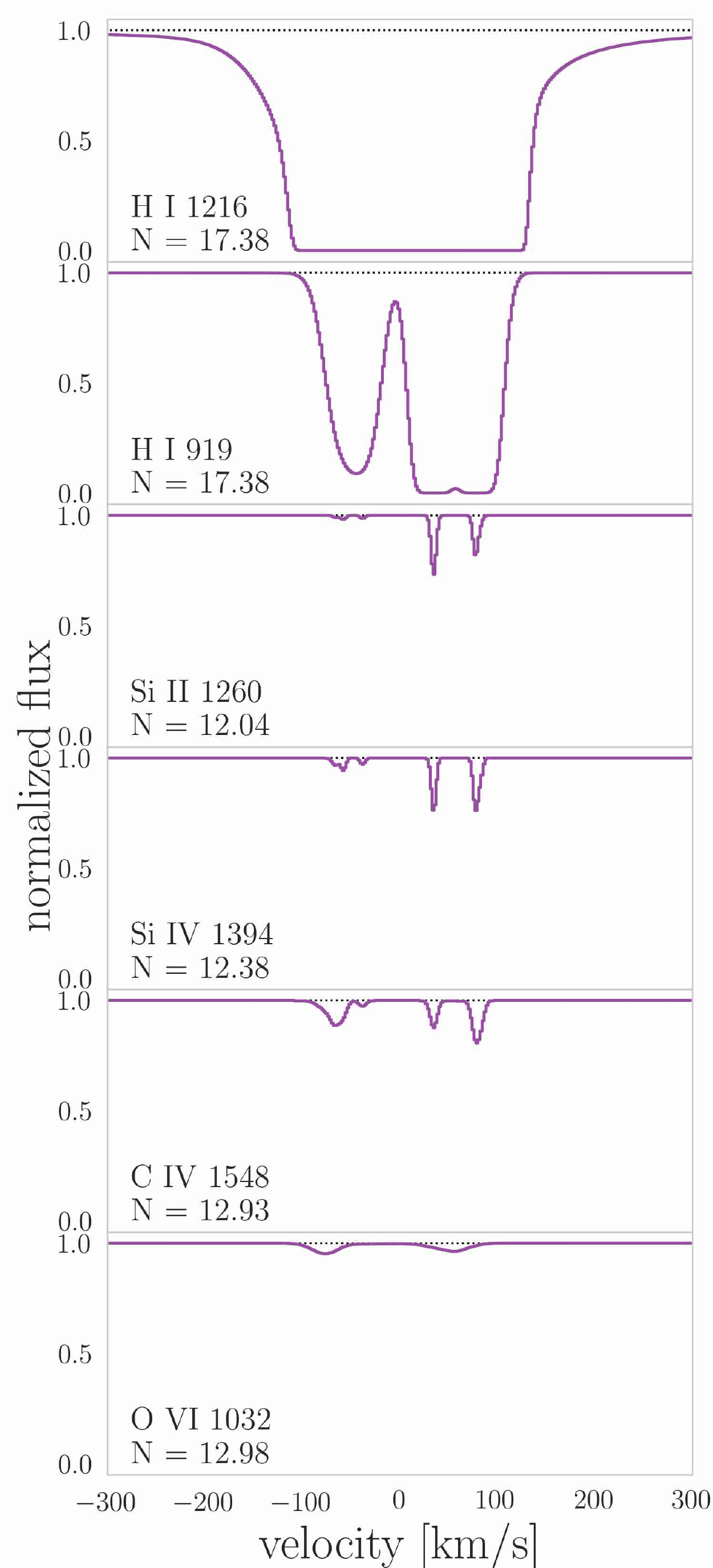}\hfill
    \includegraphics[width=0.24\textwidth]{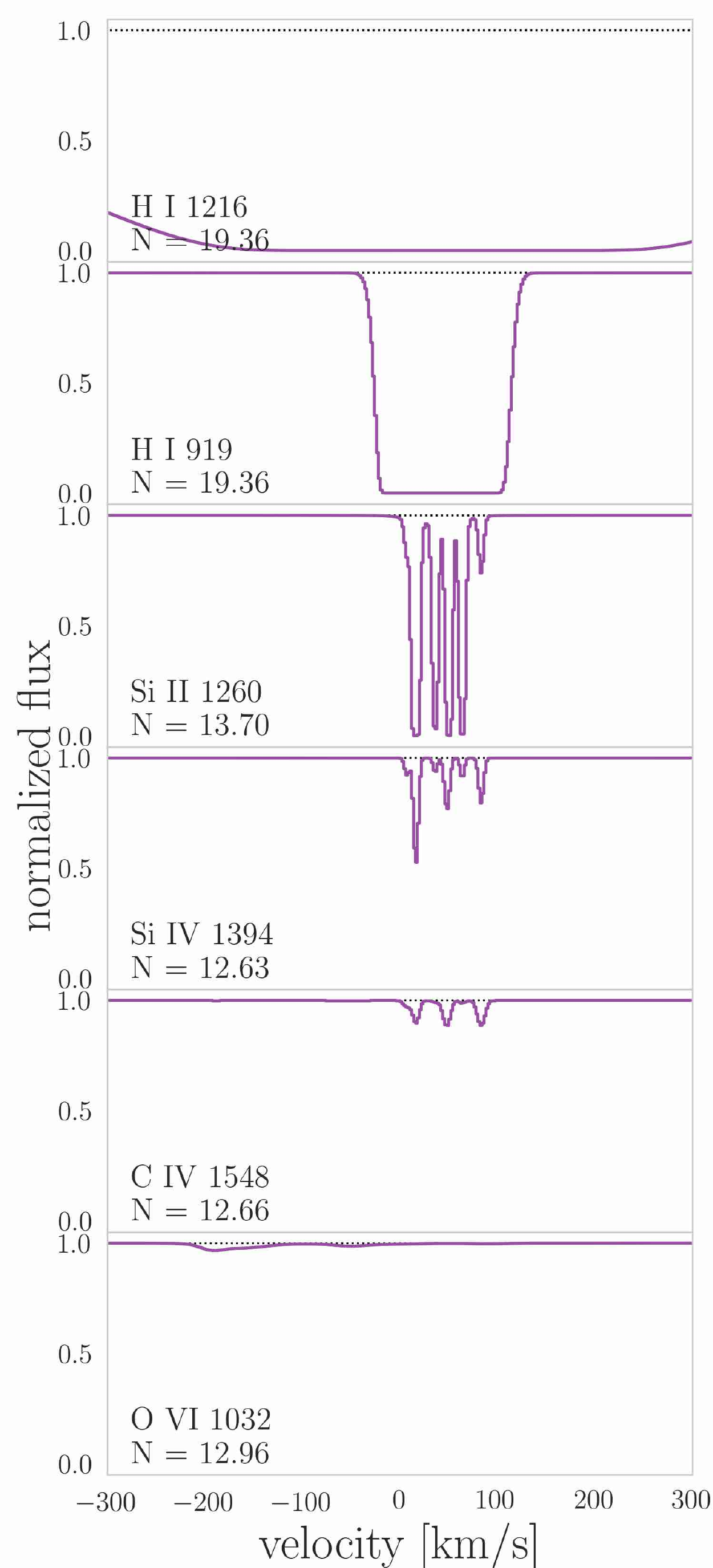}\hfill
    \includegraphics[width=0.24\textwidth]{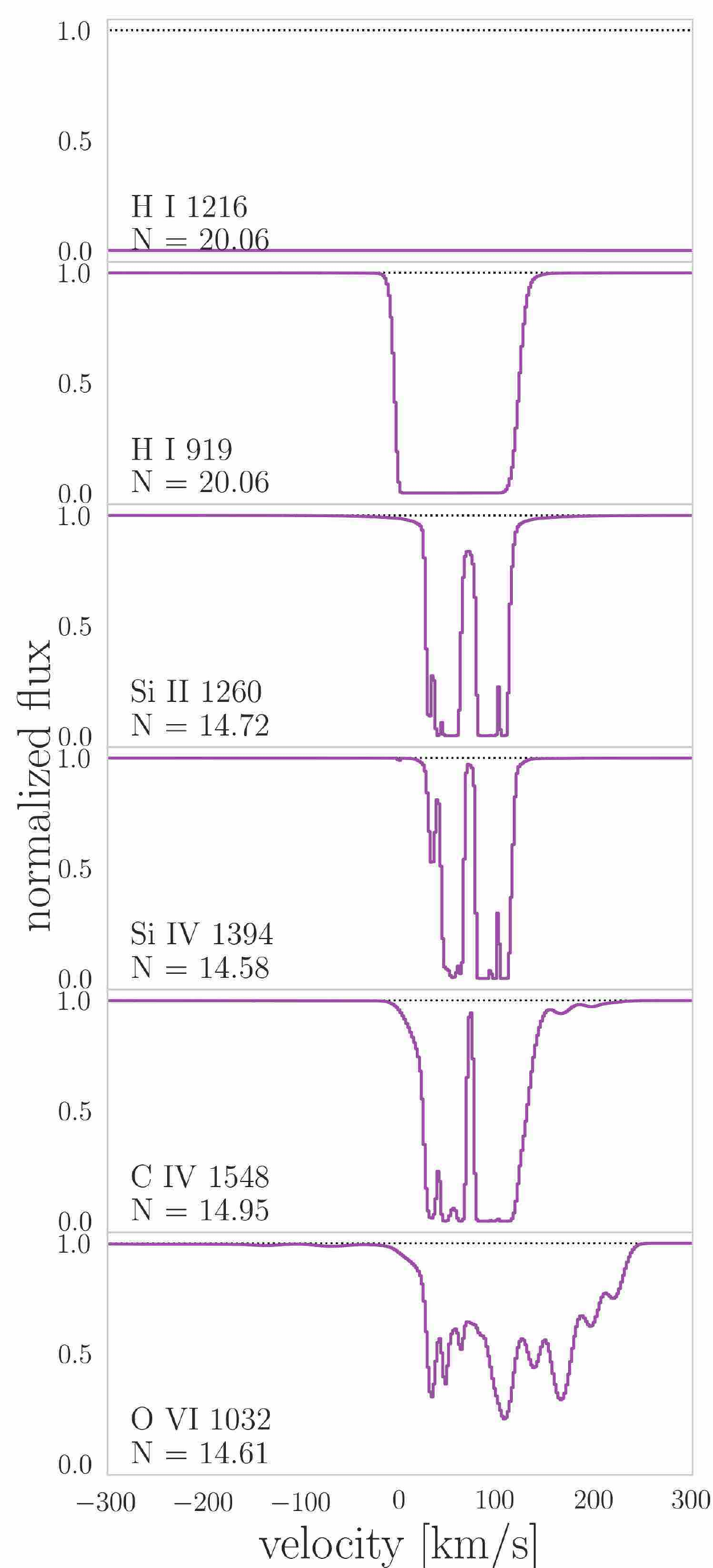}
    \caption{Typical absorbers from the standard-resolution simulation with $\mlnhi$ increasing to the right. Each panel gives the species and rest-frame wavelength and the log column density in cm$^{-2}$ as ``$N =$''; we include \hi$\lambda 919$\AA\ to highlight some of the structure that is lost owing to the saturation of \lya. \label{fig:natspectra}}
\end{figure*}

\begin{figure*}
    \includegraphics[width=0.24\textwidth]{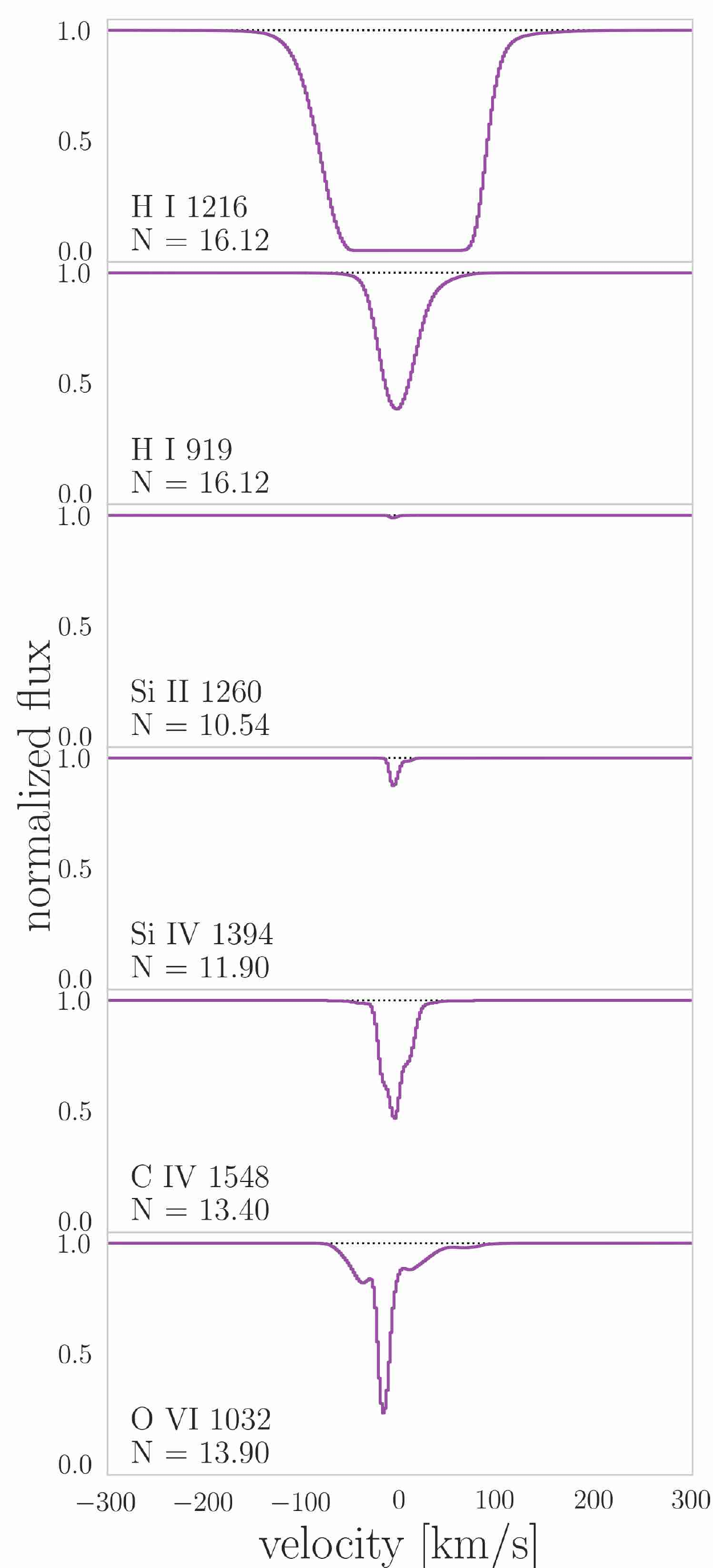}\hfill
    \includegraphics[width=0.24\textwidth]{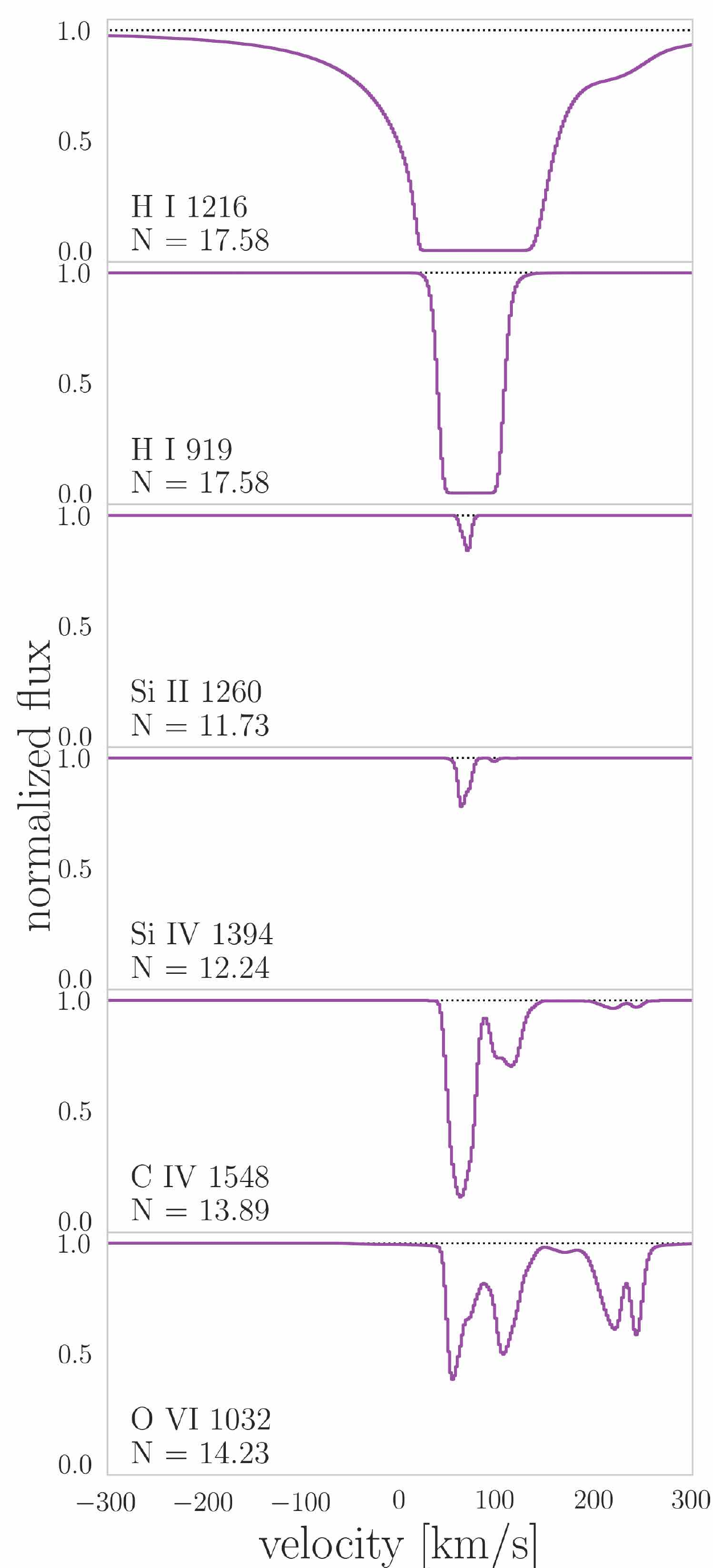}\hfill
    \includegraphics[width=0.24\textwidth]{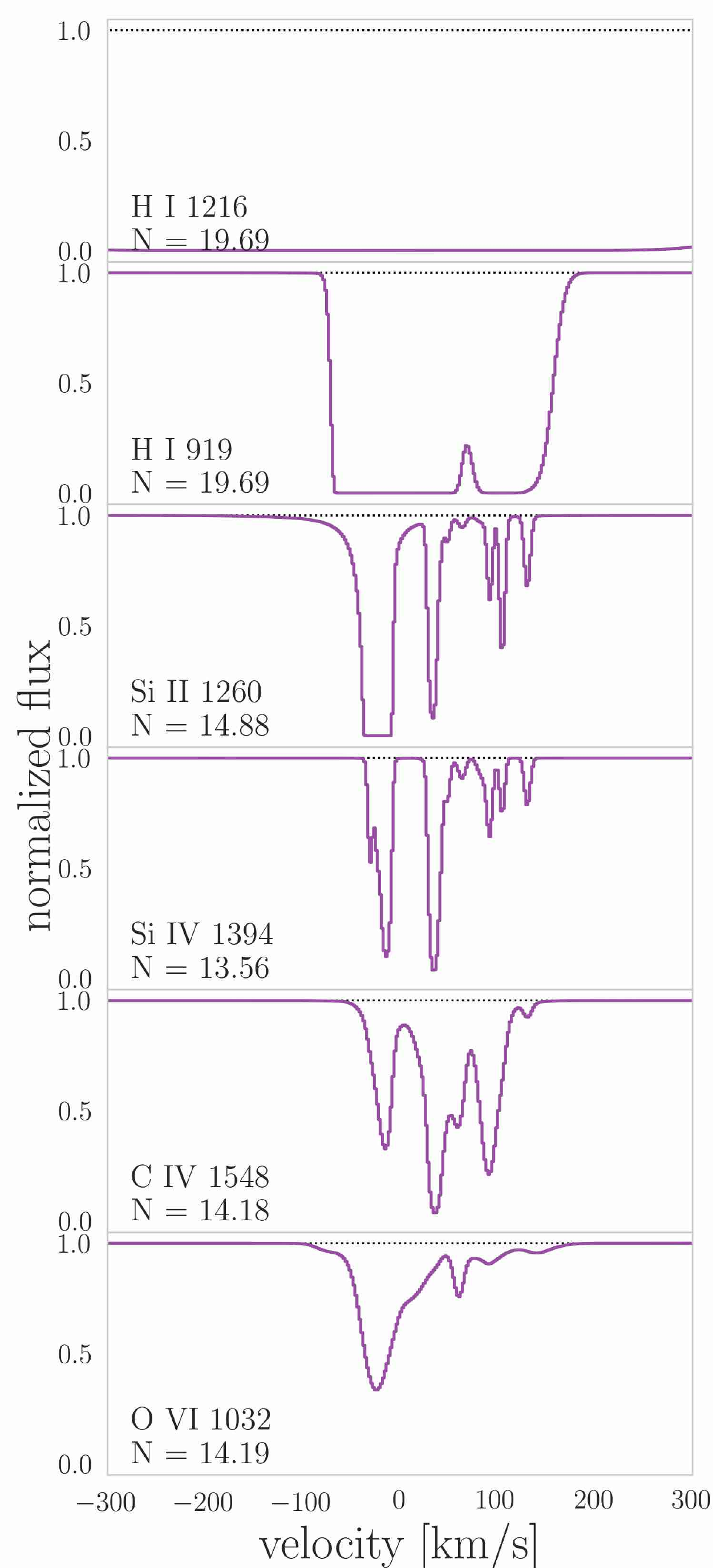}\hfill
    \includegraphics[width=0.24\textwidth]{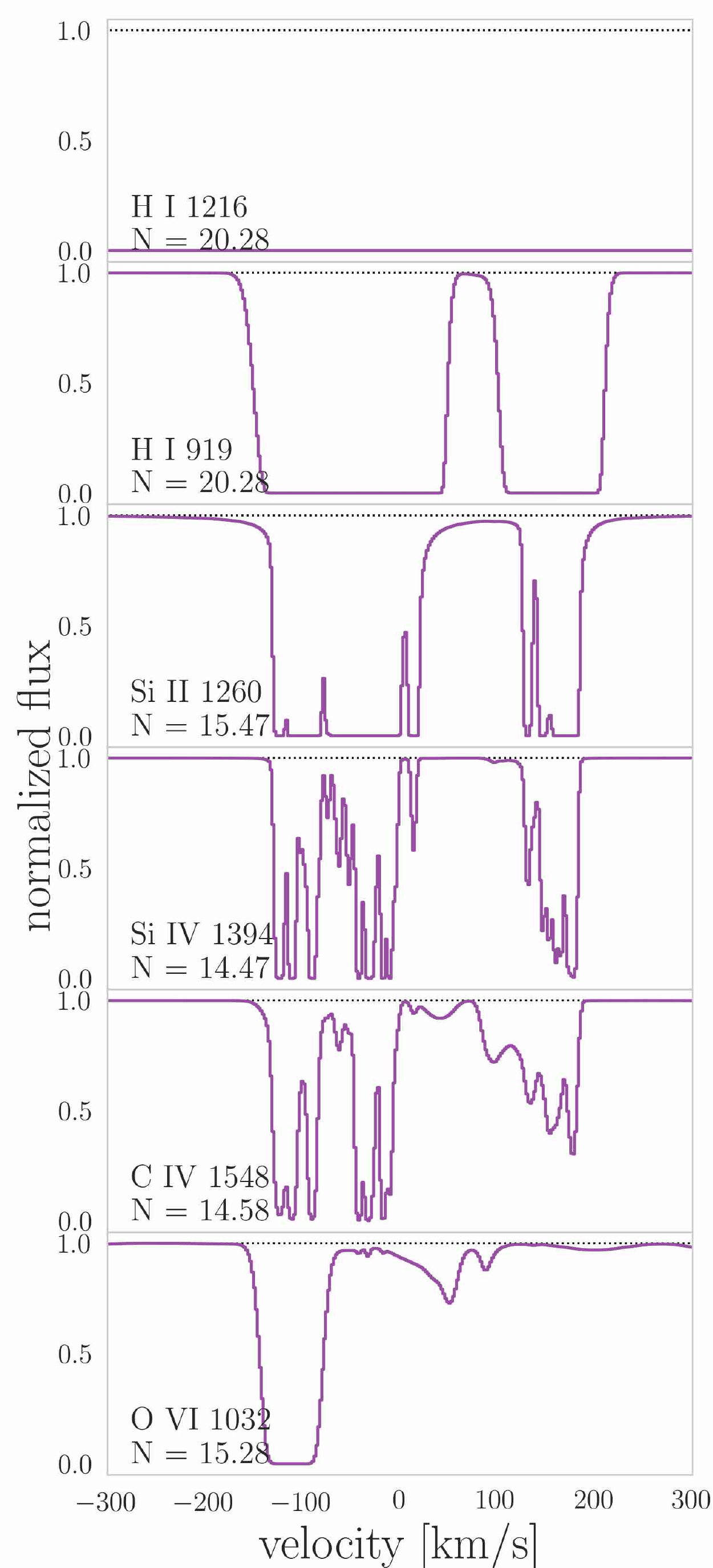}\\
    \caption{Typical absorbers from the high-resolution simulation with $\mlnhi$ increasing to the right. Each panel gives the species and rest-frame wavelength and the log column density in cm$^{-2}$ as ``$N =$''; we include \hi$\lambda 919$\AA\ to highlight some of the structure that is lost owing to the saturation of \lya.  \label{fig:refspectra}}
\end{figure*}

In the observable universe, we of course cannot directly measure number-density weighted velocity profiles, but instead observe number density weighted by the oscillator strength and wavelength of the relevant atomic transition. The effects of saturation in particular wash out the underlying velocity structure, as is especially obvious when using the locations of low-ionization components to guide fitting saturated \hi\ lines. Nonetheless, as the velocity phase diagrams shown in Figures~\ref{fig:velphaseref} and \ref{fig:velphasenat} (and in Appendix~\ref{app:velphase}) demonstrate, how well resolved---or not---the underlying gas flow is can have a dramatic impact on the kinematic structure in the resulting absorption-line spectrum. We explore these effects and how they complicate comparisons to the KODIAQ data in \S\,\ref{sec:comparison}.

%%%%%%%%%%%%%%%%%%%%%%%%%%%%%%%%%%%%%%%%%%%%%%%%%%%%%%%%%%%%%%%%%%%%
%%%%%%%%%%%%%%%%%%%%%%%%%%%%%%%%%%%%%%%%%%%%%%%%%%%%%%%%%%%%%%%%%%%%
%%%%%%%%%%%%%%%%%%%%%%  spectral comparisons %%%%%%%%%%%%%%%%%%%%%%%
%%%%%%%%%%%%%%%%%%%%%%%%%%%%%%%%%%%%%%%%%%%%%%%%%%%%%%%%%%%%%%%%%%%%
\section{Quantifying Kinematic Structure and Comparisons to KODIAQ}\label{sec:comparison}
We turn now to comparing the kinematic structure of the synthetic absorbers from the two simulations to observed absorption-line systems in high-resolution spectra at $2\lesssim z \lesssim 2.5$.

\begin{figure*}[!ht]
    \centering
    \includegraphics[width=0.49\textwidth]{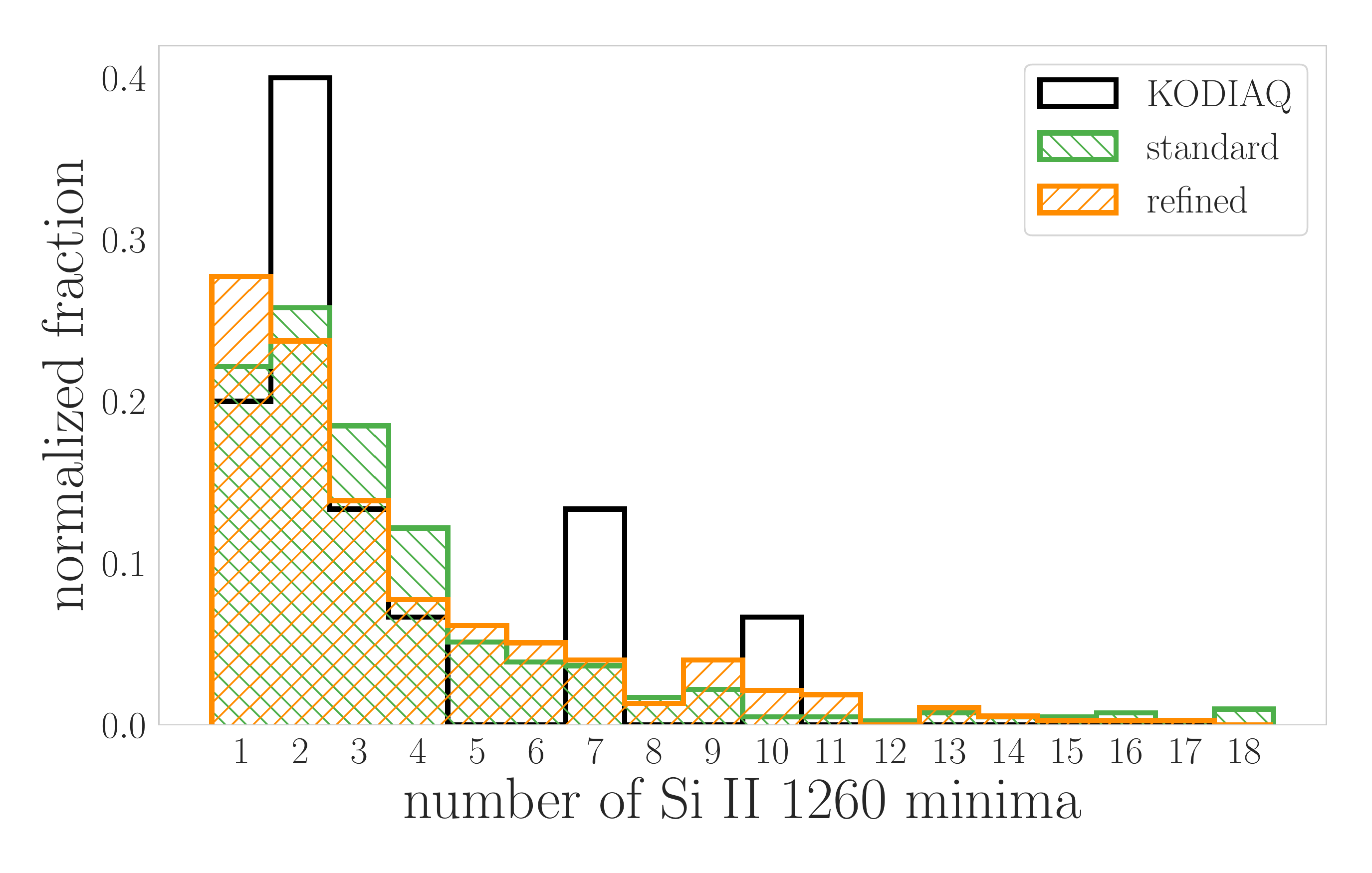}\hfill
    \includegraphics[width=0.49\textwidth]{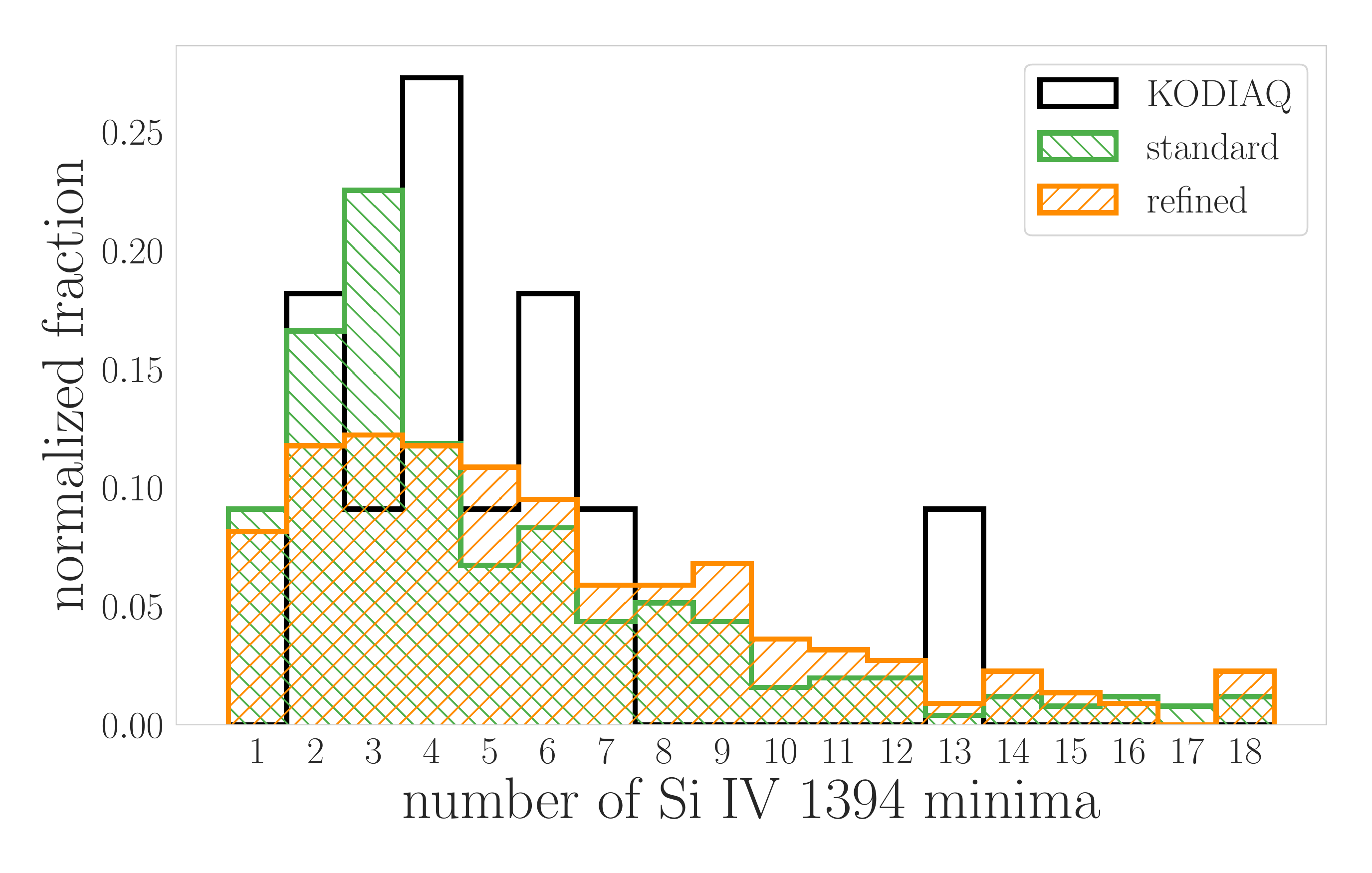}\\
    \includegraphics[width=0.49\textwidth]{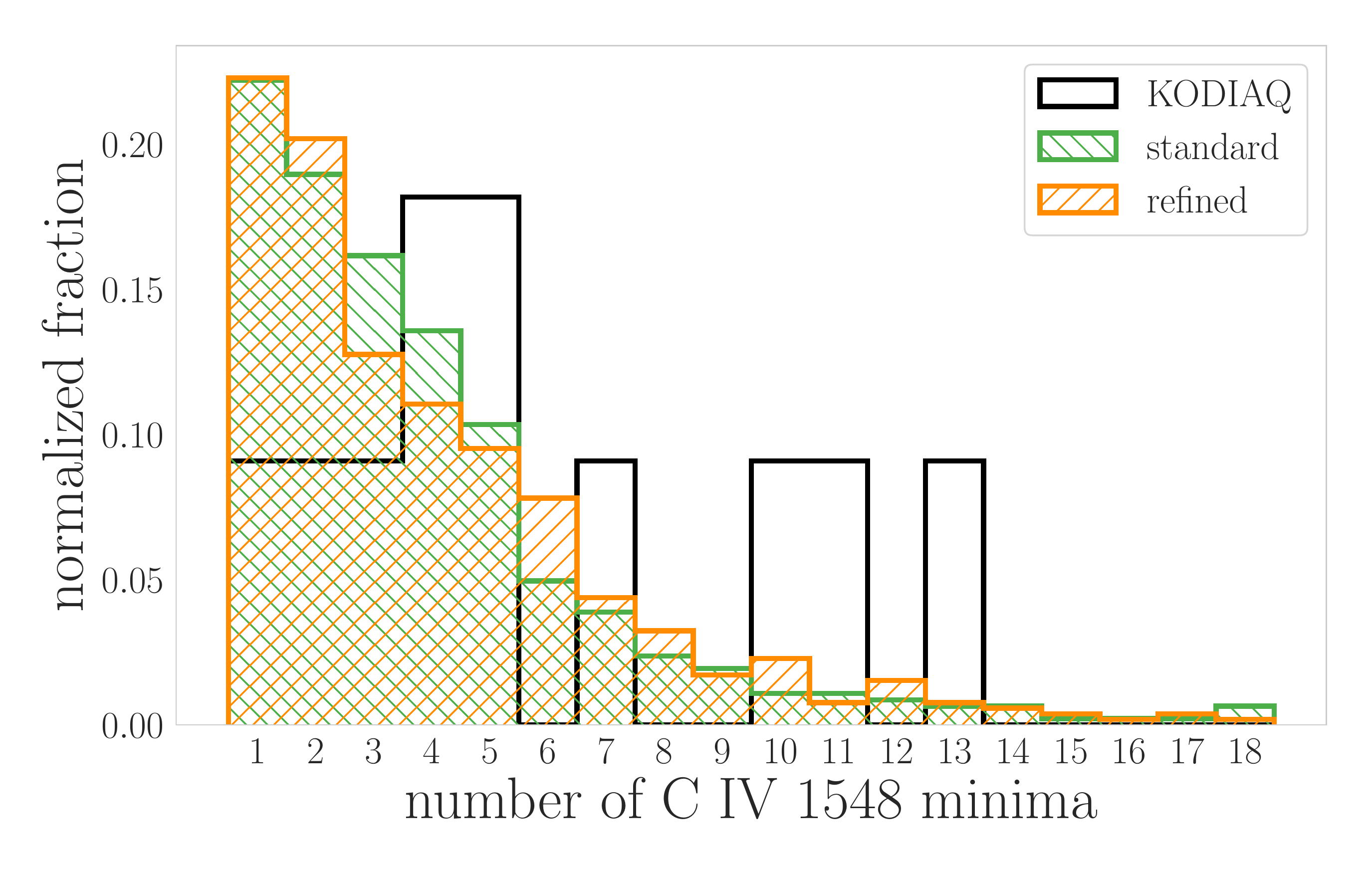}\hfill
    \includegraphics[width=0.49\textwidth]{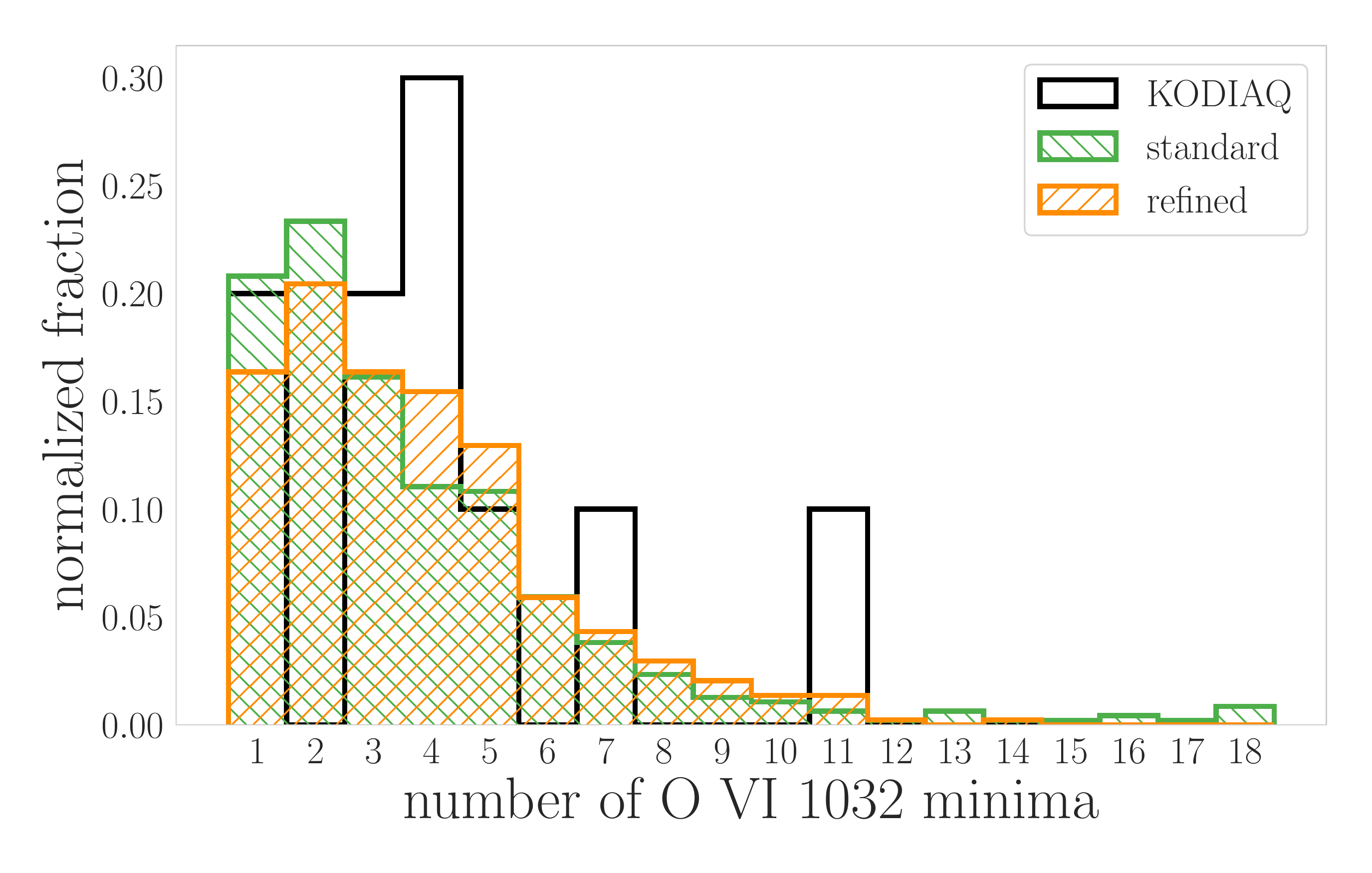}
    \caption{The relative number of minima with flux $<0.95$ in LLS-selected synthetic sightlines chosen to have at least the minimum column as seen in the KODIAQ data.
    \label{fig:minimahist}}
\end{figure*}

The larger spreads of physical properties emergent in the higher resolution simulation also correspond to larger spreads in the line-of-sight velocities. Because of the changes to the underlying structure of the gas, we compare sightlines in the two simulations at fixed \hi\ column density rather than fixed position in the box. We show in Figures~\ref{fig:refspectra} and \ref{fig:natspectra} sample spectra, at fixed \hi\ column density, in the two runs, as compared to a KODIAQ set of absorbers at similar \hi\ column. By eye, the high-resolution simulation yields a much richer kinematic structure. First, the absorbers from the high-resolution simulation are much more likely to have many components, whereas the standard resolution simulation yields absorbers that are more frequently a single component. Second, when the standard resolution simulation does have multiple components, they are very often unblended, with the flux reaching or nearly reaching the continuum between minima. While the first effect may be in part because of the different covering fractions (and conditional covering fractions) in the two simulations, both effects are likely caused by the artificial discretization of velocity from the larger resolution elements in the standard resolution simulation.

By eye, the spectra from the high-resolution simulation seem to have many more components than those from the standard-resolution simulation. Moreover, when the standard-resolution spectra {\em do} have many components, they tend to be fairly unblended, i.e., they have a comb-like structure with distinct lines that go back up near the continuum before absorbing again. We postulated in Section~\ref{sec:cloudvel} that these unblended components are predominantly caused by single grid cells, and while these cases do occur in spectra generated from the high-resolution simulation, they are less frequent. 

When the absorption-line systems in the KODIAQ data have many components, they tend to be more blended than even found in the high-resolution simulation. We stress that the comparisons for this generation of the FOGGIE simulations to the KODIAQ data should be taken only qualitatively: the real spectra are probing pathlengths of $\gg 200$\ckpch\ and a much wider range of galaxy masses, environments, and impact parameters than offered by our single halo. As we show in Figure~\ref{fig:metalcover}, the column density parameter space traced by our synthetic absorbers and the KODIAQ data is partially non-overlapping---there is simply a small parameter space of LLS-gas probed by this one halo. 
Also, as we analyze the synthetic spectra without adding noise, we are able to ``detect'' more minima in these spectra than we would be able to fit for in even the highest signal-to-noise KODIAQ spectra.
Nonetheless, we find it instructive to compare the synthetic data to the real universe where possible.

In Figure~\ref{fig:minimahist} we show the distribution of the number of minima in the two simulations for \siii, \siiv, \civ, and \ovi\ as compared to the KODIAQ sample. While the differences between the standard- and high-resolution simulations are subtle, there are generally more systems with $4\leq\nmin\leq 10$ in the high-resolution simulation than the standard resolution simulation. 
Two-sided Kolmogorov-Smirnov tests rule out the \siiv\ distributions as being drawn from the same sample at a $9.6\times 10^{-3}$ level; the other distributions cannot be ruled out as being drawn from the same parent sample. 

\begin{figure*}
    \centering
    \includegraphics[width=0.49\textwidth]{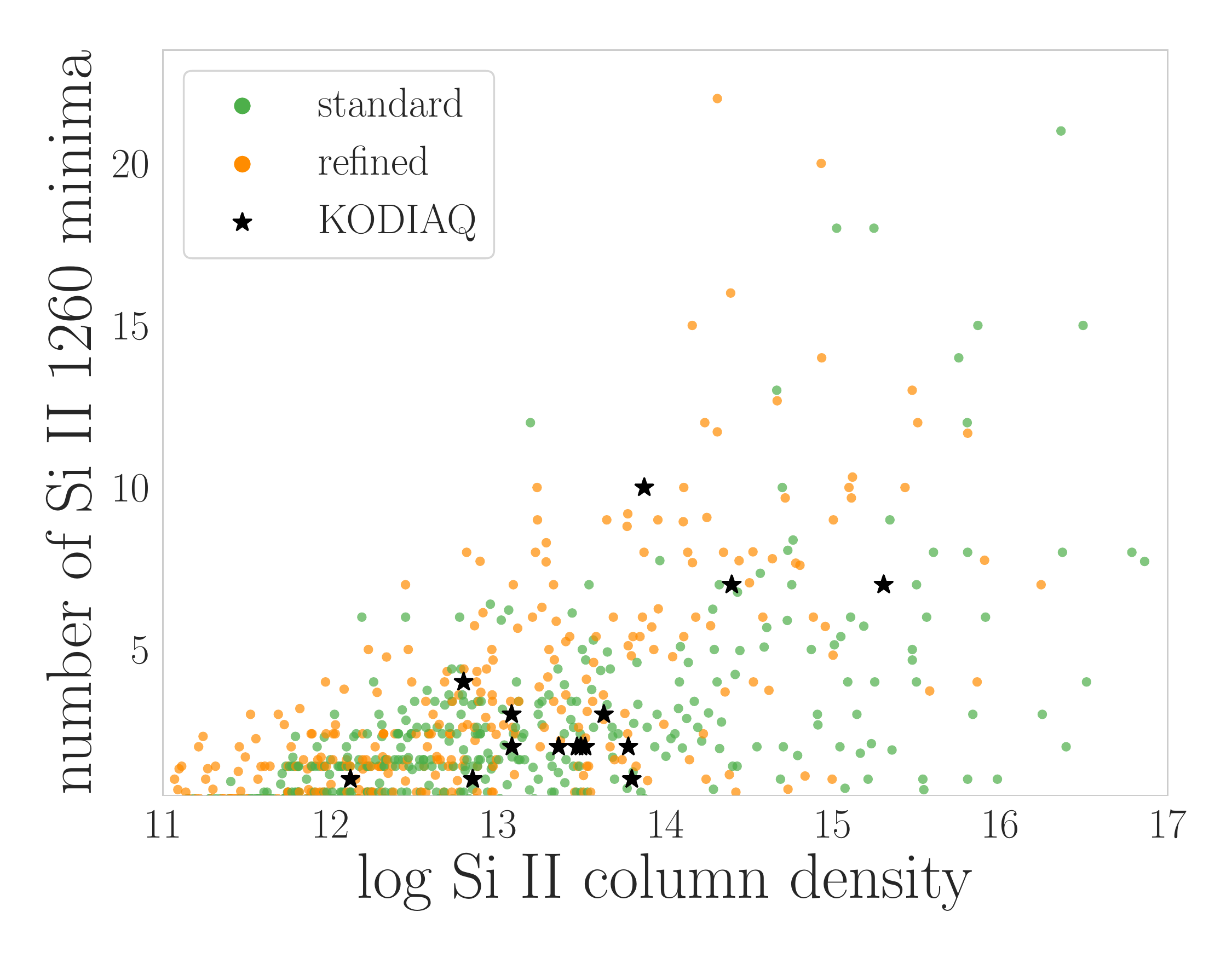}\hfill
    \includegraphics[width=0.49\textwidth]{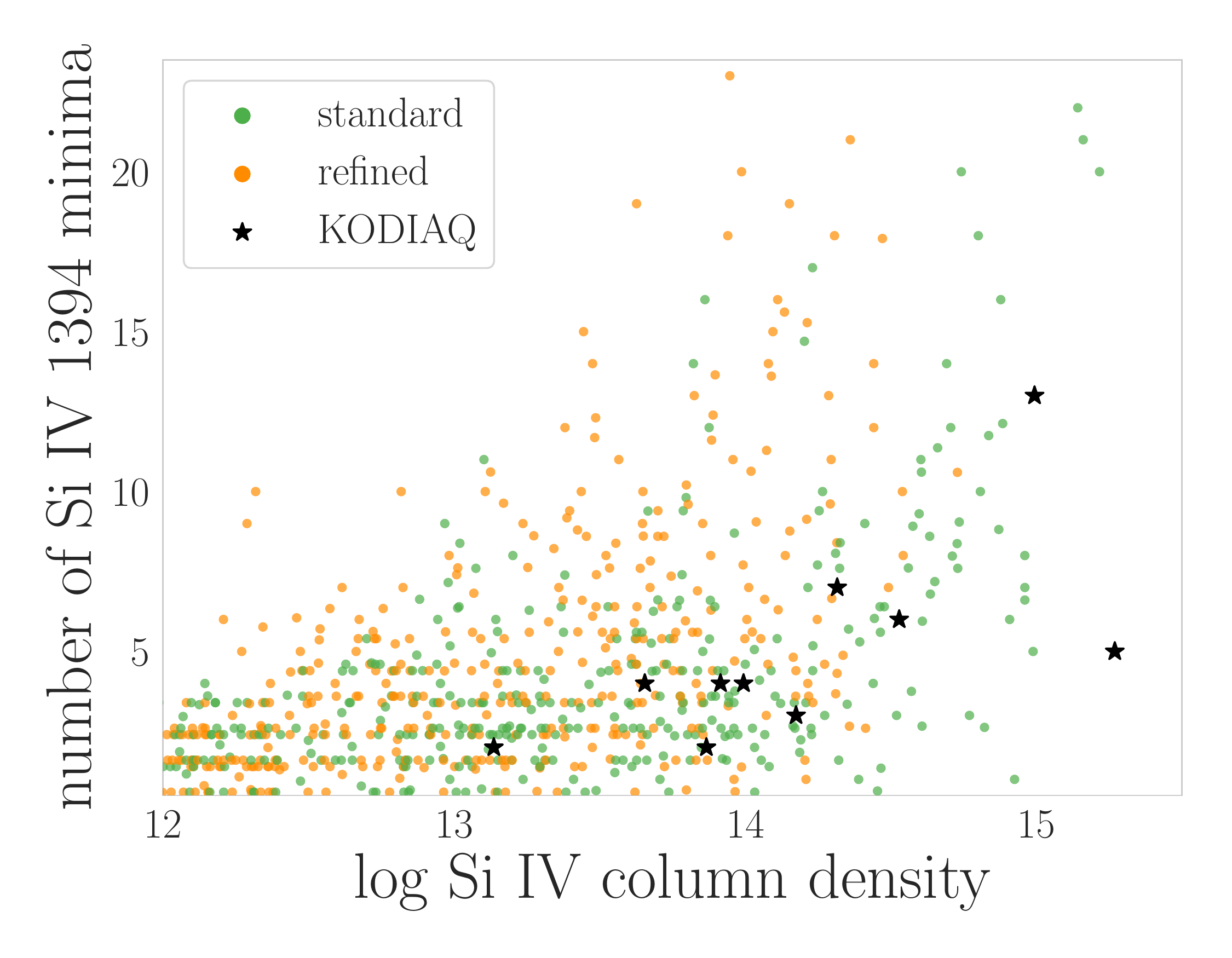}\\
    \includegraphics[width=0.49\textwidth]{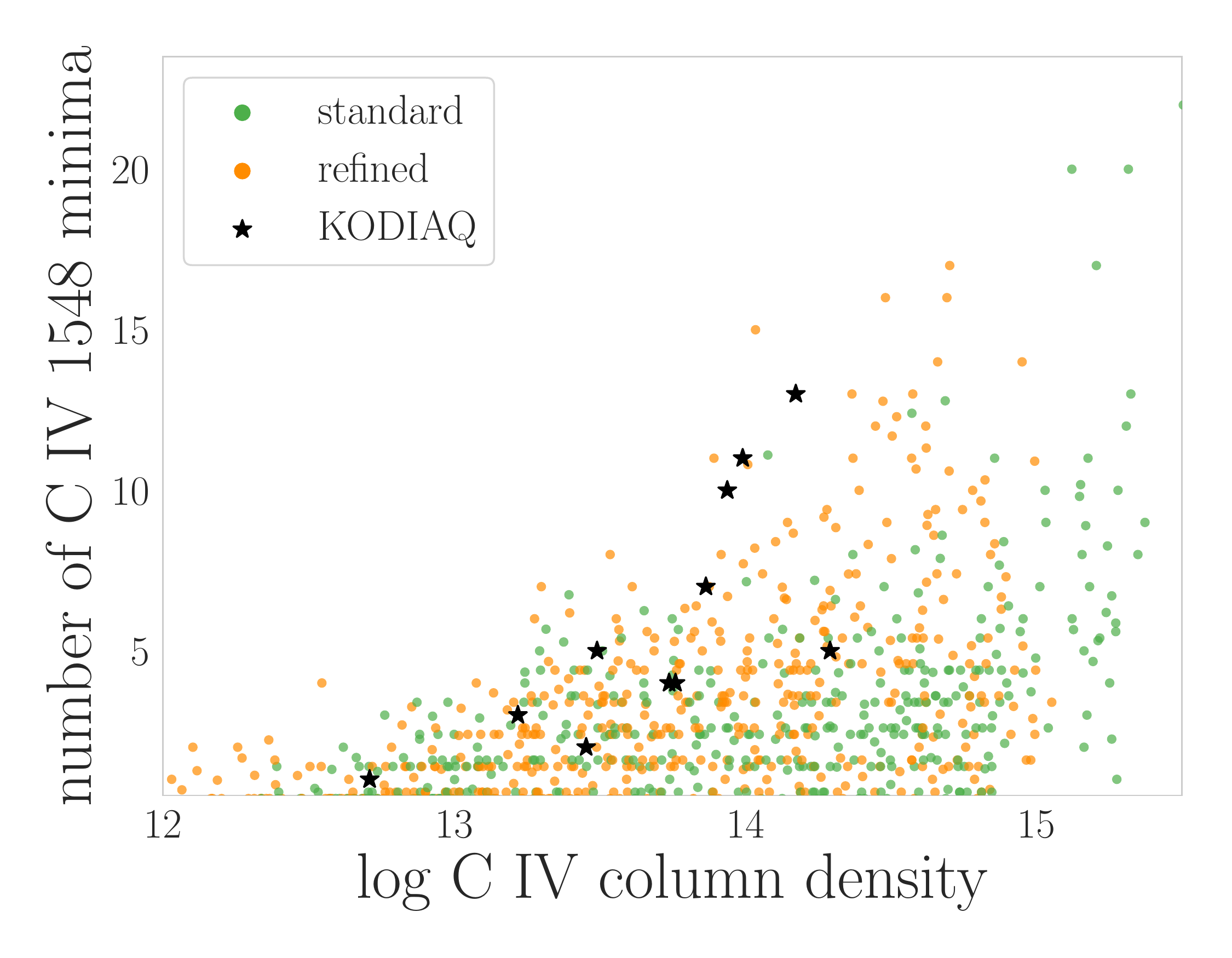}\hfill
    \includegraphics[width=0.49\textwidth]{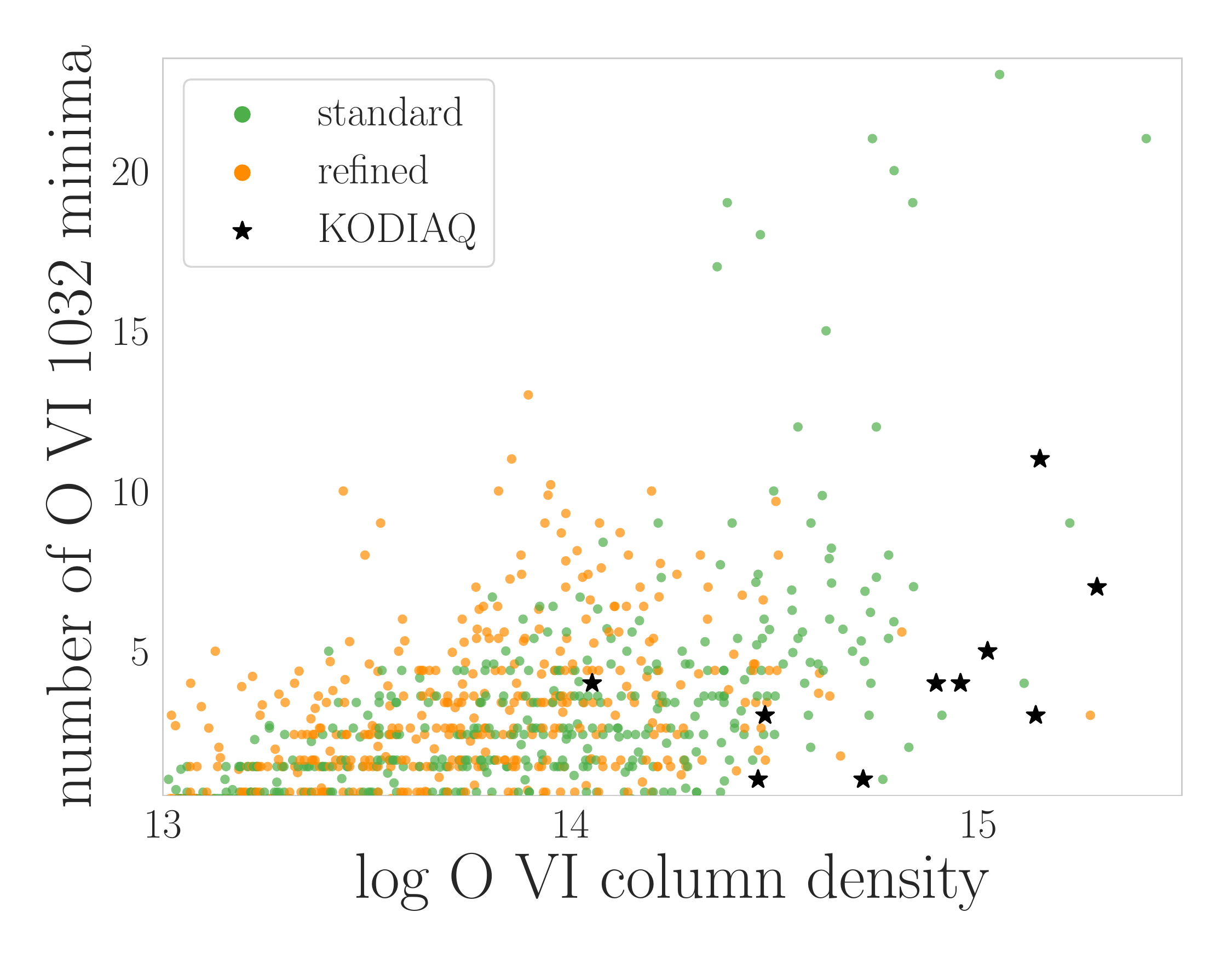}
    \caption{The number of minima with flux $<0.95$ for LLS sightlines versus column density. The lines are linear regressions to guide the eye; in general, higher column densities have more absorption minima and for most ions at most column densities, the high-resolution ``refined'' simulation has more minima than the standard-resolution simulation.
    \label{fig:colmin}}
\end{figure*}

The distributions of $\nmin$ in Figure~\ref{fig:minimahist} are complicated by the fact that higher column density systems tend to have more minima, as we show in Figure~\ref{fig:colmin}, and that the high-resolution simulation has lower metal column densities for LLS systems (Figures~\ref{fig:metalcover} and \ref{fig:cddf}). Figure~\ref{fig:colmin} also shows that, at fixed column density, the high-resolution simulation tends to have more minima than the standard resolution simulation, and for \siii\ and \civ, the high-resolution simulation is in better agreement with the KODIAQ data.

\begin{figure}
    \centering
    \includegraphics[width=0.47\textwidth]{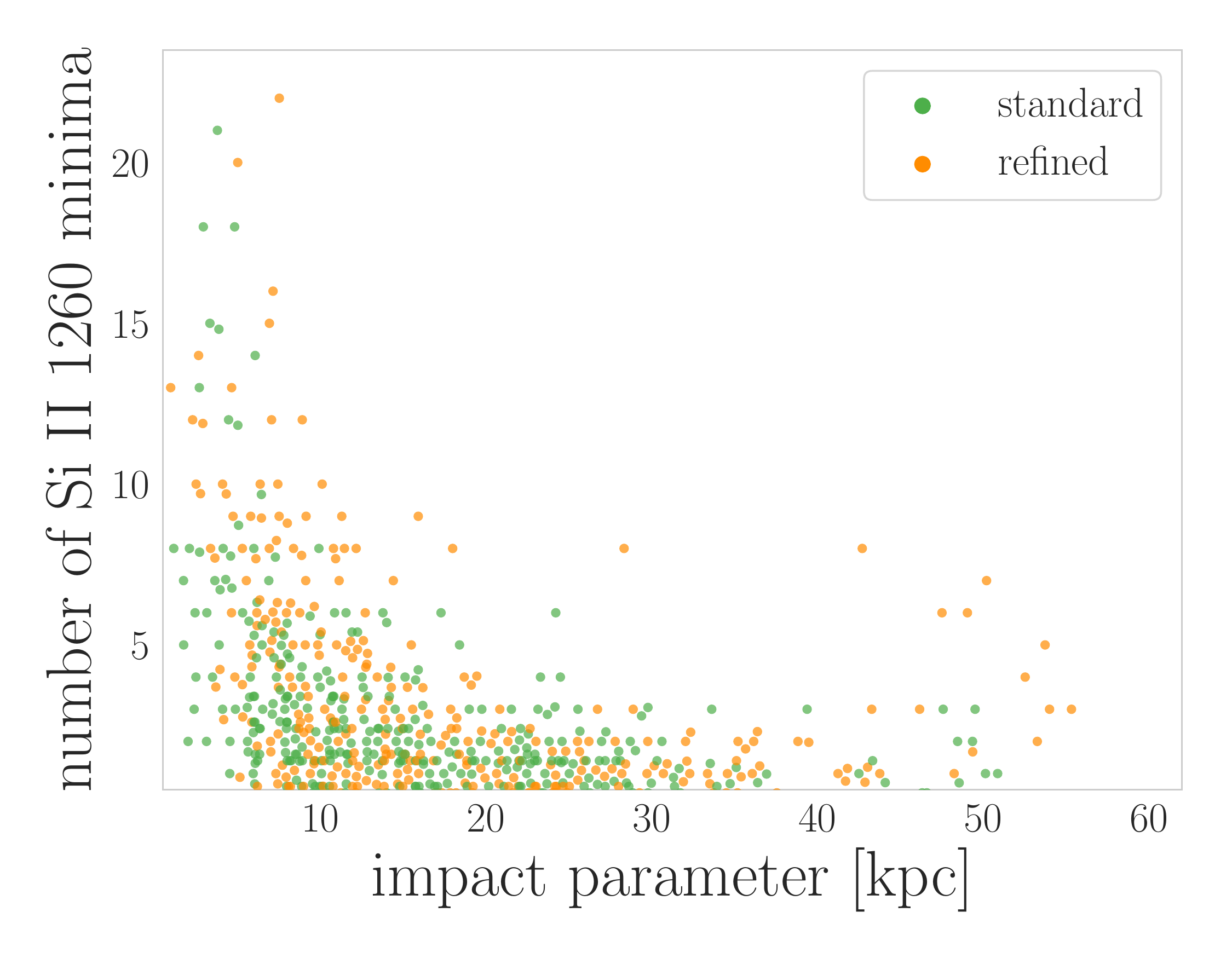}\\
    \includegraphics[width=0.47\textwidth]{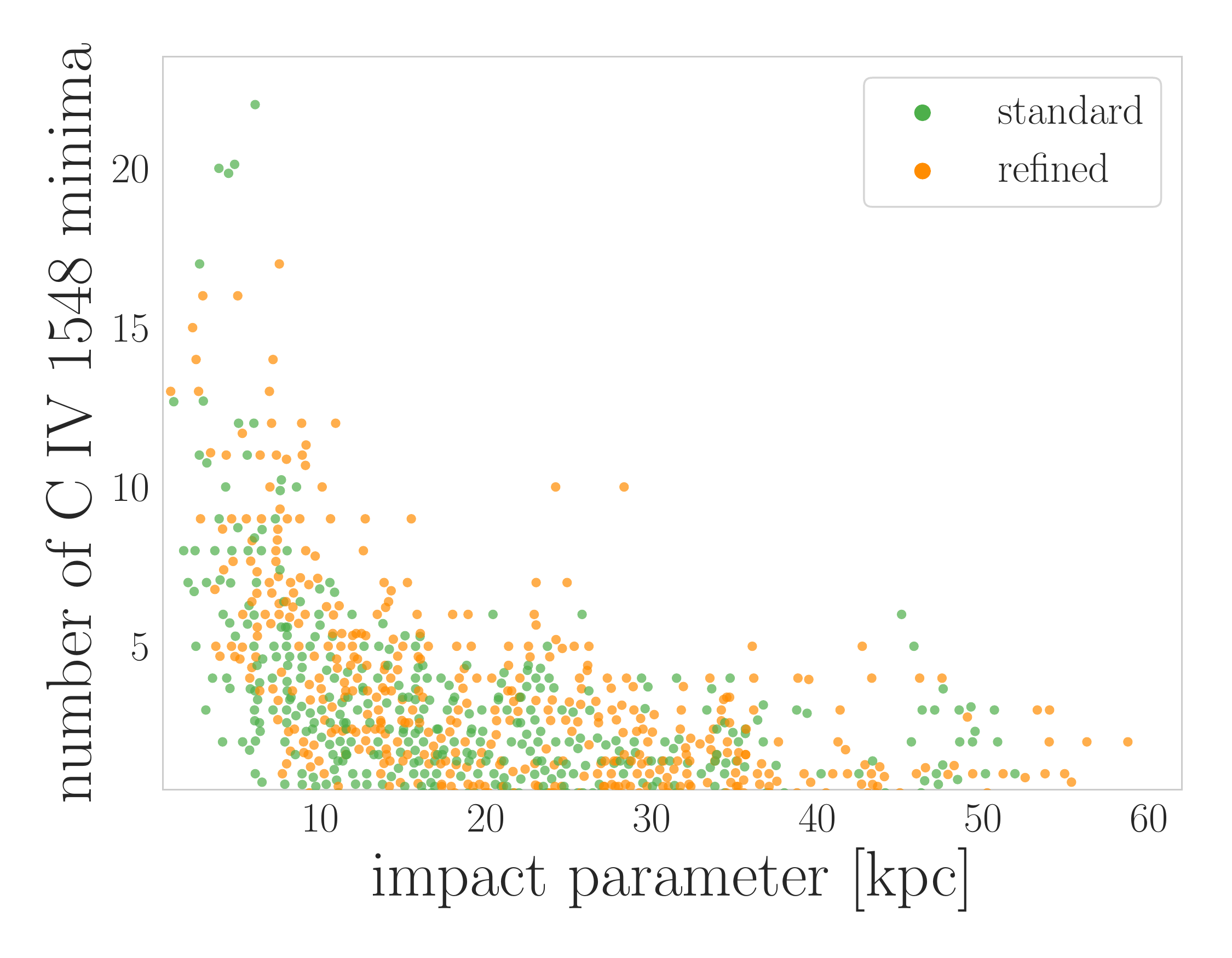}
    \caption{The number of \siii\ and \civ\ minima versus impact parameter for LLS sightlines in the two simulations. The high-$\nmin$\ systems tend to be found at smaller impact parameters.
    \label{fig:impact}}
\end{figure}

Finally, in both simulations we find that the absorbers with the highest number of minima are preferentially found at low impact parameters, as shown in Figure \ref{fig:impact}. This prediction is consistent with the COS-Halos data of $L^*$ galaxies at $z\sim 0.25$ (\citealp{werk14, werk16}; c.f.\ \citealp{borthakur16}). As we gather the associated galaxy information for the KODIAQ absorbers in the future, it will be interesting to see if the more kinematically complex absorbers have galaxies at small impact parameters.

%%%%%%%%%%%%%%%%%%%%%%%%%%%%%%%%%%%%%%%%%%%%%%%%%%%%%%%%%%%%%%%%%%%%
%%%%%%%%%%%%%%%%%%%%%%%%%%%%%%%%%%%%%%%%%%%%%%%%%%%%%%%%%%%%%%%%%%%%
%%%%%%%%%%%%%%%%%%%% conclusions   %%%%%%%%%%%%%%%%%%%%%%%%%%%%%%%%%
%%%%%%%%%%%%%%%%%%%%%%%%%%%%%%%%%%%%%%%%%%%%%%%%%%%%%%%%%%%%%%%%%%%%
\section{Conclusions and Future Directions} \label{sec:conc}
Testing theories of galaxy evolution with inter- and circumgalactic absorption-line spectroscopy is a longstanding goal of extragalactic astrophysics that has motivated many substantial investments in observing time and computing resources. Both observations and physical arguments point to the CGM having physical structure on scales smaller than those typically resolved by cosmological hydrodynamic simulations aimed at understanding galaxy evolution. Simulations using a wide range of numerical techniques applied at a range of scales have addressed this problem with mixed success. By forcing the circumgalactic resolution to sub-kpc scales, we have shown here that the gas structures responsible for most of the circumgalactic absorption are resolved---and that this is not the case in a standard-resolution density-refined simulation. 

Many studies have assumed that readily observable quantities and statistics derived from them---such as column densities, covering fractions, and system-level velocities---are adequate to provide detailed constraints on models. Under this assumption, feedback prescriptions and other input physics are varied to match the adopted data, and recovery of the observables is taken as an indication that the input physics is ``right'' \citep[e.g.,][]{hummels13,ford13,oppenheimer18a}. FOGGIE shows that the resolution of a simulation has a major effect on the physical configuration of the CGM gas---cloud sizes, masses, and kinematics---even if the underlying ``subgrid'' prescriptions do not change. Figure~\ref{fig:metalcover} shows that FOGGIE's covering fractions of several commonly observed ions do not change dramatically when moving from standard to forced resolution. Cumulative covering fractions are typically within 10--20\% of one another for \siii, \civ, and \ovi. However, the ``cloud'' sizes that produce these covering fractions (Figures~\ref{fig:cloudsizehist}, \ref{fig:cloudcellshist}, and \ref{fig:cloudmasses}) change by orders of magnitude. The forced refinement simulations yield more structure in the temperature, density, and metallicity phase space than standard refinement, yet somehow the bulk, projected quantities such as mass and ion surface densities change only marginally, at a level often within observational uncertainties. This result suggests that such integrated quantities are poor diagnostics of the actual physical processes that give rise to the detected absorption. 

FOGGIE's findings on the kinematic structure of simulated absorbers points to the same conclusion: low resolution simulations---that is, the currently cutting-edge cosmological simulations with Lagrangian-like fixed mass resolution---cannot adequately reproduce the kinematic complexity of real CGM absorption. To quantify the changes in kinematic resolution, we have defined a new figure of merit (Equation~\ref{eqn:velratio}) that expresses how well simulations resolve the velocity field of detected absorption. Figure~\ref{fig:velratio} shows that a standard resolution simulations fail this criterion at rates 2--5 times higher than the high-resolution simulations. These improvements are driven by better sampling of the true velocity fields and by placing more resolution elements across physical structures, with the net effect being that there are more---and better resolved---velocity components. 
Since bulk quantities such as projected column densities have proven to have relatively little diagnostic power over small scale gas structure, the detailed kinematic structure of the multiphase ions and their interrelationships may prove our best hope of actually constraining the small scale physics of the CGM.
Quantifying the kinematic structure of absorption-line measurements in a meaningful and robust way is therefore an open challenge for both future simulations and future observational analyses.

This study is, however, limited in that we have simulated a single, low-mass halo, and have a limited set of observational data from KODIAQ. For future FOGGIE comparisons to KODIAQ and $z\geq2$ galaxy samples, we will simulate a larger sample of halos from a broad range of halo masses and analyze a larger sample of real LLS spectra.
Moreover, the feedback scheme used here only includes the thermal feedback from supernovae, which may lead to stellar-to-halo mass ratios that are too high \citep{hopkins14}; likewise, outflows from purely thermal feedback are by design not multiphase, so any cool gas in the CGM must arise in situ in this current generation of simulations (though, in principle, cold ISM gas could be entrained by the hot outflows, this is less likely; \citealp{schneider17}).
For these reasons, future FOGGIE studies will improve the star formation and AGN feedback schemes in Enzo. Sources of non-thermal pressure from, e.g., cosmic rays \citep{butsky18}, have been shown in lower-resolution simulations to have a dramatic impact on the structure of the CGM \citep*{salem16} such that the models are brought more in line with the observations; it will be important to quantify how these processes are affected by better resolving the CGM.

Finally, the forced refinement scheme used here, while simple, is both computationally costly and limited. Future FOGGIE runs will take advantage of a new scheme we are calling ``cooling refinement,'' where in addition to the forced refinement region used here, we additionally allow the gas to refine based on its cooling length. Our initial tests show that this hybrid approach allows the denser gas to cool more efficiently while not introducing unnatural mixing in the low-density phases seen in more standard refinement schemes.  

In \citetalias{corlies18}, we show that resolving gas structures in the circumgalactic medium is vital for predicting signatures of the CGM in emission. 
We conclude that future simulations aimed at understanding the co-evolution of galaxies and their gaseous halos should attempt to resolve the circumgalactic medium in addition to the denser interstellar gas. 

\acknowledgments
This study was primarily funded by the National Science Foundation via NSF AST-1517908, which helped support the contributions of LC, BWO, NL, JOM, and JCH. LC was additionally supported in part by HST AR \#15012. BWO was supported in part by NSF grants PHY-1430152, AST-1514700, 
OAC-1835213, by NASA grants  NNX12AC98G, NNX15AP39G, and by HST AR \#14315. NL was also supported  by NASA ADAP grant NNX16AF52G. 
NE was supported by HST AR \#13919, HST GO \#14268, and HST AR \#14560. BDS was supported by NSF AST-1615848. JHW was supported by NSF grants
AST-1614333 and OAC-1835213, NASA grant NNX17AG23G, and HST-AR-14326. 
CBH was supported by HST AR \#13917, HST AR \#13919, and an NSF AAPF. 
Resources supporting this work were provided by the NASA High-End Computing (HEC) Program through the NASA Advanced Supercomputing (NAS) Division at Ames Research Center and were sponsored by NASA's Science Mission Directorate; we are grateful for the superb user-support provided by NAS. Resources were also provided by the Blue Waters sustained-petascale computing project, which is supported by the NSF (award number ACI 1238993 and ACI-1514580) and the state of Illinois. Blue Waters is a joint effort of the University of Illinois at Urbana-Champaign and its NCSA.  This work benefited from the dancing penguin and all the things emojis on Slack. The data presented herein were obtained at the W.\ M.\ Keck Observatory, which is operated as a scientific partnership among the California Institute of Technology, the University of California and the National Aeronautics and Space Administration. The Observatory was made possible by the generous financial support of the W.\ M.\ Keck Foundation. This research has made use of the Keck Observatory Archive (KOA), which is operated by the W.\ M.\ Keck Observatory and the NASA Exoplanet Science Institute (NExScI), under contract with the National Aeronautics and Space Administration. The authors wish to recognize and acknowledge the very significant cultural role and reverence that the summit of Maunakea has always had within the indigenous Hawaiian community. We are most fortunate to have the opportunity to conduct observations from this mountain.  We have made extensive use of the python libraries Astropy, a community-developed core Python package for Astronomy (\citealp{astropy2}, \url{http://www.astropy.org}), datashader and holoviews from Anaconda (\url{datashader.org}), pandas \citep{pandas}, and seaborn. Computations described in this work were performed using the publicly-available Enzo code, which is the product of a collaborative effort of many independent scientists from numerous institutions around the world.

\vspace{5mm}
\facilities{NASA Pleiades, NCSA Blue Waters, Keck (HIRES)}

\software{astropy \citep{astropy2},  
          Cloudy \citep{ferland13}, 
          Enzo \citep{bryan14},
          grackle \citep{smith17},
          yt \citep{turk11},
          Trident \citep{hummels17},
          ytree \citep{ytree}
          }

\bibliographystyle{aasjournal}

%%%%%%%%%%%%%%%%%%%%%%%%%%%%%%%%%%%%%%%%%%%%%%%%%%%%%%%%%%%%%%%%%%%%
%%%%%%%%%%%%%%%%%%%%%   appendices        %%%%%%%%%%%%%%%%%%%%%%%%%%
%%%%%%%%%%%%%%%%%%%%%%%%%%%%%%%%%%%%%%%%%%%%%%%%%%%%%%%%%%%%%%%%%%%%

\appendix 
\section{Resolution Test: The CGM at $<100$ parsecs}\label{app:resolution}
In addition to the $\nref=10$ ``high resolution'' simulation we evolved to $z=0$ and discuss in the main text, we also evolved a simulation enforcing $\nref=11$ within the same forced refinement volume to $z=2.5$. (More detailed properties of this simulation at $z=3$ are provided in \citetalias{corlies18}.) We show here the same analyses as in the main text, but for $z=2.5$ only. At this redshift, the $\nref=11$ simulation has a fixed circumgalactic (and interstellar) resolution of 78.4\,physical parsecs; the $\nref=10$ simulation has a CGM resolution of 157\,pc; we refer in this Appendix to these simulations by these physical resolutions.
Given the complex multiphysics nature of the simulations a formal convergence test is difficult to define precisely, so instead we choose to examine more meaningful metrics, i.e., convergence as it applies to the physical phenomena that are relevant to this investigation.
Broadly speaking, we find that improving the resolution by another factor of two spatially (and thus $8\times$ in volume) does result in yet smaller clouds, but the change is much less dramatic than between the standard resolution simulation and the $\nref=10$ simulation.

\begin{figure*}[htb]
    \centering
    \includegraphics[height=0.45\textwidth]{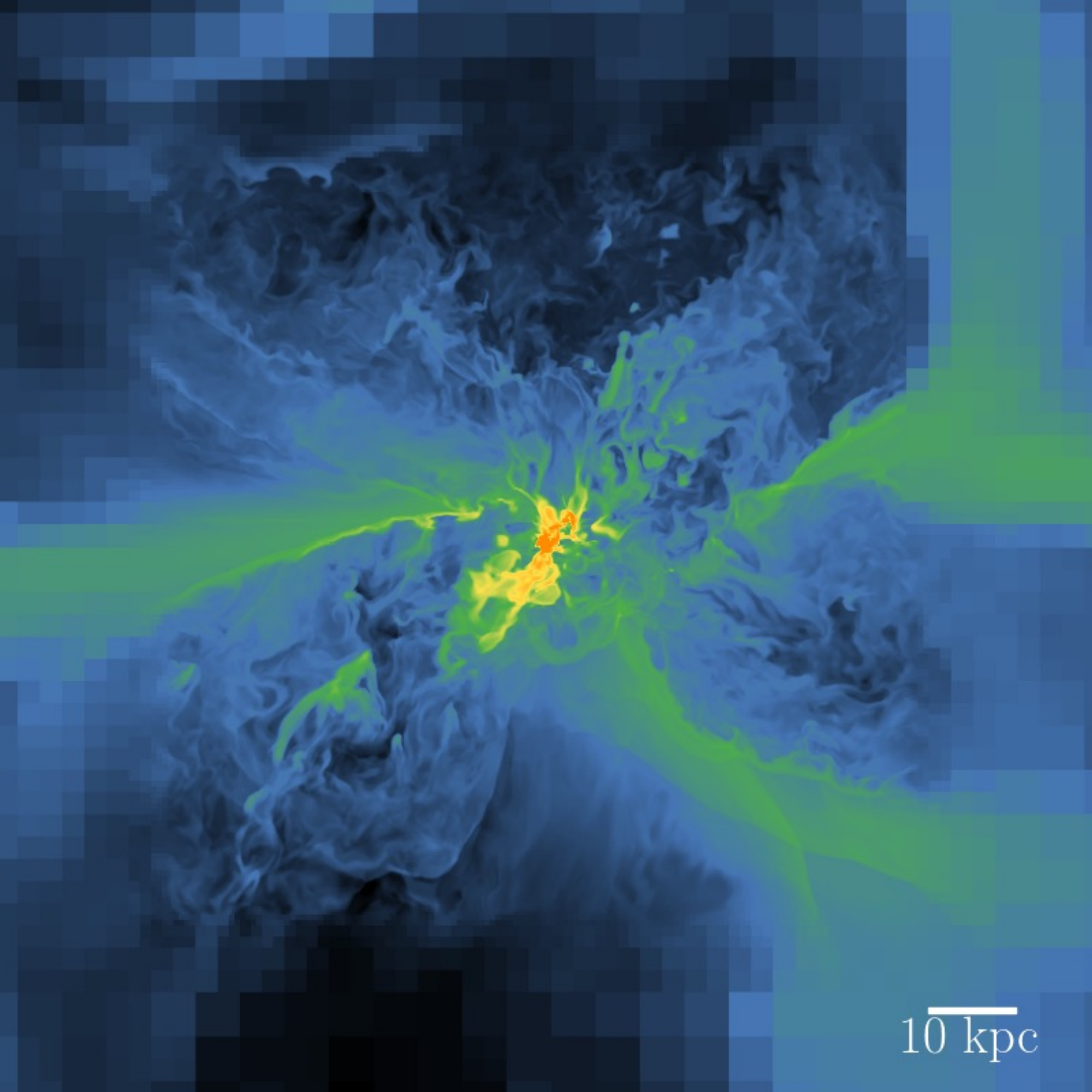}\hfill
    \includegraphics[height=0.45\textwidth]{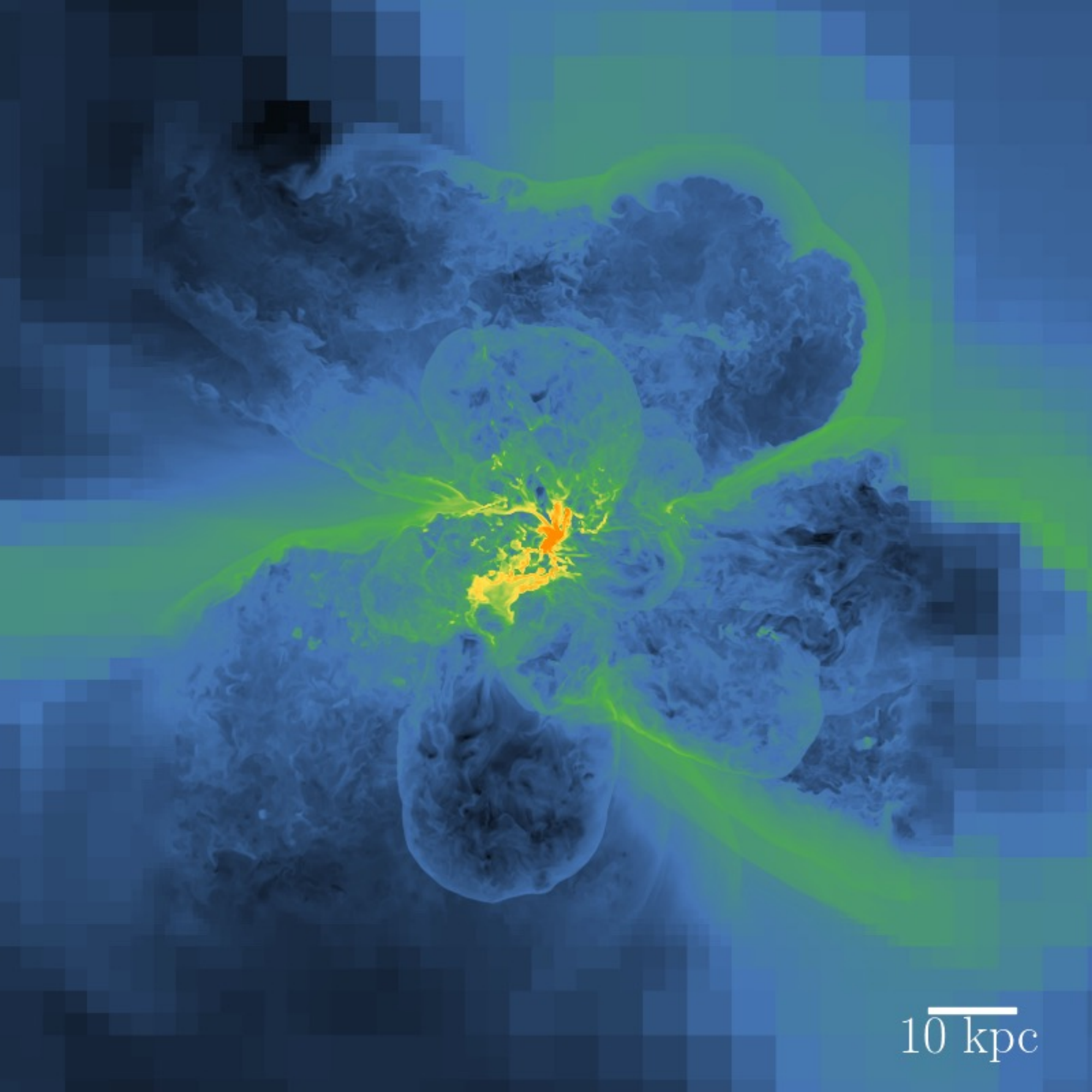}\hfill
    \includegraphics[height=0.45\textwidth]{density_slice_colorbar.pdf}\\
    \includegraphics[height=0.45\textwidth]{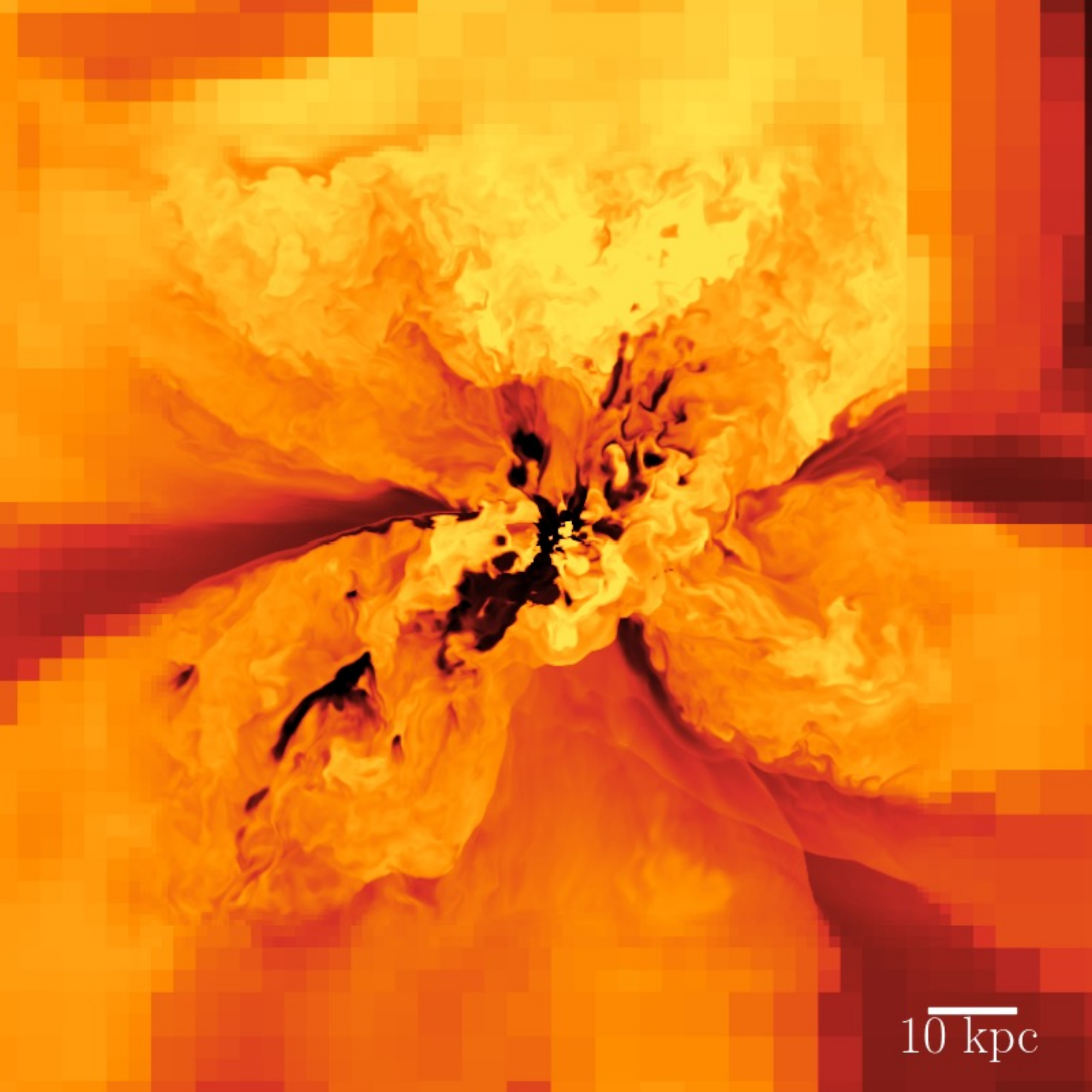}\hfill
    \includegraphics[height=0.45\textwidth]{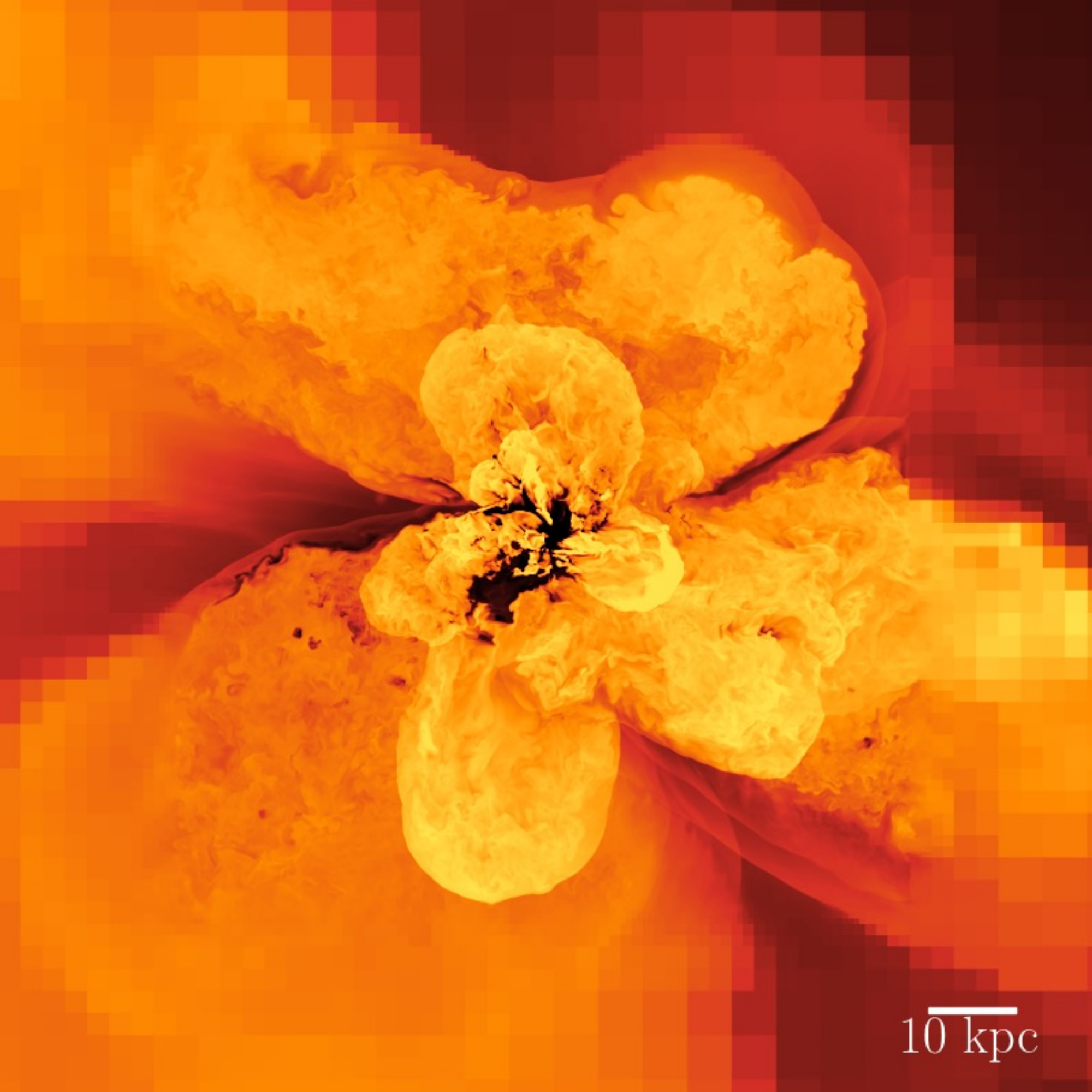}\hfill
    \includegraphics[height=0.45\textwidth]{temperature_colorbar.pdf}\\
    \includegraphics[height=0.45\textwidth]{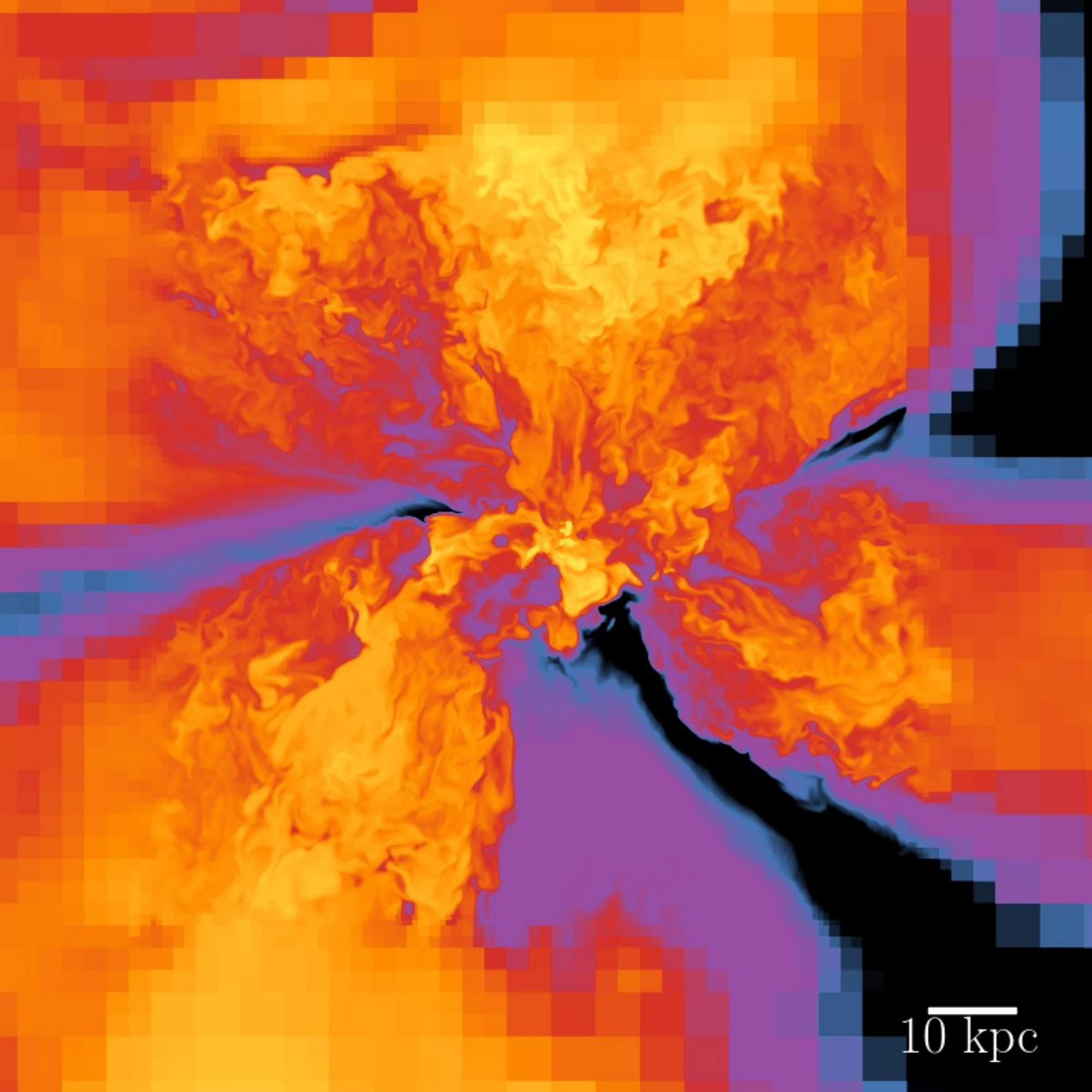}\hfill
    \includegraphics[height=0.45\textwidth]{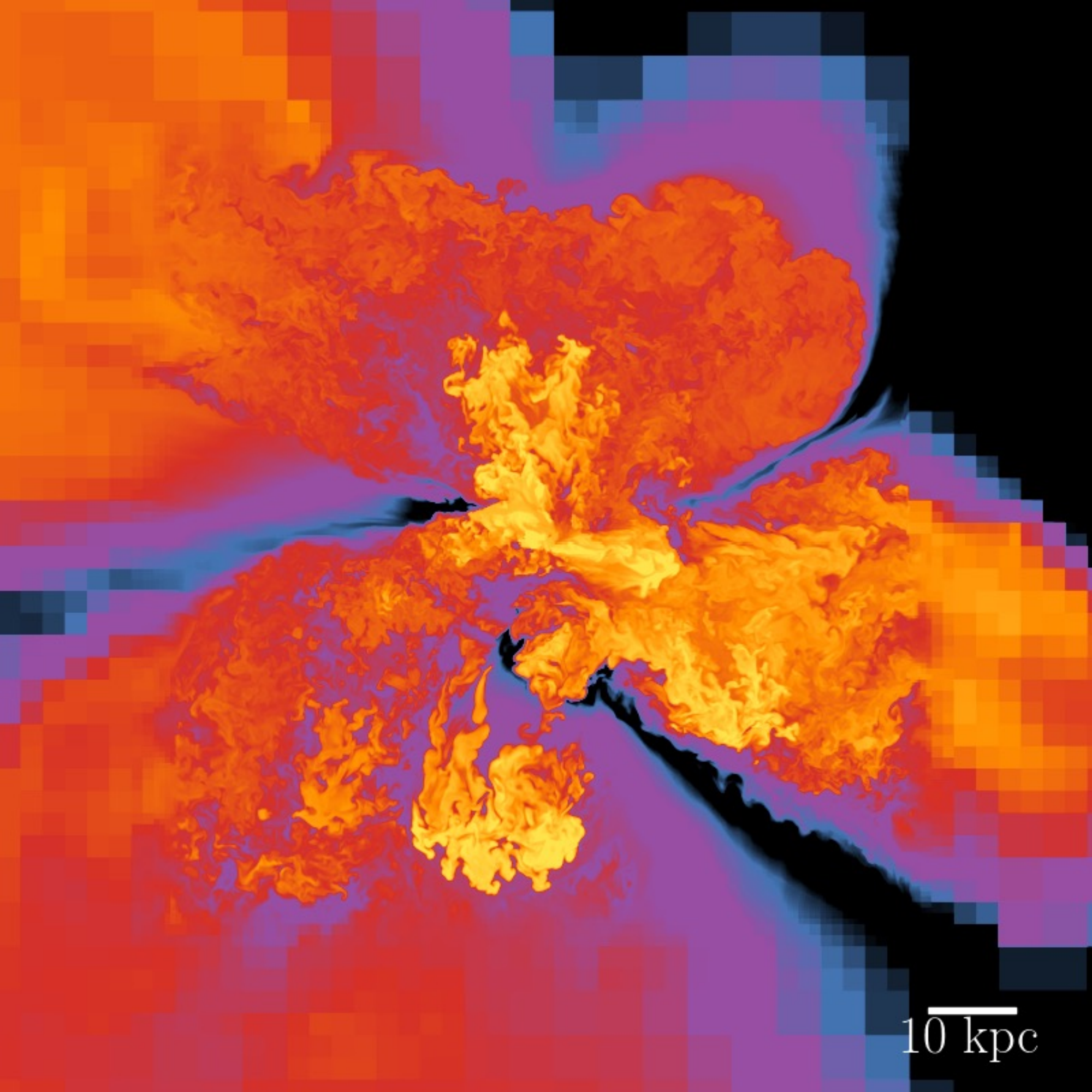}\hfill
    \includegraphics[height=0.45\textwidth]{metallicity_colorbar.pdf}
    \caption{Slices of density ({\em top}), temperature ({\em middle}), and metallicity ({\em bottom}) through central halo in the 157\,pc ({\em left}) and 78\,pc ({\em right}) resolution simulations at $z=2.5$.
    \label{fig:app-physical-slices}}
\end{figure*}

\begin{figure*}[htb]
\centering
    \includegraphics[angle=90,trim={7cm 1cm 10.5cm 1cm}, clip,height=0.142\textheight]{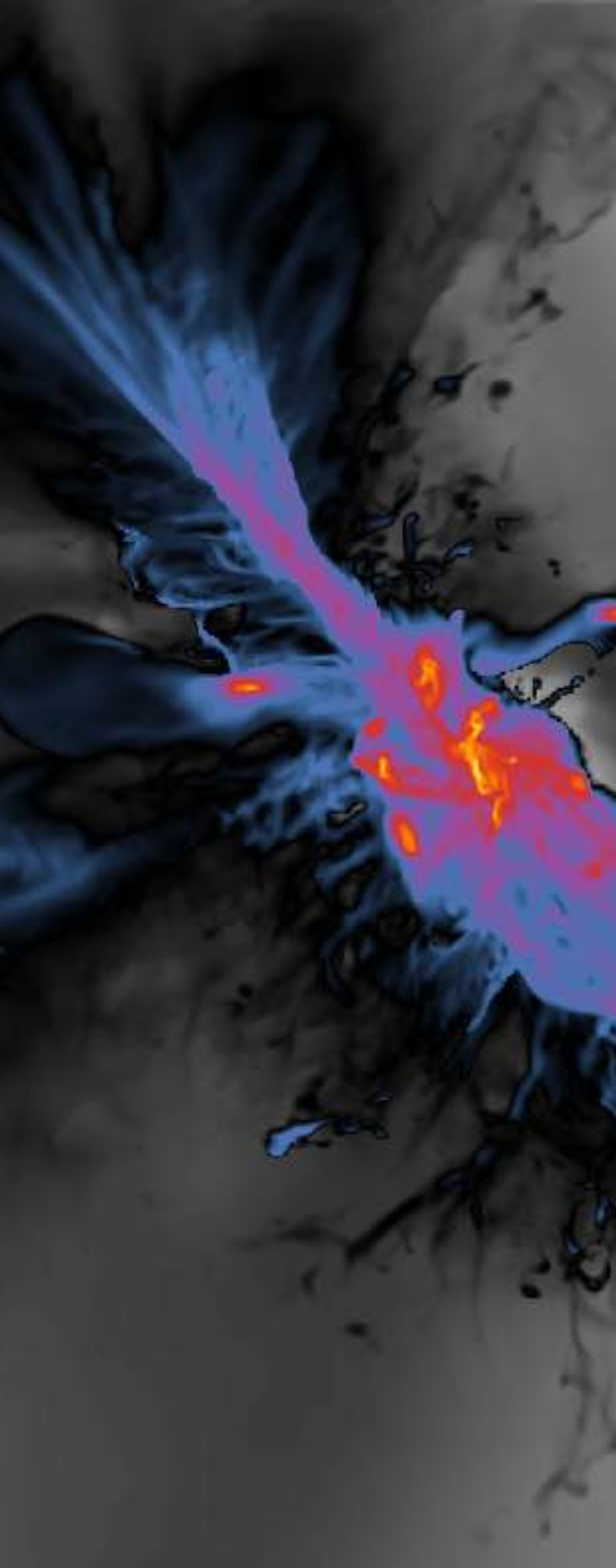}\hfill
    \includegraphics[angle=90,trim={7cm 1cm 10.5cm 1cm} ,clip,height=0.142\textheight]{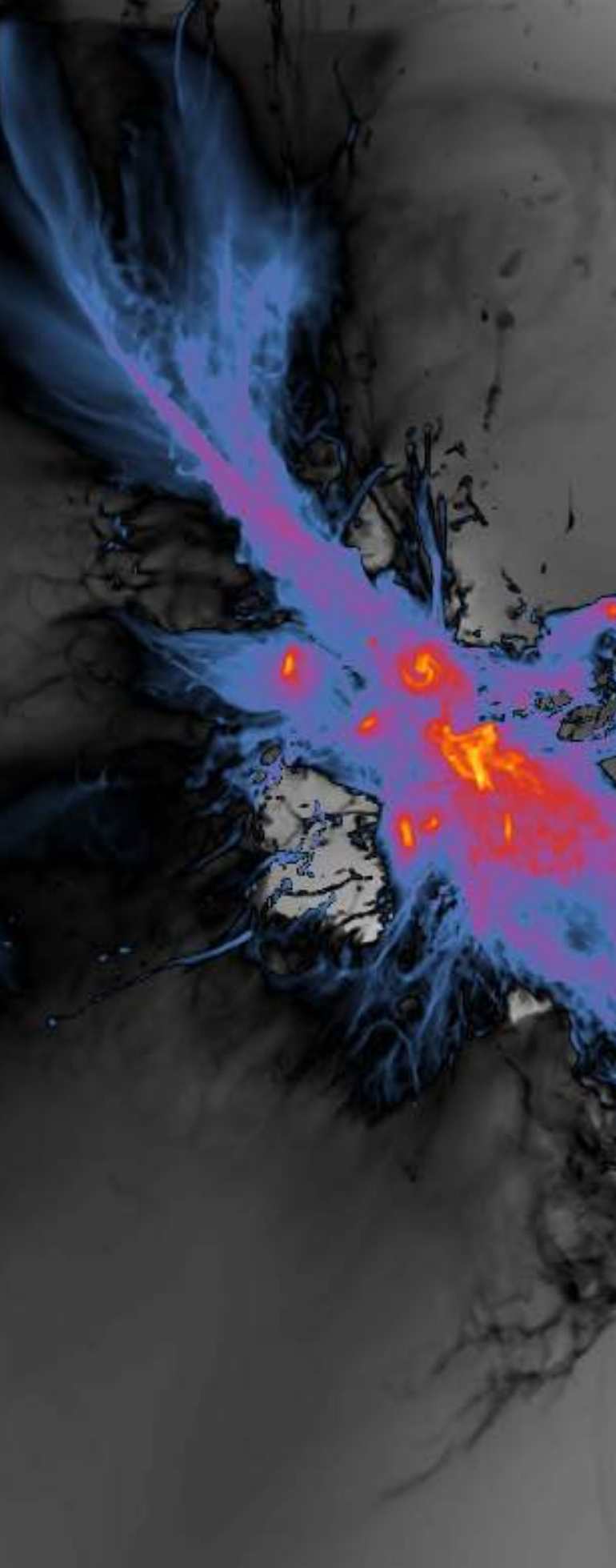}\hfill
    \includegraphics[height=0.142\textheight]{h1_colorbar_small.pdf}\\
    \includegraphics[angle=90,trim={7cm 1cm 10.5cm 1cm}, clip,height=0.142\textheight]{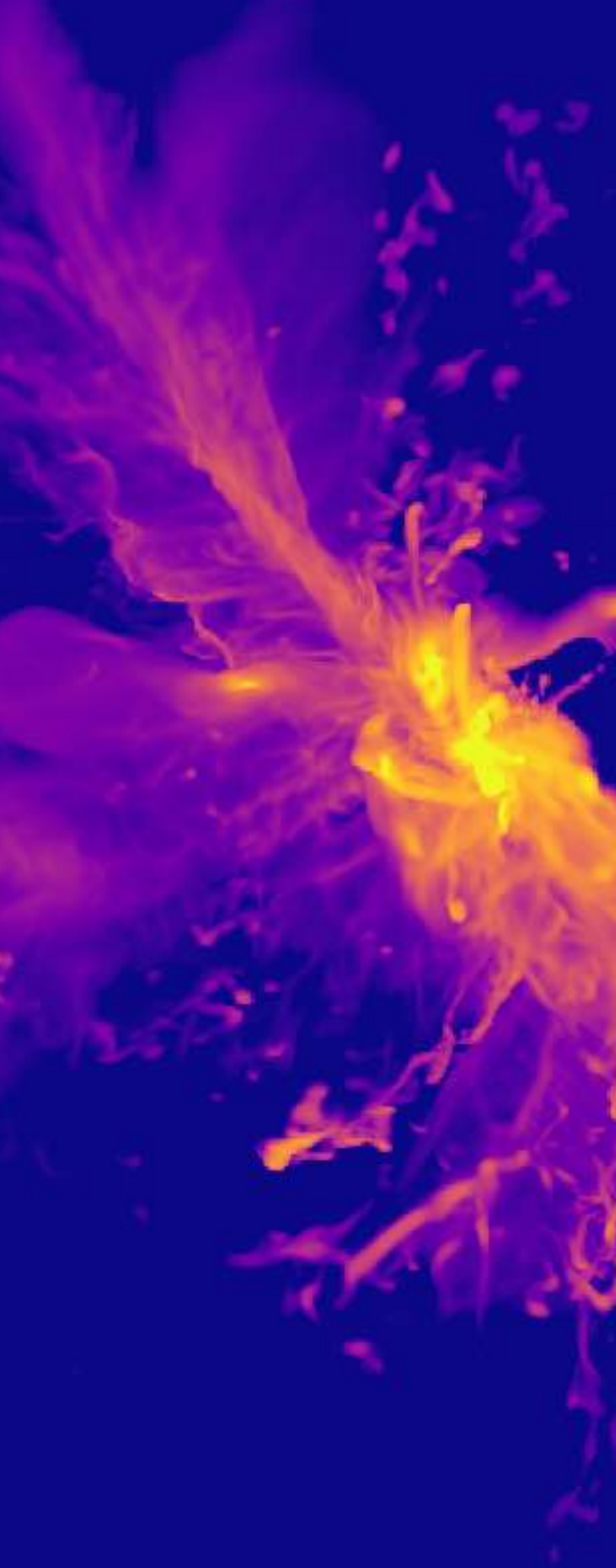}\hfill
    \includegraphics[angle=90,trim={7cm 1cm 10.5cm 1cm} ,clip,height=0.142\textheight]{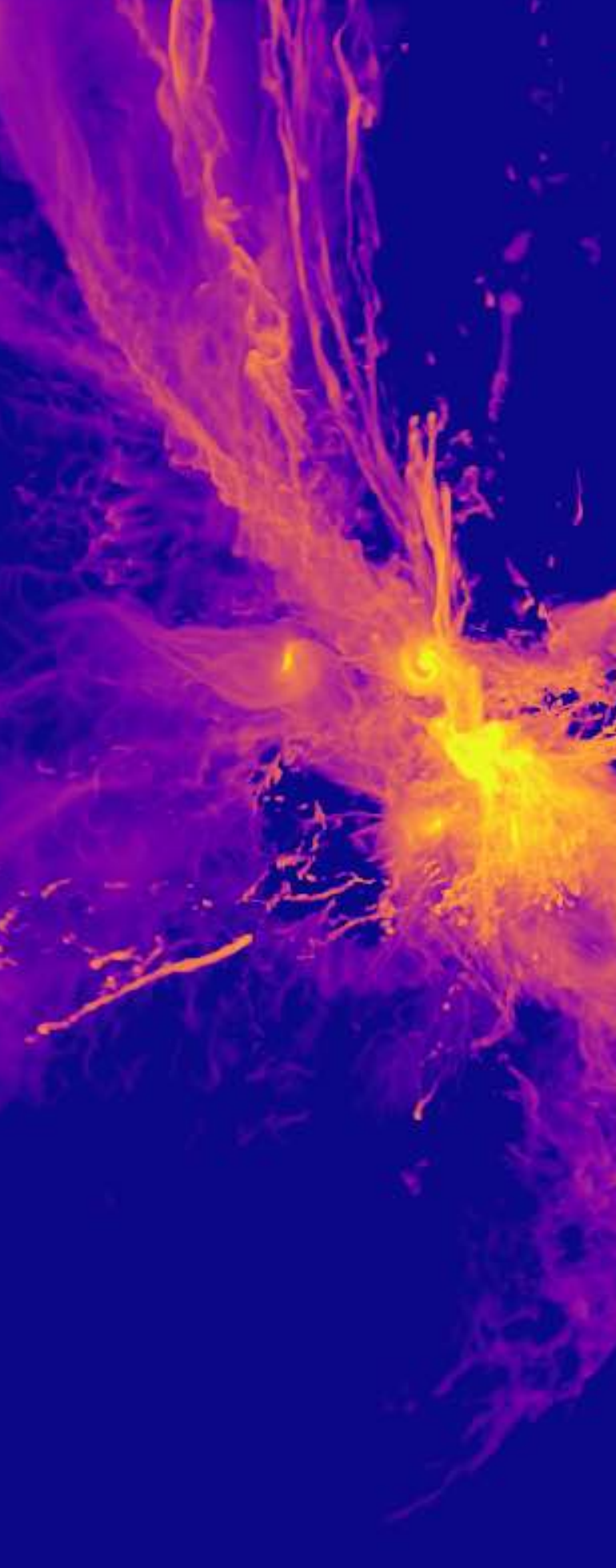}\hfill
    \includegraphics[height=0.142\textheight]{si2_colorbar_small.pdf}\\
    \includegraphics[angle=90,trim={7cm 1cm 10.5cm 1cm}, clip,height=0.142\textheight]{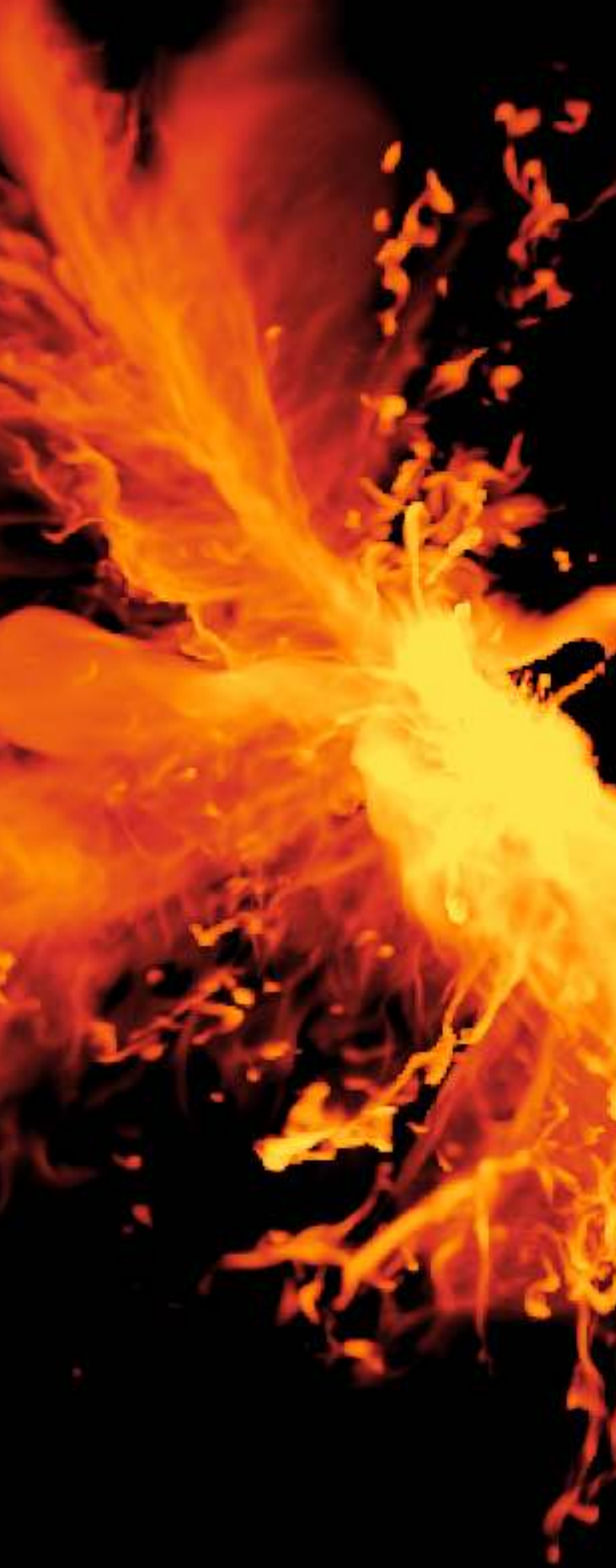}\hfill
    \includegraphics[angle=90,trim={7cm 1cm 10.5cm 1cm} ,clip,height=0.142\textheight]{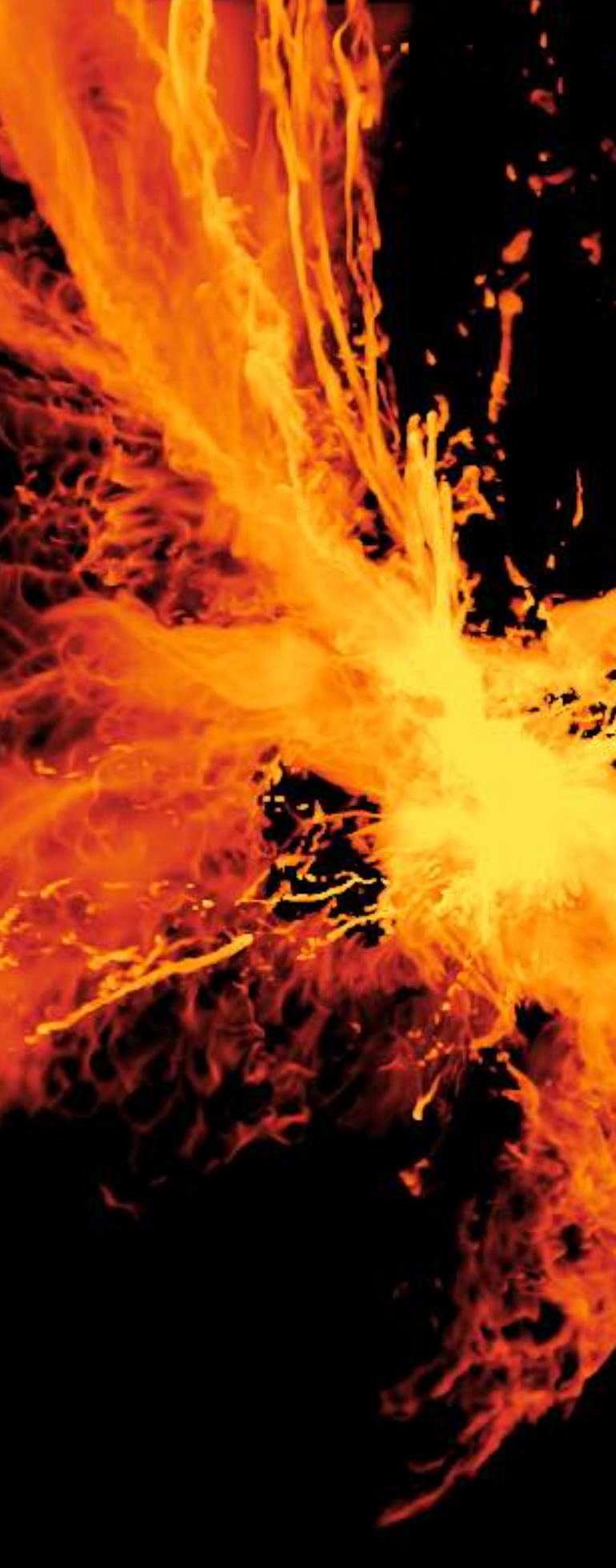}\hfill
    \includegraphics[height=0.142\textheight]{si3_colorbar_small.pdf}\\
    \includegraphics[angle=90,trim={7cm 1cm 10.5cm 1cm}, clip,height=0.142\textheight]{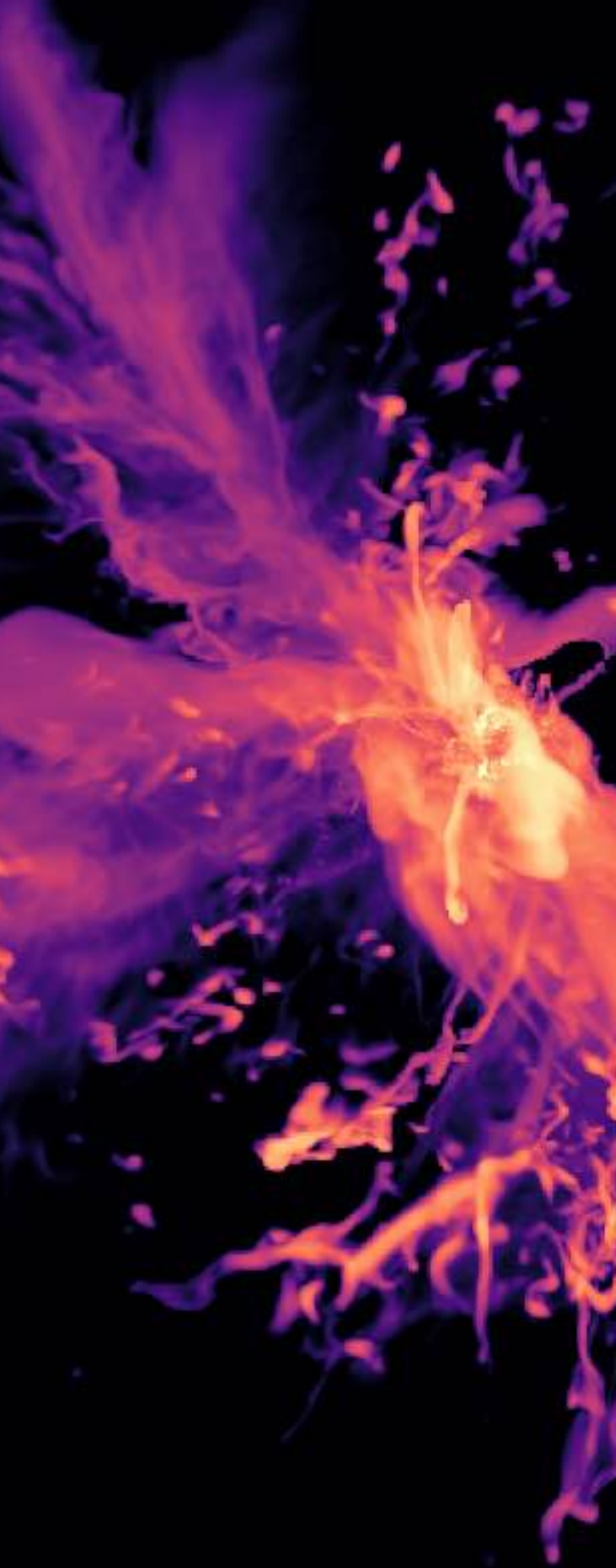}\hfill
    \includegraphics[angle=90,trim={7cm 1cm 10.5cm 1cm} ,clip,height=0.142\textheight]{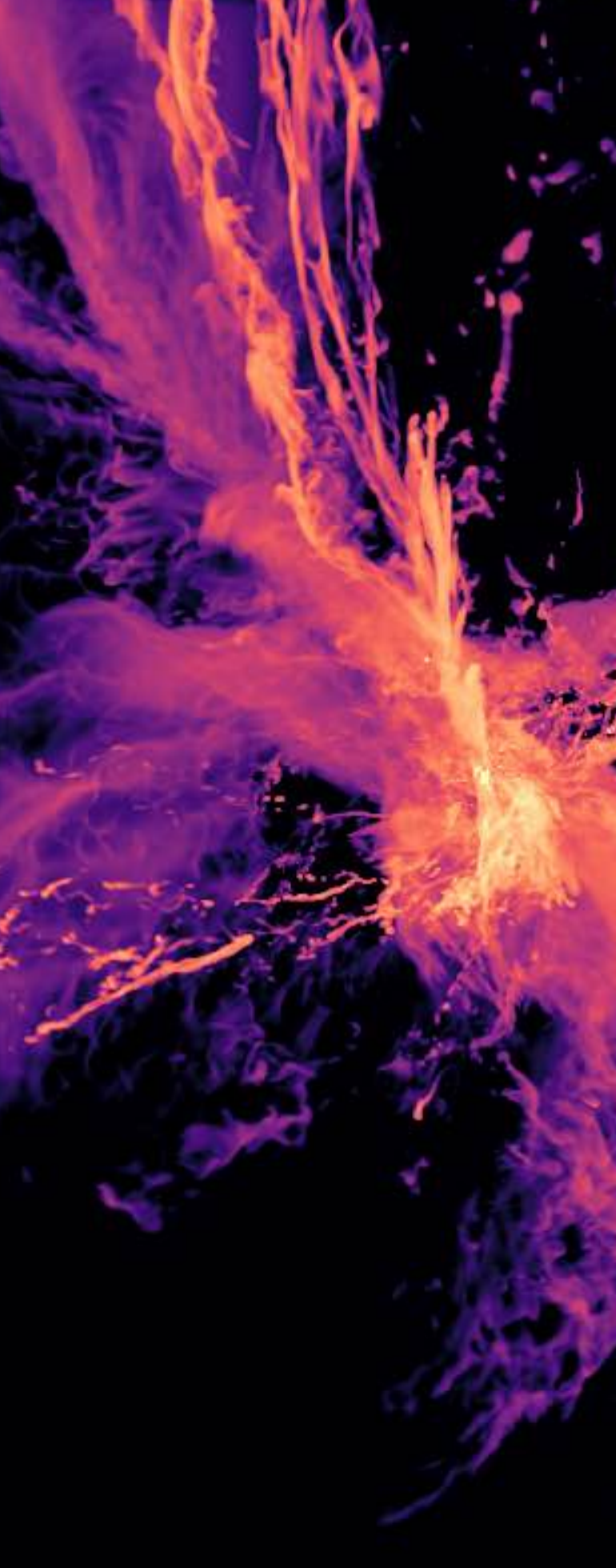}\hfill
    \includegraphics[height=0.142\textheight]{si4_colorbar_small.pdf}\\
    \includegraphics[angle=90,trim={7cm 1cm 10.5cm 1cm}, clip,height=0.142\textheight]{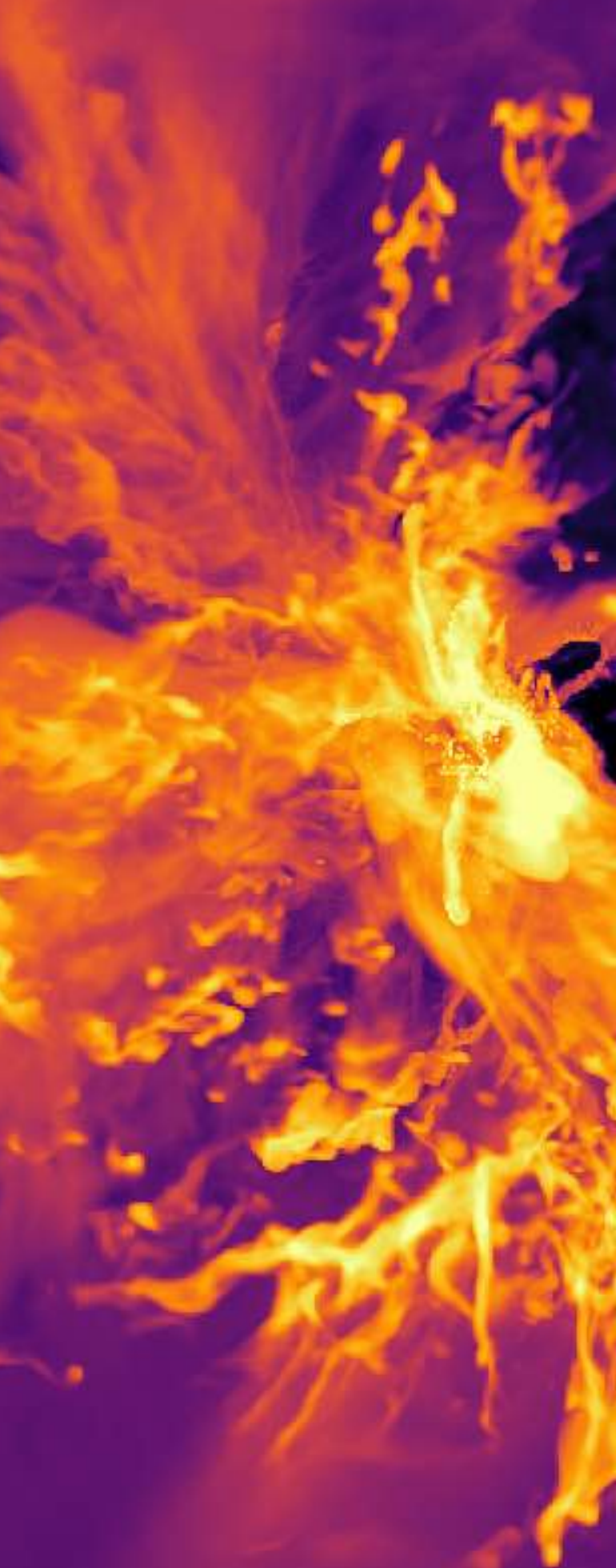}\hfill
    \includegraphics[angle=90,trim={7cm 1cm 10.5cm 1cm} ,clip,height=0.142\textheight]{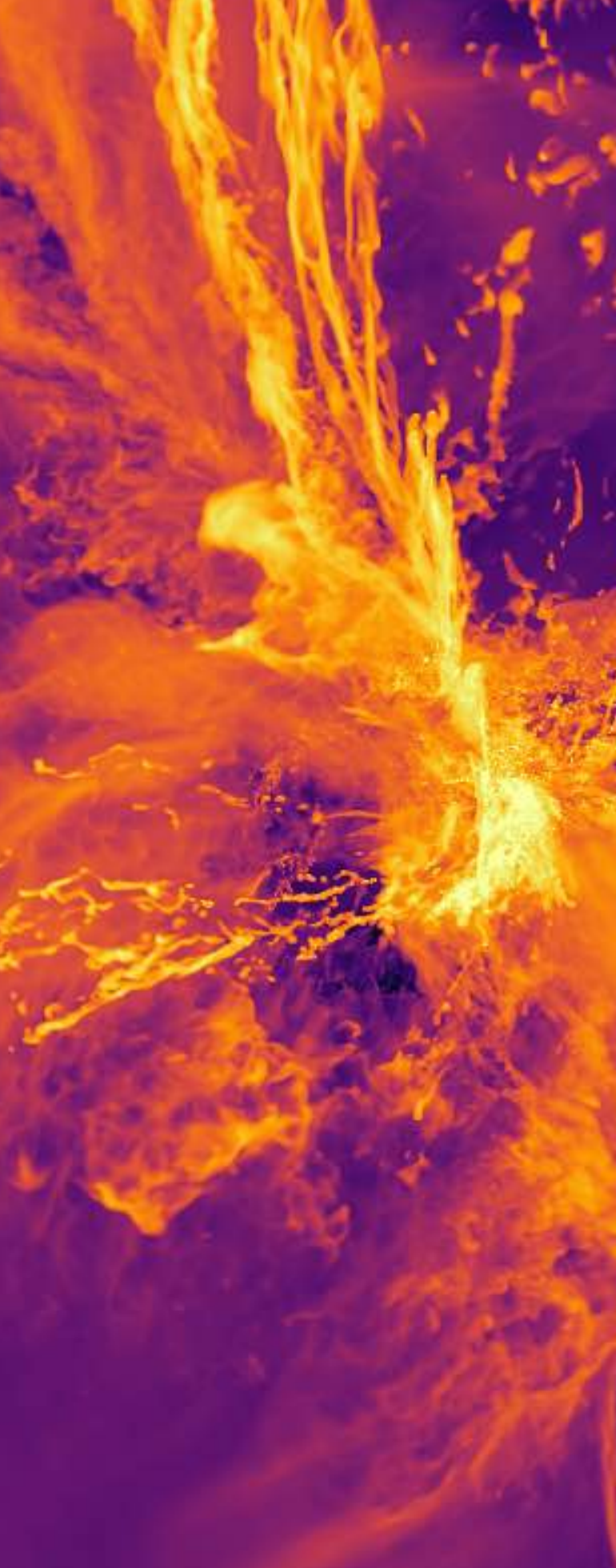}\hfill
    \includegraphics[height=0.142\textheight]{c4_colorbar_small.pdf}\\
    \includegraphics[angle=90,trim={7cm 1cm 10.5cm 1cm}, clip,height=0.142\textheight]{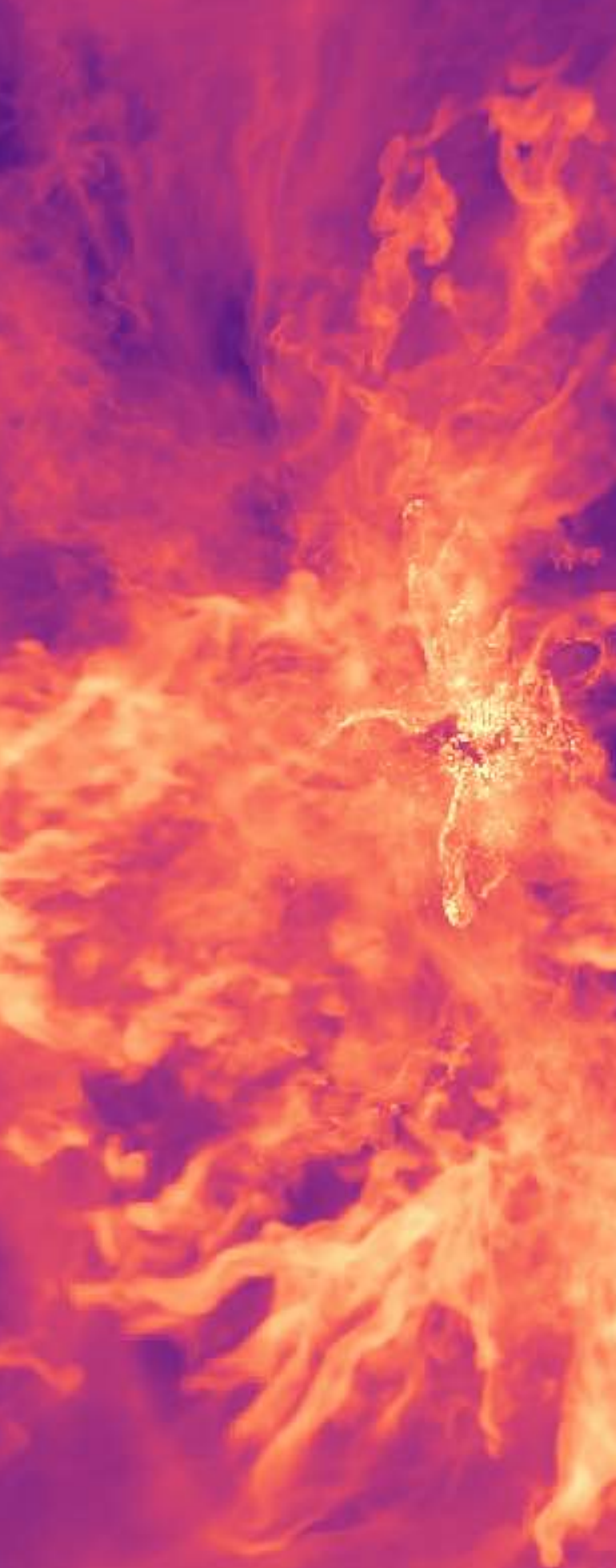}\hfill
    \includegraphics[angle=90,trim={7cm 1cm 10.5cm 1cm} ,clip,height=0.142\textheight]{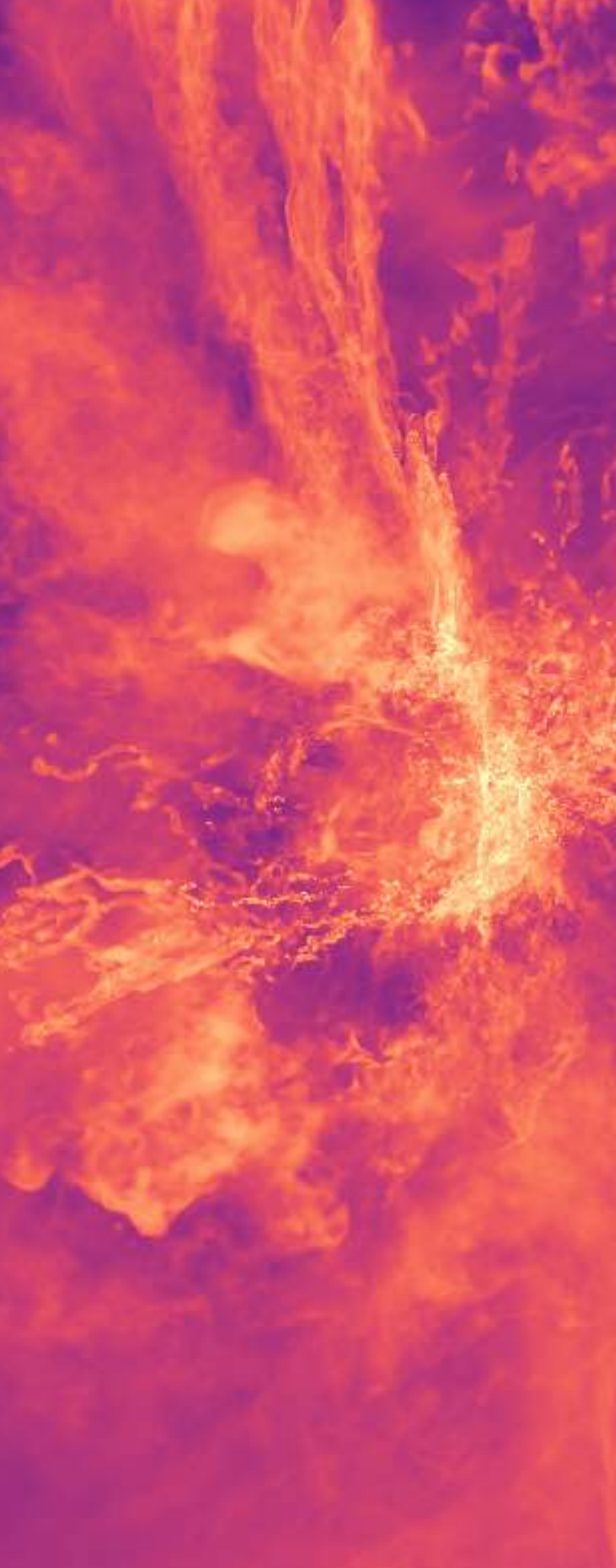}\hfill
    \includegraphics[height=0.142\textheight]{o6_colorbar_small.pdf}
    \caption{\hi, \siii, \siiii, \siiv, \civ, and \ovi\ projections at $z=2.5$ in the 157\,pc- ({\em left}) and 78\,pc-resolution ({\em right}) simulations. Each panel is 200\ckpch\ ($\sim 82$\,pkpc) across and deep. The central galaxy is in the top-center of each panel. The colormap for \hi\ is chosen such that Lyman-limit gas is black/blue and DLA gas is red/orange, with optically-thin gas in greyscale. Both the bulk and detailed properties of these two simulations are much more similar than the 157\,pc and standard resolution simulations are (Figure~\ref{fig:columns}).
    \label{fig:columnsapp}}
\end{figure*}

\begin{figure*}[thb]
\centering
    \includegraphics[width=0.475\textwidth]{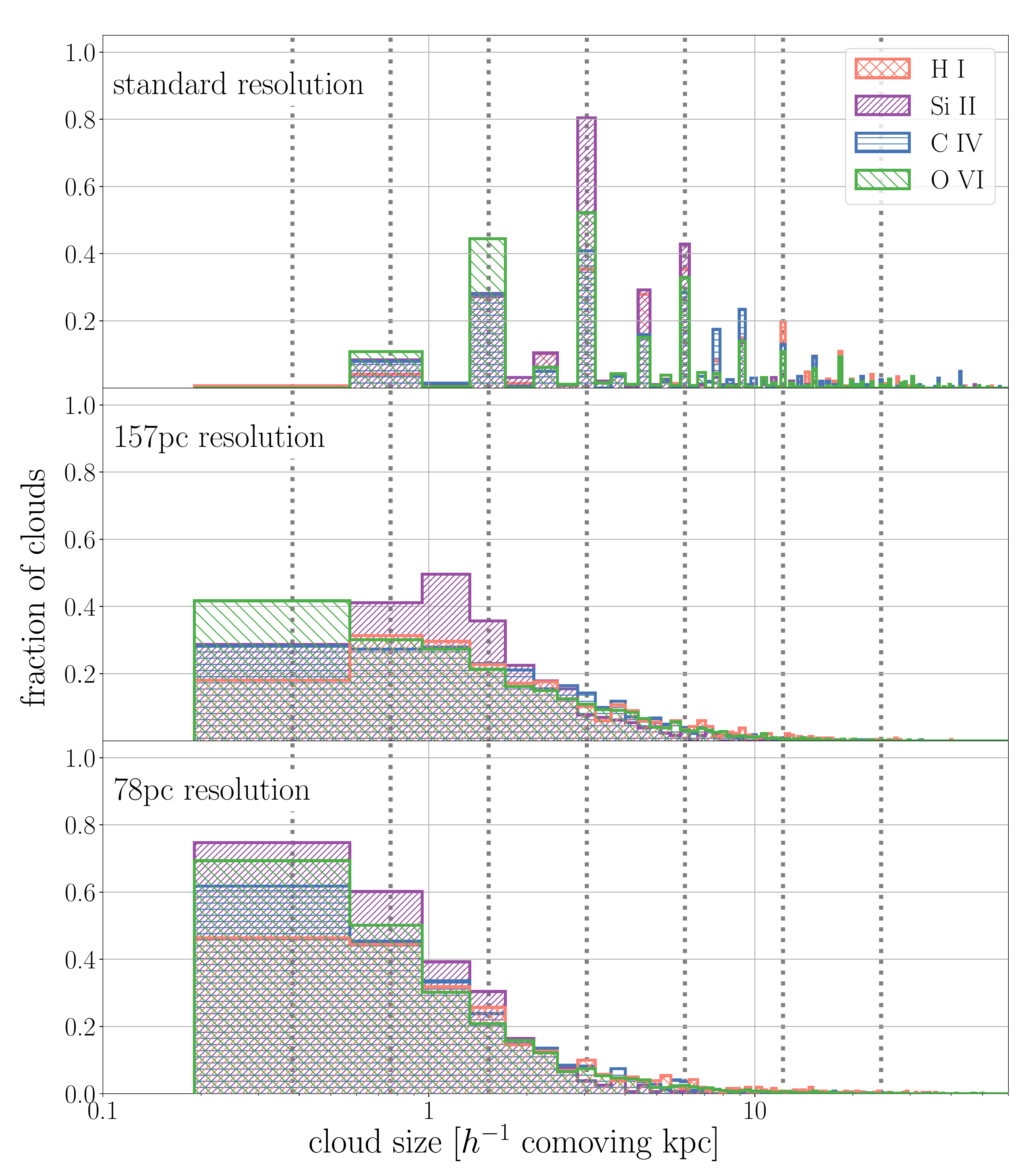}\hfill
    \includegraphics[width=0.475\textwidth]{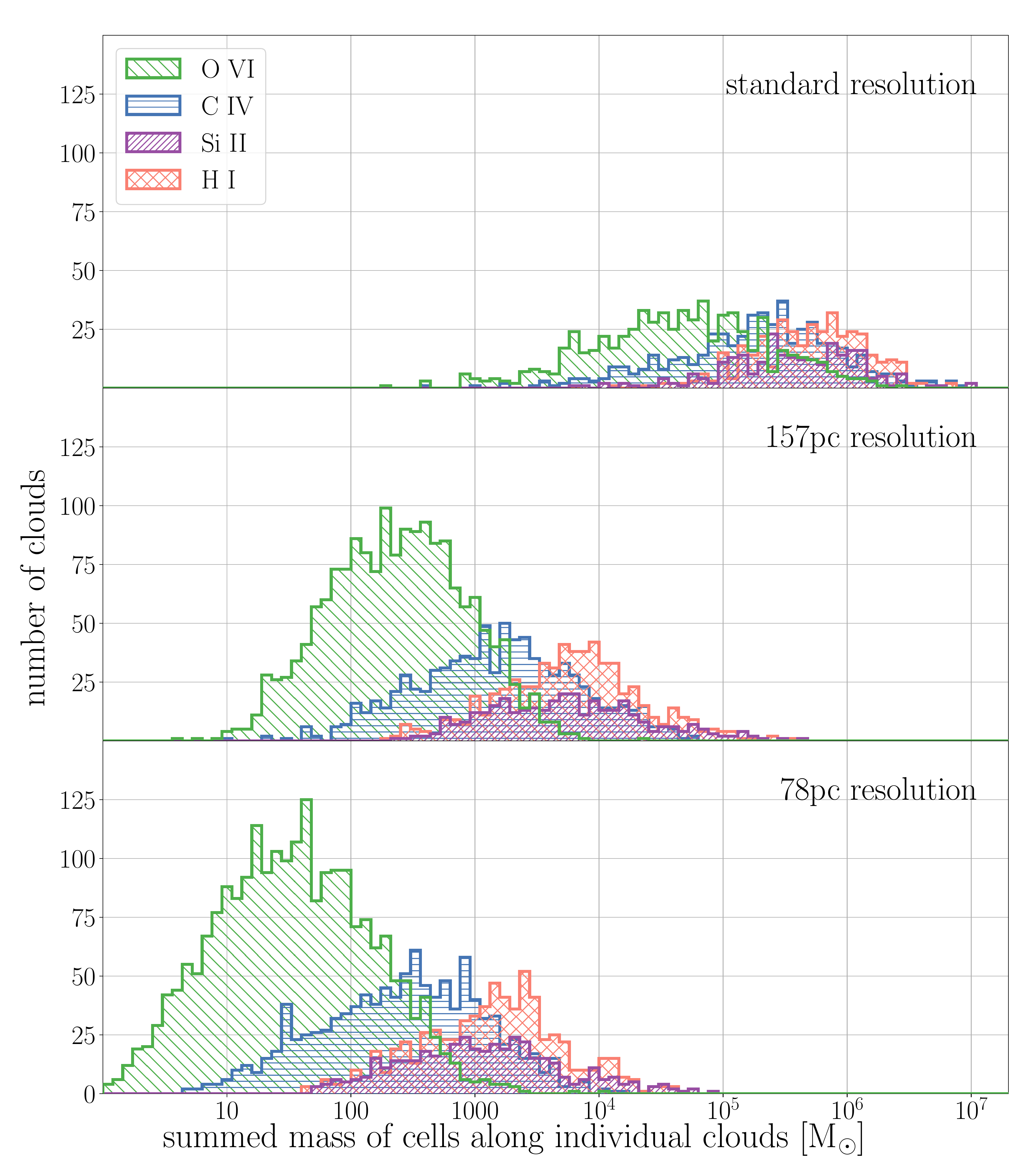}
    \caption{Distributions of 1D cloud sizes ({\em left}) and masses ({\em right}) for \hi\ (pink), \siii\ (purple), \civ\ (blue) and \ovi\ (green) in the standard- ({\em top}) and 157\,pc- ({\em middle}), and 78\,pc-resolution ({\em bottom}) simulations at $z=2.5$.
    \label{fig:appclouds}}
\end{figure*}

\begin{figure*}[hbt]
\centering
    \includegraphics[width=0.475\textwidth]{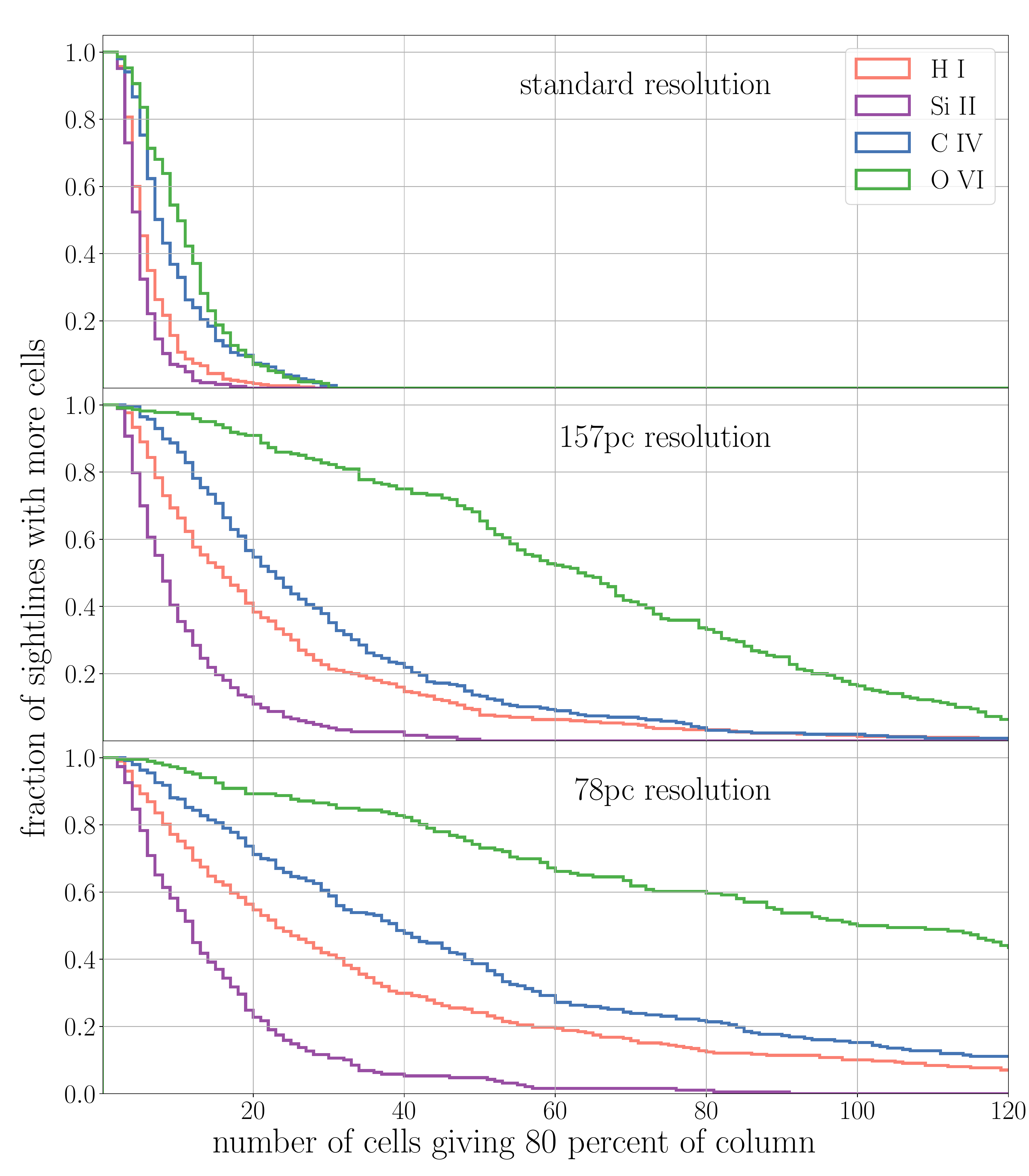}\hfill
    \includegraphics[width=0.475\textwidth]{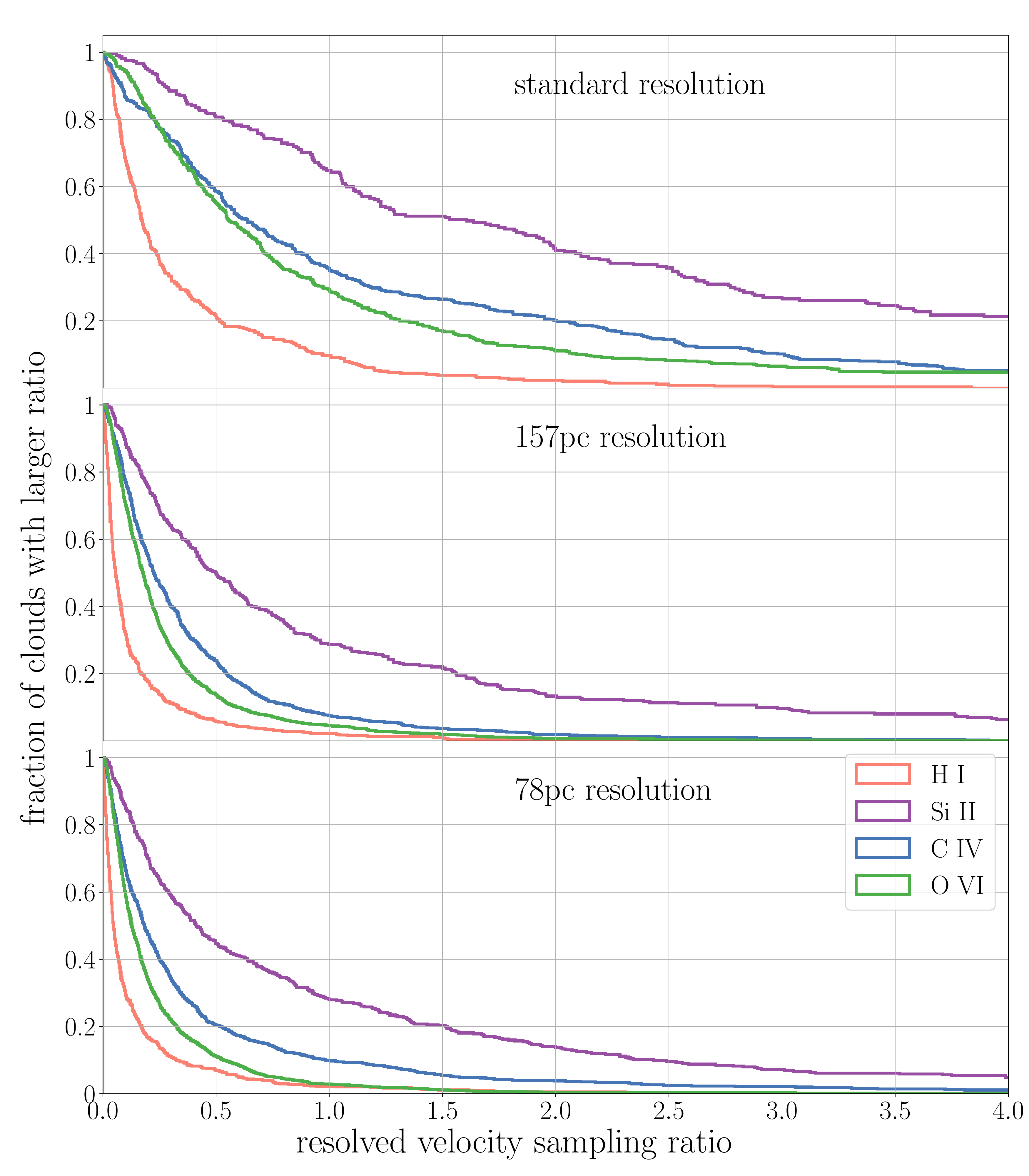}
    \caption{Distributions of number of cells in 1D clouds ({\em left}) and the resolved velocity sampling ratio ({\em right}) for \hi\ (pink), \siii\ (purple), \civ\ (blue) and \ovi\ (green) in the standard- ({\em top}) and 157\,pc- ({\em middle}), and 78\,pc-resolution ({\em bottom}) simulations at $z=2.5$. Note that for the left panels, the $x$-axis has been extended by a factor of two compared to Figure~\ref{fig:cloudcellshist}.
    \label{fig:appcloudstwo}}
\end{figure*}

Figure~\ref{fig:app-physical-slices} shows slices of density, temperature, and metallicity and
Figure~\ref{fig:columnsapp} shows projected column densities at $z=2.5$ for both the 78\,pc and 157\,pc resolution simulations. These large-scale structures are much more similar to one another than to the standard resolution simulation (Figure~\ref{fig:columns}), though finer structure, smaller-scale turbulence, and more well-defined bubbles are visible in the 78\,pc resolution projections. 
We give the total masses of stars, gas, and the ions of interest within the forced refinement volume at $z=2.5$ in Table~\ref{tbl:appmass}; these values are more converged for the high-resolution simulations than for the standard resolution simulation. 
Following the analysis techniques outlined in \S\,\ref{sec:clouds}, we show the 1D cloud sizes and masses in Figure~\ref{fig:appclouds}; while the clouds are somewhat smaller in the 78\,pc resolution simulation, they are much more similar in both extent and mass to those from the 157\,pc resolution simulation than either are to the standard resolution simulation. While the fraction of clouds larger than an $\nref=10$ cell ($\sim 157$\,pc) does not significantly change, many of the smallest clouds fragment to smaller scales, again as expected in a turbulent medium. The masses of the low-ionization clouds do not shift much, but essentially all of the \ovi\ clouds in the 78\,pc resolution simulation have masses $\lesssim 1000$\,\Msun.  Figure~\ref{fig:appcloudstwo} shows how this improved resolution translates to many more cells contributing to 80\% of the column density along the line of sight, but that the combination of these changes do {\em not} translate to a significant change in the {\em kinematic} structure of the absorbing clouds.

\begin{table}
\centering
\begin{tabular}{ ccccr } 
 Species & 78\,pc & 157\,pc & Standard & Ratio\\ 
 \hline
stars & $5.012\times 10^{9}$ &  $4.967\times 10^{9}$ & $6.731\times 10^{9}$ & $1.009$ \\
 all gas & $1.207\times 10^{10}$  & $1.180\times 10^{10}$ & $1.093\times 10^{10}$ & $1.023$\\
\hi\ & $3.527\times 10^{9}$ & $3.431\times 10^{9}$ & $2.528\times 10^{9}$ & $1.028$ \\
\siii\ &  $3.228\times 10^{6}$ & $3.257\times 10^{6}$ & $3.446\times 10^{6}$ & $0.991$ \\ 
\siiv\ &  $1.973\times 10^{4}$ & $1.628\times 10^{4}$ &  $1.497\times 10^{3}$  & $1.212$\\ 
\civ & $2.785\times 10^{4}$  & $2.645\times 10^{4}$ & $5.770\times 10^{4}$ & $1.053$\\
\ovi & $2.674\times 10^{4}$  & $3.416\times 10^{4}$  & $4.788\times 10^{4}$ & $0.783$ \\
\end{tabular} 
\caption{Mass of all gas and different ionic species within the $(200\,{\rm ckpc}/h)^3$ forced refinement volume in the three simulations at $z=2.5$; masses are given in \Msun. The rightmost column gives the ratio of the mass in the 78\,pc simulation to that in the 157\,pc resolution simulation. Relative to the changes from the standard resolution simulation to the forced refinement 157\,pc simulation, the 78\,pc simulation has very small differences in these bulk properties, though overall all three simulations have more similar masses than the standard- and high-resolution simulations do at $z=2$ (Table~\ref{tbl:mass}).} \label{tbl:appmass}
\end{table}

\section{KODIAQ data Voigt Profile Fits}\label{app:kodiaq}
We provide in Table~\ref{tbl:kodiaq} the Voigt profile fits to the KODIAQ data as described in \S\,\ref{sec:kodiaq}. For each absorber, we give the redshift, \hi\ column density, and 1-$\sigma$ logarithmic uncertainty on the \hi\ column density. Where available, we also include the fitted absorber metallicity [X/H] with the upper- and lower-  1-$\sigma$ logarithmic uncertainties; the metallicity determinations are described in detail in \citet{lehner14} and \citet{lehner16}. For each metal species fitted component $i$, we fit provide its velocity offset $v_i$ relative to the \hi\ redshift and uncertainty $\sigma_{v_i}$, the Voigt $b$-parameter and uncertainty $\sigma_{b_i}$, and component column density $\log N_i$ and uncertainty $\sigma_{\log N_i}$. All velocities are giving in \kms\ and column densities in cm$^{-2}$. The last column provides the original source for each of the absorbers \citep{lehner14,lehner16,burns14}.

\begin{deluxetable}{ccccccclcrcccccc}[hp]
\tablewidth{\textwidth}
\tabletypesize{\footnotesize}

 \tablecaption{ Fitted KODIAQ Components\label{tbl:kodiaq}}
\tablehead{
\colhead{Target} & \colhead{$z_{\rm abs}$} & \colhead{$\log N_{\rm HI}$} & 
\colhead{$\sigma_{\log N_{\rm H\,I}}$} & \colhead{[X/H]} & 
\colhead{$\sigma^{\rm up}_{\rm metal}$} &  \colhead{$\sigma^{\rm low}_{\rm metal}$} &
\colhead{Ion}  & \colhead{$i$}  &  \colhead{$v_i$} &  \colhead{$\sigma_{v_i}$} &
\colhead{$b_i$} & \colhead{$\sigma_{b_i}$} & \colhead{$\log N_i$} &   \colhead{$\sigma_{\log N_i}$} & \colhead{Ref}
}
\startdata
Q1009+2956 & $2.42903$ & $17.75$ & $0.15$ & $-2.10$ & $0.20$ & $0.20$ & \siii & $1$ & $-31.7$ & $0.6$ & $11.5$ & $0.6$ & $12.28$ & $0.03$ & L14\\
Q1009+2956 & $2.42903$ & $17.75$ & $0.15$ & $-2.10$ & $0.20$ & $0.20$ & \siii & $2$ & $-19.9$ & $0.2$ & $5.2$ & $0.4$ & $11.85$ & $0.07$ & L14\\
Q1009+2956 & $2.42903$ & $17.75$ & $0.15$ & $-2.10$ & $0.20$ & $0.20$ & \siii & $3$ & $-2.1$ & $0.1$ & $4.3$ & $0.1$ & $12.48$ & $0.01$ & L14\\
Q1009+2956 & $2.42903$ & $17.75$ & $0.15$ & $-2.10$ & $0.20$ & $0.20$ & \siii & $4$ & $25.2$ & $0.9$ & $18.7$ & $1.4$ & $11.79$ & $0.02$ & L14\\
Q1009+2956 & $2.42903$ & $17.75$ & $0.15$ & $-2.10$ & $0.20$ & $0.20$ & \siiv & $1$ & $-166.6$ & $0.3$ & $7.4$ & $0.4$ & $12.05$ & $0.02$ & L14\\
Q1009+2956 & $2.42903$ & $17.75$ & $0.15$ & $-2.10$ & $0.20$ & $0.20$ & \siiv & $2$ & $-40.9$ & $0.3$ & $5.6$ & $0.3$ & $12.75$ & $0.05$ & L14\\
Q1009+2956 & $2.42903$ & $17.75$ & $0.15$ & $-2.10$ & $0.20$ & $0.20$ & \siiv & $3$ & $-28.2$ & $0.3$ & $9.4$ & $0.3$ & $13.30$ & $0.01$ & L14\\
Q1009+2956 & $2.42903$ & $17.75$ & $0.15$ & $-2.10$ & $0.20$ & $0.20$ & \siiv & $4$ & $-2.8$ & $0.1$ & $4.2$ & $0.1$ & $13.29$ & $0.01$ & L14\\
Q1009+2956 & $2.42903$ & $17.75$ & $0.15$ & $-2.10$ & $0.20$ & $0.20$ & \siiv & $5$ & $7.0$ & $0.5$ & $5.3$ & $1.5$ & $11.99$ & $0.07$ & L14\\
Q1009+2956 & $2.42903$ & $17.75$ & $0.15$ & $-2.10$ & $0.20$ & $0.20$ & \siiv & $6$ & $30.3$ & $0.1$ & $10.9$ & $0.1$ & $12.88$ & $0.01$ & L14\\
Q1009+2956 & $2.42903$ & $17.75$ & $0.15$ & $-2.10$ & $0.20$ & $0.20$ & \civ & $1$ & $-166.6$ & $0.1$ & $7.7$ & $0.2$ & $12.79$ & $0.01$ & L14\\
Q1009+2956 & $2.42903$ & $17.75$ & $0.15$ & $-2.10$ & $0.20$ & $0.20$ & \civ & $2$ & $-30.8$ & $0.1$ & $12.5$ & $0.1$ & $13.54$ & $0.01$ & L14\\
Q1009+2956 & $2.42903$ & $17.75$ & $0.15$ & $-2.10$ & $0.20$ & $0.20$ & \civ & $3$ & $-2.6$ & $0.1$ & $5.6$ & $0.1$ & $13.34$ & $0.01$ & L14\\
Q1009+2956 & $2.42903$ & $17.75$ & $0.15$ & $-2.10$ & $0.20$ & $0.20$ & \civ & $4$ & $29.9$ & $0.1$ & $10.4$ & $0.1$ & $13.29$ & $0.01$ & L14\\
Q1009+2956 & $2.42903$ & $17.75$ & $0.15$ & $-2.10$ & $0.20$ & $0.20$ & \ovi & $1$ & $-89.8$ & $1.8$ & $20.6$ & $2.7$ & $13.08$ & $0.06$ & L14\\
Q1009+2956 & $2.42903$ & $17.75$ & $0.15$ & $-2.10$ & $0.20$ & $0.20$ & \ovi & $2$ & $-70.8$ & $0.3$ & $6.0$ & $0.6$ & $12.77$ & $0.07$ & L14\\
Q1009+2956 & $2.42903$ & $17.75$ & $0.15$ & $-2.10$ & $0.20$ & $0.20$ & \ovi & $3$ & $-18.5$ & $0.8$ & $20.9$ & $2.7$ & $13.12$ & $0.13$ & L14\\
Q1009+2956 & $2.42903$ & $17.75$ & $0.15$ & $-2.10$ & $0.20$ & $0.20$ & \ovi & $4$ & $23.3$ & $3.5$ & $50.7$ & $3.8$ & $13.73$ & $0.04$ & L14\\
Q1009+2956 & $2.42903$ & $17.75$ & $0.15$ & $-2.10$ & $0.20$ & $0.20$ & \ovi & $5$ & $151.2$ & $1.9$ & $44.1$ & $3.4$ & $13.45$ & $0.03$ & L14\\\hline
J1343+5721 & $2.83437$ & $17.78$ & $0.20$ & $-0.60$ & $0.20$ & $0.20$ & \siii & $1$ & $-61.3$ & $1.2$ & $10.5$ & $1.8$ & $12.58$ & $0.06$ & L14\\
J1343+5721 & $2.83437$ & $17.78$ & $0.20$ & $-0.60$ & $0.20$ & $0.20$ & \siii & $2$ & $-0.7$ & $0.3$ & $9.6$ & $0.4$ & $13.44$ & $0.02$ & L14\\
J1343+5721 & $2.83437$ & $17.78$ & $0.20$ & $-0.60$ & $0.20$ & $0.20$ & \siiv & $1$ & $-136.2$ & $1.3$ & $7.6$ & $3.2$ & $12.74$ & $0.29$ & L14\\
J1343+5721 & $2.83437$ & $17.78$ & $0.20$ & $-0.60$ & $0.20$ & $0.20$ & \siiv & $2$ & $-113.2$ & $13.8$ & $32.0$ & $24.9$ & $13.07$ & $0.33$ & L14\\
J1343+5721 & $2.83437$ & $17.78$ & $0.20$ & $-0.60$ & $0.20$ & $0.20$ & \siiv & $3$ & $-67.4$ & $1.3$ & $17.0$ & $1.5$ & $13.52$ & $0.06$ & L14\\
J1343+5721 & $2.83437$ & $17.78$ & $0.20$ & $-0.60$ & $0.20$ & $0.20$ & \siiv & $4$ & $-12.3$ & $1.6$ & $16.2$ & $2.3$ & $12.86$ & $0.05$ & L14\\
J1343+5721 & $2.83437$ & $17.78$ & $0.20$ & $-0.60$ & $0.20$ & $0.20$ & \civ & $1$ & $-132.5$ & $1.3$ & $17.3$ & $1.9$ & $13.41$ & $0.04$ & L14\\
J1343+5721 & $2.83437$ & $17.78$ & $0.20$ & $-0.60$ & $0.20$ & $0.20$ & \civ & $2$ & $-70.9$ & $0.9$ & $25.8$ & $1.4$ & $14.06$ & $0.02$ & L14\\
J1343+5721 & $2.83437$ & $17.78$ & $0.20$ & $-0.60$ & $0.20$ & $0.20$ & \civ & $3$ & $-12.5$ & $3.5$ & $17.1$ & $5.0$ & $12.95$ & $0.11$ & L14\\
J1343+5721 & $2.83437$ & $17.78$ & $0.20$ & $-0.60$ & $0.20$ & $0.20$ & \ovi & $1$ & $-125.0$ & $4.5$ & $25.2$ & $8.3$ & $13.64$ & $0.15$ & L14\\
J1343+5721 & $2.83437$ & $17.78$ & $0.20$ & $-0.60$ & $0.20$ & $0.20$ & \ovi & $2$ & $-31.2$ & $2.0$ & $58.7$ & $3.0$ & $14.68$ & $0.02$ & L14\\\hline
SDSSJ1023+5142 & $3.10586$ & $19.85$ & $0.15$ & $-1.62$ & $0.11$ & $0.15$ & \siii & $1$ & $-2.2$ & $0.2$ & $7.5$ & $0.2$ & $13.32$ & $0.01$ & L14\\
SDSSJ1023+5142 & $3.10586$ & $19.85$ & $0.15$ & $-1.62$ & $0.11$ & $0.15$ & \siii & $2$ & $17.3$ & $0.3$ & $8.7$ & $0.4$ & $13.09$ & $0.02$ & L14\\\hline
J143316+313126 & $2.90116$ & $16.16$ & $0.01$ & $-1.80$ & $0.15$ & $0.15$ & \siii & $1$ & $2.9$ & $2.4$ & $10.2$ & $3.8$ & $11.27$ & $0.11$ & L16\\\hline
J172409+531405 & $2.48778$ & $16.20$ & $0.03$ & $+0.20$ & $0.10$ & $0.10$ & \siii & $1$ & $-13.0$ & $0.2$ & $6.1$ & $0.3$ & $12.38$ & $0.02$ & L16\\
J172409+531405 & $2.48778$ & $16.20$ & $0.03$ & $+0.20$ & $0.10$ & $0.10$ & \siii & $2$ & $-0.1$ & $0.1$ & $2.9$ & $0.2$ & $12.49$ & $0.02$ & L16\\
J172409+531405 & $2.48778$ & $16.20$ & $0.03$ & $+0.20$ & $0.10$ & $0.10$ & \siii & $3$ & $10.7$ & $0.3$ & $7.0$ & $0.4$ & $12.49$ & $0.02$ & L16\\
J172409+531405 & $2.48778$ & $16.20$ & $0.03$ & $+0.20$ & $0.10$ & $0.10$ & \siii & $4$ & $49.4$ & $0.1$ & $5.0$ & $0.1$ & $12.54$ & $0.01$ & L16\\\hline
Q2126-158 & $2.90731$ & $16.45$ & $0.25$ & $-999$ & $999$ & $999$ & \siii & $1$ & $-23.6$ & $4.2$ & $34.9$ & $7.5$ & $12.23$ & $0.07$ & B14\\
Q2126-158 & $2.90731$ & $16.45$ & $0.25$ & $-999$ & $999$ & $999$ & \siiv & $1$ & $-25.3$ & $1.7$ & $24.7$ & $2.1$ & $12.82$ & $0.03$ & B14\\
Q2126-158 & $2.90731$ & $16.45$ & $0.25$ & $-999$ & $999$ & $999$ & \siiv & $2$ & $-3.6$ & $0.6$ & $5.0$ & $1.4$ & $11.99$ & $0.14$ & B14\\
Q2126-158 & $2.90731$ & $16.45$ & $0.25$ & $-999$ & $999$ & $999$ & \siiv & $3$ & $19.8$ & $1.2$ & $9.0$ & $1.7$ & $12.06$ & $0.08$ & B14\\
Q2126-158 & $2.90731$ & $16.45$ & $0.25$ & $-999$ & $999$ & $999$ & \civ & $1$ & $-87.8$ & $1.5$ & $5.3$ & $2.3$ & $12.02$ & $0.14$ & B14\\
Q2126-158 & $2.90731$ & $16.45$ & $0.25$ & $-999$ & $999$ & $999$ & \civ & $2$ & $-30.9$ & $2.4$ & $14.4$ & $2.4$ & $12.90$ & $0.16$ & B14\\
Q2126-158 & $2.90731$ & $16.45$ & $0.25$ & $-999$ & $999$ & $999$ & \civ & $3$ & $-0.0$ & $2.1$ & $23.6$ & $2.5$ & $13.48$ & $0.04$ & B14\\
Q2126-158 & $2.90731$ & $16.45$ & $0.25$ & $-999$ & $999$ & $999$ & \civ & $4$ & $4.2$ & $0.9$ & $4.2$ & $1.8$ & $12.33$ & $0.19$ & B14\\
Q2126-158 & $2.90731$ & $16.45$ & $0.25$ & $-999$ & $999$ & $999$ & \ovi & $1$ & $-44.1$ & $1.5$ & $15.4$ & $1.9$ & $13.50$ & $0.07$ & B14\\
Q2126-158 & $2.90731$ & $16.45$ & $0.25$ & $-999$ & $999$ & $999$ & \ovi & $2$ & $0.8$ & $1.8$ & $29.0$ & $2.7$ & $13.96$ & $0.03$ & B14\\
\enddata
    \tablecomments{The full dataset is available as an electronic table.
    The references are \citet{lehner14} as L14, \citet{lehner16} as L16, and \citet{burns14} as B14.
    }
\end{deluxetable}

\section{Additional High-Resolution Phase Space Diagrams}\label{app:velphase}
We show here additional examples of the velocity phase-space diagrams introduced in \S\,\ref{sec:physical} and Figures~\ref{fig:velphaseref} and \ref{fig:velphasenat}. Figures~28--33 are sightlines from the high-resolution simulation, highlighting examples of where the additional resolution either increases or decreases the number of apparent spectral components and where the low- and high-ions are aligned or mis-aligned in interesting ways. Figures~34--39 are from the standard resolution simulation, highlighting the effects of coarse or variable resolution and cases where the bulk of the absorption for some of the ions is from individual cells.  {\em These figures are omitted here to fit within the arXiv space limits. The full version of this appendix is included in the published version of the paper or available upon request.}

\end{document}